%% file: prd.tex
\documentclass[twocolumn,showpacs,aps,prd,floatfix]{revtex4}
\usepackage{graphicx}
\usepackage{dcolumn}
\usepackage{amsmath}
\usepackage{epsfig}
\usepackage{multirow}

\graphicspath{{figures/}}

\newcommand{\BABARPubYear}{11}
\newcommand{\BABARPubNumber}{024}
\newcommand{\SLACPubNumber}{14858}

\input babarsym

\long\def\inst#1{\par\nobreak\kern 4pt\nobreak
    {\it #1}\par\vskip 10pt plus 3pt minus 3pt}

\begin{document}

\begin{flushleft}
SLAC-PUB-\SLACPubNumber \\
\babar-PUB-\BABARPubYear/\BABARPubNumber
\end{flushleft}

\title{
\large \bf
\boldmath
Study of \CP violation in Dalitz-plot analyses of \bkkks, \bkkkboth, and 
\bkksks
} 

\input authors_nov2011_bad2400

\date{June 22, 2012}

\begin{abstract}
   \noindent
   \input{abstract}
\end{abstract}


\maketitle

\input{introduction}

\input{overview}

\input{DetectorAndData}
\input{EventSelection}

\input{Likelihood}

\input{ModelDetermination}

\input{FitResults}

\input{Systematics}
\input{summary}

\section{ACKNOWLEDGMENTS}
\label{sec:acknowledgments}
\input pubboard/acknowledgements

\clearpage
\bibliography{prd}
\bibliographystyle{apsrev}


\input{Appendix}

\end{document}

%% file: authors_nov2011_bad2400.tex
\author{J.~P.~Lees}
\author{V.~Poireau}
\author{V.~Tisserand}
\affiliation{Laboratoire d'Annecy-le-Vieux de Physique des Particules (LAPP), Universit\'e de Savoie, CNRS/IN2P3,  F-74941 Annecy-Le-Vieux, France}
\author{J.~Garra~Tico}
\author{E.~Grauges}
\affiliation{Universitat de Barcelona, Facultat de Fisica, Departament ECM, E-08028 Barcelona, Spain }
\author{D.~A.~Milanes$^{a}$}
\author{A.~Palano$^{ab}$ }
\author{M.~Pappagallo$^{ab}$ }
\affiliation{INFN Sezione di Bari$^{a}$; Dipartimento di Fisica, Universit\`a di Bari$^{b}$, I-70126 Bari, Italy }
\author{G.~Eigen}
\author{B.~Stugu}
\affiliation{University of Bergen, Institute of Physics, N-5007 Bergen, Norway }
% ...
\collaboration{The \babar\ Collaboration}
\noaffiliation

%% file: abstract.tex
We perform amplitude analyses of the decays  \bkkks, \bkkkboth, and \bkksks, and 
measure \CP-violating parameters and partial branching fractions.  The results 
are based on a data sample of approximately $470\times 10^6$ \BB decays, 
collected with the \babar\ detector at the PEP-II asymmetric-energy
\B factory at the SLAC National Accelerator Laboratory.  For \bkkkboth, we find 
a direct \CP asymmetry in $\Bp\to\phiI\Kp$ of $\Acp= (12.8\pm 4.4 \pm 1.3)\%$,
which differs from zero by $2.8 \sigma$.  For \bkkks, we measure the \CP-violating
phase $\betaeff(\phiI\KS) = (21\pm 6 \pm 2)^\circ$.  For
\bkksks, we measure an overall direct \CP asymmetry of $\Acp = (4 ^{+4}_{-5} \pm 2)\%$.
We also perform an angular-moment analysis of the three channels, and determine that
the $f_X(1500)$ state can be described well by the sum of the resonances $\fII$, $\ftwop$, and
$\fIII$.

%% file: introduction.tex
\section{INTRODUCTION}
\label{sec:introduction}

In the Standard Model (SM), \CP violation in the quark sector
is entirely described by a single weak phase in the 
CKM quark-mixing matrix.  Studies of time-dependent \CP violation 
in $\Bz\to(c\overline{c})K^0$ decay~\cite{ChargeConj}
have yielded precise measurements~\cite{Aubert:2009yr,Chen:2006nk}
of $\sin{2\beta}$, where
$\beta \equiv {\rm arg}[-(V^*_{cb}V_{cd})/(V^*_{tb}V_{td})]$
and $V_{ij}$ are the elements of the CKM matrix.  
Measurements of time-dependent \CP violation in
$b\to q\overline{q}s\ (q=u,d,s)$ decays offer an alternative method for
measuring $\beta$. Such decays are dominated by $b\to s$ 
loop diagrams, and therefore are sensitive to possible
new physics (NP) contributions appearing in the loops of
these diagrams.  As a result, the effective 
$\beta$ ($\betaeff$) measured in such decays could differ from 
the $\beta$ measured in $\Bz\to(c\overline{c})K^0$.  Deviations
of \betaeff from $\beta$ are also possible in the SM, due
to additional amplitudes from $b \to u$ tree diagrams, loop diagrams 
containing different CKM factors (``$u$-penguins''), and 
final-state interactions.  

The decay mode $\Bz\to\phi\KS$ is particularly suited for 
a NP search, as $\betaeff$ for this mode is expected to be very
near in value to $\beta$ in the SM, with $\sin{2\betaeff} - \sin{2\beta}$ 
in the range $(-0.01 , 0.04)$~\cite{Beneke:2005pu,Cheng:2005bg,Li:2006jv}.
However, the measurement of $\betaeff$ is complicated due to other
$\bkkks$ decays that interfere with $\Bz\to\phi\KS$.  In general,
$\kkks$ is not a \CP eigenstate: the \kkks system
is \CP even (odd) if the $\KpKm$ system has even (odd)
angular momentum. Thus, one must account for the (mostly 
S-wave) $\kkks$ states that interfere with $\phi\KS$. This
can be done by measuring $\betaeff$ 
using a Dalitz-plot (DP) analysis of $\bkkks$.
A further benefit of a DP analysis is that it allows
both $\sin{2\betaeff}$ and  $\cos{2\betaeff}$ to be determined, through
the interference of odd and even partial waves, which eliminates a
trigonometric ambiguity between $\betaeff$ and $90^\circ - \betaeff$.

The related decay mode $\Bp\to\phi\Kp$ is another interesting channel
in which to search for NP.  This decay is also dominated
by a $b \to s$ penguin amplitude, and its direct \CP asymmetry, 
$\Acp$, is predicted to be small in the SM, 
$(0.0 $-$ 4.7) \%$~\cite{Beneke:2003zv,Li:2006jv}, so a significant deviation from zero could be a
signal of NP.

In addition to measuring $\betaeff$ in $\Bz\to\phi\KS$, it is 
possible to measure it for the other resonant
and nonresonant  \bkkks decays.  However, these decays may 
contain a mixture of even and odd partial-waves, so the
final state is not guaranteed to be a \CP eigenstate, thus
posing a challenge to a measurement of $\betaeff$.
A DP analysis can reveal which partial waves are present, thus 
eliminating a source of systematic uncertainty affecting 
the extraction of $\betaeff$,  without 
having to rely on theoretical predictions. 

Previous analyses of \bkkkboth~\cite{Aubert:2006nu,Garmash:2004wa} 
and \bkkks~\cite{Nakahama:2010nj,Aubert:2007sd} have revealed a 
complex
DP structure that is poorly understood.  Both modes 
exhibit a large peak around $m(\Kp\Km) \sim 1500\mevcc$, which has 
been dubbed the $f_X(1500)$.  Both \babar\ and Belle have modeled it 
as a scalar resonance. The recent DP analysis of $\bksksks$ by 
\babar~\cite{Lees:2011nf} does
not yield evidence for this resonance. It is important to clarify 
the properties of the $f_X(1500)$ with a larger data sample, 
and in particular to determine its spin, as that affects the 
$\betaeff$ measurement in \bkkks.  

An additional feature seen in \bkkks and \bkkkboth decays is
a large broad ``nonresonant'' (NR) contribution.  Previous analyses
have found that a uniform-phase-space model is insufficient to describe the NR term,
and have instead parameterized it with an empirical 
model.  The NR term has been taken to be purely $\KpKm$ S-wave in 
\bkkkboth~\cite{Aubert:2006nu,Garmash:2004wa}, 
while smaller $\Kp\KS$ and
 $\Km\KS$ S-wave terms have been seen in 
\bkkks~\cite{Nakahama:2010nj,Aubert:2007sd}, 
which correspond 
effectively to higher-order $\Kp\Km$ partial waves.  Because
the NR contribution dominates much of the available phase space,
it is crucial to study its angular distribution if one wishes to
accurately measure $\betaeff$ over the entire \bkkks DP.

Because of the importance of understanding the DP structure in \bkkks, we
study the related modes \bkkkboth and \bkksks along with \bkkks.  
The mode \bkkkboth is valuable because it has the most
signal events by far of any $\bkkkgeneric$ mode.
Far fewer events are expected in $\bkksks$, but its
DP has a simplified 
 spin-structure  due to the fact that the two $\KS$ mesons in the final state 
are forbidden (by Bose-Einstein statistics) to be in an odd angular
momentum configuration. 
This implies that the $f_X(1500)$ can decay to $\KS\KS$ only if it has even
spin, and it also ensures that the nonresonant component in $\bkksks$ 
does not contain any $\KS\KS$ P-wave contribution.

In this paper we report the results of DP analyses of \bkkkboth and \bkksks, 
and a time-dependent DP analysis of \bkkks. In Sec.~\ref{sec:ana_overview},
we introduce the formalism used for the DP amplitude analyses.
In Sec.~\ref{sec:DetectorAndData}, we briefly describe
the \babar\ detector and datasets used, and Sec.~\ref{sec:selection}
describes the event selection and backgrounds.
Section~\ref{sec:ML} describes the maximum likelihood (ML) fit 
parameterization and implementation.  
In Sec.~\ref{sec:modelDetermination}, we present studies of 
the DP structure in the three
modes, which enable us to determine the nominal DP models.
In Sec.~\ref{sec:fitResults}, we then present the final fit results including 
measurements of \CP violation.  We discuss systematic uncertainties
in Sec.~\ref{sec:Systematics} and summarize our results in
Sec.~\ref{sec:summary}.

%% file: overview.tex
\section{DECAY MODEL FORMALISM}
\label{sec:ana_overview}

Taking advantage of the interference pattern in the DP, we measure 
the magnitudes and phases of the different resonant decay modes using an
unbinned maximum-likelihood fit. 

We consider the decay of a $\B$ meson with four-momentum
$p_B$ into the three daughters $K_1$, $K_2$, and $K_3$,
with corresponding four-momenta $p_1$, $p_2$, and $p_3$. 
The
squares of the invariant masses are given by $s_{ij} = m^2_{ij} = (p_i + p_j)^2$.

We will
use the following convention for the $K$ indices:
\bei
\item For \bkkkbothCharge, $K_1\equiv\Kpm$, $K_2\equiv\Kmp$, and 
$K_3\equiv\Kpm$.  The indices for the two like-sign kaons are defined
such that $s_{12}\le s_{23}$.
\item For \bkksksboth, $K_1\equiv\KS$, $K_2\equiv\KS$, and 
$K_3\equiv\Kpm$.  The indices for the two \KS are defined
such that $s_{13}\le s_{23}$.
\item For \bkkksboth, $K_1\equiv\Kp$, $K_2\equiv\Km$, and 
$K_3\equiv\KS$. 
\eei

The $s_{ij}$ obey the relation
\beq
\label{eq:magicSum}
    s_{12} + s_{13} + s_{23} \;=\; m_{\B}^2 + m_{K_1}^2 + m_{K_2}^2 + m_{K_3}^2\,.
\eeq

The DP distribution of the $\Bpm$ decays 
is given by
\beq
\label{eq:partialWidth}
        \frac{d\Gamma}{d\sab d\sbc} \;=\; 
\frac{1}{(2\pi)^3}\frac{1}{32 m_{\Bp}^3}  |\AorAbar|^2\,,
\eeq
where $\Amp$ ($\Ampbar$) is the Lorentz-invariant amplitude
of the $\Bp$ ($\Bm$) three-body decay, and is a function of $\sab$ 
and $\sbc$.

For \bkkks, the time-dependence of the decay rate is a function of DP location.
With $\deltat \equiv t_{\rm sig} - t_{\rm tag}$ defined as the proper 
time interval between the decay of the fully reconstructed $\bkkks$
($B^0_{{\rm sig}}$)
and that of the  other meson ($\Bz_{\rm tag}$) from the \FourS,  the time-dependent decay
rate over the DP is given by
\begin{eqnarray}
\frac{d\Gamma}{d\sab d\sbc d \deltat} & = &
       \frac{1}{(2\pi)^3}\frac{1}{32 m_{\Bz}^3}  \frac{e^{-|\deltat|/\tau_{\Bz}}}{4\tau_{\Bz}}  
       \bigg[ \absAmp^2 + \absAmpbar^2 \nonumber \\
   &  & -~ \Qtag~(1-2 w) ~\left( \absAmp^2 - \absAmpbar^2 \right) \cos\deltamd\deltat \nonumber \\
   &  & + \Qtag~(1-2 w) ~2\I \left[ e^{-2i \beta} \Ampbar \Amp^*  \right] \sin\deltamd\deltat 
	 \bigg ]\,, \nonumber \\
&  & 
\label{eq:dalitz_plot_rate}
\end{eqnarray}
where $\tau_{B^0}$ is the  neutral $B$ meson lifetime and $\deltamd$ is the \Bz-\Bzb 
mixing frequency.
$\Amp$ ($\Ampbar$) is the amplitude of the $B^0_{{\rm sig}}$ ($\Bzb_{{\rm sig}}$) decay
and $\Qtag = +1(-1)$ when the $\Bz_{\rm tag}$ is identified as a \Bz(\Bzb).
The parameter $w$ is the fraction of events in which the $\Bz_{\rm tag}$ is tagged with 
the incorrect flavor.

We describe the distribution of signal events 
in the DP using an isobar approximation,
which models the total amplitude as
a coherent sum of amplitudes from $N$ individual decay channels (``isobars''):
\beq
  \label{eq:isobar}
\AorAbar = \sum_{j=1}^N {\AorAbar}_j~,
\eeq
where 
\beqn
\label{eq:aj_Bpm}
\Amp_j       & \equiv  & a_j {F}_j(\sab,\sbc)\,,  \nonumber \\ 
\Ampbar_j & \equiv  & \overline{a}_j \overline{F}_j(\sab,\sbc)\,.
\eeqn
The $F_j$ are DP-dependent dynamical amplitudes described below, 
and $a_j$ are complex coefficients describing the relative
magnitude and phase of the different decay channels.
All the weak phase dependence is contained in $a_j$, and $F_j$
contains strong dynamics only. 

The amplitudes must be symmetric under exchange of identical bosons,
so for \bkkkboth, $F_j(\sab,\sbc)$ is replaced by 
$F_j(\sab,\sbc) + F_j(\sbc,\sab)$.  Similarly, in \bkksks, $F_j(\sab,\sbc)$
 is replaced by $F_j(\sab,\sbc) + F_j(\sab,\sac)$.  

We parameterize the complex coefficients as
\beqn
\label{eq:isobarPars}
a_j       & = & c_j (1 + b_j) e^{i(\phi_j + \delta_j)} \,, \nonumber \\
\overline{a}_j & = & c_j (1 - b_j) e^{i(\phi_j - \delta_j)}\,,
\eeqn
where $c_j$, $b_j$, $\phi_j$, and $\delta_j$ are real numbers.
We define the fit fraction ($\fitfrac_j$) for an intermediate state
as
\beq
\label{eq:fitfrac}
\fitfrac_j \equiv \frac{ \int\int  \big( |\Amp_j|^2 + |\Ampbar_j|^2 \big) \,d\sab d\sbc}
	{\int\int  \big( |\Amp|^2 + |\Ampbar|^2 \big)  \,d\sab d\sbc}\,.
\eeq
Note that the sum of the fit fractions is not necessarily unity, due to 
interference between states.  This interference can be quantified by the 
interference fit fractions $\fitfrac_{jk}$, defined as
\beq
\label{eq:interfitfrac}
\fitfrac_{jk} \equiv  2~\R  \frac{ \int\int  \big( \Amp_j\Amp^*_k + \Ampbar_j\Ampbar^*_k \big) \,d\sab d\sbc}
	{\int\int  \big( |\Amp|^2 + |\Ampbar|^2 \big)  \,d\sab d\sbc}\,.
\eeq
With this definition, 
\beq
\sum_{j} \fitfrac_j + \sum_{j<k} \fitfrac_{jk} = 1\,.
\eeq

In the \Bp modes, the direct \CP asymmetry $\Acp(j)$ for a
particular intermediate state is given by
\beq
\label{eq:directAcp}
\Acp(j) \equiv \frac{ \int\int \big(  |\Ampbar_j|^2 - |\Amp_j|^2 \big) \,d\sab d\sbc}
	{\int\int \big(  |\Ampbar_j|^2 + |\Amp_j|^2 \big)  \,d\sab d\sbc} = \frac{-2 b_j}{1+b_j^2}\,,
\eeq
while there can also be a \CP asymmetry in the interference
between two intermediate states, which depends on both the $b$'s and $\delta$'s of the
interfering states.   We define the \CP-violating phase difference as
\beq
\label{eq:deltaphi}
\Delta\phi_j \equiv \mbox{arg}( a_j \overline{a}_j^{*}) = 2 \delta_j\,.
\eeq

For \bkkks, we can define the direct \CP asymmetry as in Eq.~\eqref{eq:directAcp}, while we can 
also compute the effective $\beta$ for an intermediate state
as
\beq
\betaeffj \equiv \frac{1}{2} \mbox{arg}(e^{2 i \beta} a_j \overline{a}_j^{*}) = \beta + \delta_j \,,
\eeq
which quantifies the \CP violation due to the interference between mixing and decay.

The resonance dynamics are contained within the $F_j$ terms, which are 
the product of the invariant mass and angular distributions,
\begin{equation}
\label{eq:ResDynEqn}
F_j^L(\sab,\sbc) = R_j(m) X_L(|\vec{p}\,^{\star}|\,r') X_L(|\vec{q}\,|\,r) T_j(L,\vec{p},\vec{q}\,)\,,
\end{equation}
where
\begin{itemize}
\item $L$ is the spin of the resonance.
\item $m$ is the invariant mass of the decay products of the resonance.
\item $R_j(m)$ is the resonance mass term or ``lineshape'' (\eg~Breit-Wigner).
\item $\vec{p}\,^{\star}$ is the momentum of the ``bachelor'' particle,
      \ie, the particle not belonging to the resonance,
      evaluated in the rest frame of the $\B$.
\item $\vec{p}$ and $\vec{q}$ are the momenta of the bachelor particle and
      one of the resonance daughters, respectively, both evaluated in the
      rest frame of the resonance. For $\Kp\Km$ resonances, $\vec{q}$ is assigned 
      to the momentum of the $\Kp$, except for $\bkkkm$ decays, in which case
      $\vec{q}$ is assigned to the momentum of the $\Km$.
      For $\KS\KS$ resonances, it is irrelevant to which $\KS$ we assign $\vec{q}$,
      so we arbitrarily assign $\vec{q}$ to whichever $\KS$ forms the smaller angle
      with the $\Kp$.
\item $X_L$ are Blatt-Weisskopf angular momentum barrier factors~\cite{blatt-weisskopf}:
\begin{eqnarray}
 L=0 & : & X_L(z) =  1\,,  \\
 L=1 & : & X_L(z) =  \sqrt{\frac{1+z_0^2}{1+z^2}}\,,  \\
 L=2 & : & X_L(z) =  \sqrt{\frac{9+3z_0^2+z_0^4}{9+3z^2+z^4}}\,,
\end{eqnarray}
where $z$ equals  $|\vec{q}\,|\,r$ or $|\vec{p}\,^{\star}|\,r'$, and 
$z_0$ is the value of $z$ when the invariant mass of the pair
of daughter particles equals the mass of the parent resonance.
$r$ and $r'$ are effective meson radii.
We take $r'$ as zero, while $r$ is taken to be $4\pm 2.5\,(\gevc)^{-1}$ for
each resonance.
\item $T_j(L,\vec{p},\vec{q})$ are the Zemach tensors~\cite{Zemach}, which 
describe the angular distributions:
\begin{eqnarray}
L=0 &:& T_j = 1~,\\
L=1 &:& T_j = 4\vec{p}\cdot\vec{q}~,\\
L=2 &:& T_j = \frac{16}{3} \left[3(\vec{p}\cdot\vec{q}\,)^2 - (|\vec{p}\,||\vec{q}\,|)^2\right]~.
\end{eqnarray}
\end{itemize}

The helicity angle of a resonance is defined as the angle between $\vec{p}$ and $\vec{q}$,
measured in the rest frame of the resonance.  For a $K_1 K_2$ resonance, the helicity
angle will be called $\thetaTh$, and is the angle between $K_3$ and $K_1$.
In \bkkks, because $\vec{q}$ is defined as the $\Kp$ momentum for both \Bz and \Bzb
decays, there is a sign flip between \Bz and \Bzb amplitudes for odd-$L$
$\Kp\Km$ resonances:
\beq
\label{eq:signflip}
 \overline{F}_j(\sab,\sbc) = F_j(\sab,\sac) = (-1)^L F_j(\sab,\sbc).
\eeq
In contrast, for \bkkkboth and \bkksks, $\overline{F}_j(\sab,\sbc)$ always equals $F_j(\sab,\sbc)$.

For most resonances in this analysis the $R_j$ are taken to be relativistic
Breit-Wigner (RBW)~\cite{Nakamura:2010zzi} lineshapes:
\begin{equation}
R_j(m) = \frac{1}{(m^2_0 - m^2) - i m_0 \Gamma(m)},
\label{eqn:BreitWigner}
\end{equation}
where $m_0$ is the nominal mass of the resonance and $\Gamma(m)$ is the
mass-dependent width.
In the general case of a spin-$L$ resonance, the latter can be expressed as
\begin{equation}
\Gamma(m) = \Gamma_0 \left( \frac{|\vec{q}|}{|\vec{q_0}|}\right)^{2L+1} 
\left(\frac{m_0}{m}\right) X^2_L(|\vec{q}| r)\,.
\label{eqn:resWidth}
\end{equation}
The symbol $\Gamma_0$ denotes the nominal width of the resonance.
The values of $m_0$ and $\Gamma_0$ are listed in Table~\ref{tab:model}.
The symbol $|\vec{q}_0|$ denotes the value of $|\vec{q}|$ when $m = m_0$.

For the \fI\ lineshape the Flatt\'e form~\cite{flatte} is used.
In this case 
\begin{equation}
R_j(m) = \frac{1}{(m^2_0 - m^2) - i (g_{\pi}\rho_{\pi\pi}(m) + g_{\kaon}\rho_{KK}(m) )}\,,
\label{eqn:Flatte}
\end{equation}
where
\begin{eqnarray}
\rho_{\pi\pi}(m) &=&
\sqrt{1 - 4m_{\pipm}^2/m^2}\,,\\
\rho_{KK}(m) &=&
 \sqrt{1 - 4m_{\kaon}^2/m^2}\,.
\label{eqn:FlatteW2}
\end{eqnarray}
Here, $m_{\kaon}$ is the average of the \Kpm and \KS masses, and $g_{\pi}$ 
and $g_{K}$ are coupling constants for which the values are 
given in Table~\ref{tab:model}.

In this paper, we test several different models to account 
for NR $\bkkkgeneric$ decays.  \babar's previous analysis~\cite{Aubert:2006nu}
of \bkkkboth modeled the NR decays with an exponential model given
by
\begin{equation}
{F}_{\NR}(\sab,\sbc) =  e^{\alpha \sab} +  e^{\alpha \sbc}\,,
\label{eq:NRexp}
\end{equation}
where the symmetrization is explicit.  $\alpha$ is
a parameter to be determined empirically.  This model consists
purely of \KpKm S-wave decays.

The most recently published \bkkks analyses by Belle~\cite{Nakahama:2010nj} 
and \babar~\cite{Aubert:2007sd}  
both used what we will call an {\it extended exponential model}.
This model adds $\Kp\KS$ and $\Km\KS$ exponential terms:
\beqn
\Amp_{\NR}(\sab,\sbc) =  a_{12} e^{\alpha \sab} + a_{13} e^{\alpha \sac} + a_{23} e^{\alpha \sbc}\,, \nonumber \\
\Ampbar_{\NR}(\sab,\sbc) = a_{12} e^{\alpha \sab} + a_{13} e^{\alpha \sbc} + a_{23} e^{\alpha \sac}\,.
\label{eq:NRexpext}
\eeqn

We also test a {\it polynomial model}, consisting of explicit S-wave and P-wave terms, each
of which has a quadratic dependence on $\mab$:
\beqn
\Amp_{\NR}(\sab,\sbc) = \left ( a_{S0} + a_{S1} x +  a_{S2} x^2 \right) + \nonumber \\
         \left ( a_{P0} + a_{P1} x +  a_{P2} x^2 \right) P_1(\cosHth)\,,
\label{eq:NRpoly}
\eeqn
where $x \equiv \mlo - \Omega$, and $\Omega$ is an offset that we define as
\beq
\Omega \equiv \frac{1}{2} \left (m_{\B} + \frac{1}{3}(m_{K_1} + m_{K_2} + m_{K_3}) \right )\,,
\eeq
and $P_1$ is the first Legendre polynomial.  
In this paper, we normalize the $P_\ell$ such that
\begin{equation}
\int_{-1}^{1}{P_{\ell}(x) P_{k}(x) dx} = \delta_{\ell k}\,.
\label{eq:legendreOrtho}
\end{equation}
Note that in the \bkkkboth channel, we symmetrize all terms in Eq.~\eqref{eq:NRpoly}:
\beq
\Amp_{\NR,total} = \Amp_{\NR}(\sab,\sbc) + \Amp_{\NR}(\sbc,\sab)\,.
\eeq
This results in S-wave and P-wave terms for both the $(K_1 K_2)$ and
$(K_2 K_3)$ pairs.  
In the \bkksks channel, the P-wave term is forbidden by Bose-Einstein symmetry.

In Sec.~\ref{sec:modelDetermination}, we present studies that allow 
us to determine the nominal DP model.  
The components of the nominal model are summarized in Table \ref{tab:model}.
Other components, taken into account only to estimate the 
systematic uncertainties due to the DP model,
are discussed in Sec.~\ref{sec:Systematics}.

\begin{table}[htbp]
\begin{center}
\caption{Parameters of the DP model used in the fit. Values are 
given in  $\mev{\rm (}/c^2{\rm)}$  unless 
specified otherwise.  All parameters are taken from Ref.~\cite{Nakamura:2010zzi}, except
for the $\fI$ parameters, which are taken from Ref.~\cite{valFlatte}.
\label{tab:model}}
\begin{tabular}{ccc}
\hline\hline
Resonance      & Parameters                      & Lineshape    \\
\hline\\[-9pt]
$\phiI$        & $m_0 = 1019.455 \pm 0.020$      & RBW         \\
               & $\Gamma_0= 4.26 \pm 0.04$       &             \\   \hline\\[-9pt]
$\fI$          & $m_0=965 \pm 10$                & Flatt\'e    \\
               & $g_{\pi}= (0.165 \pm 0.018) \gev^2/c^4$ &      \\
               & $g_{K}/g_{\pi}=4.21 \pm 0.33$   &              \\    \hline\\[-9pt]
$\fII$         & $m_0 = 1505 \pm 6$              & RBW          \\
               & $\Gamma_0= 109 \pm 7$           &              \\  \hline\\[-9pt]
$\fIII$        & $m_0 = 1720 \pm 6$              & RBW          \\
               & $\Gamma_0= 135 \pm 8$           &              \\  \hline\\[-9pt]
$\ftwop$       & $m_0 = 1525 \pm 5$              & RBW          \\
               & $\Gamma_0= 73^{+6}_{-5}$        &              \\  \hline\\[-9pt]
NR decays      &                                 & see text     \\  \hline\\[-9pt]
$\chiczero$    & $m_0=3414.75\pm0.31$            & RBW          \\
               & $\Gamma_0=10.3 \pm 0.6$         &              \\ 
\hline \hline
\end{tabular}
\end{center}
\end{table}

%% file: DetectorAndData.tex
\section{THE \babar\ DETECTOR AND DATA SET}
\label{sec:DetectorAndData}

The data used in this analysis were collected with the \babar\ detector at the 
\pep2\ asymmetric energy \epem\ storage rings.
The \bkkks and \bkksks modes use an integrated luminosity of 
429 \invfb\ or $(471\pm3)\times 10^{6}$ $\BB$ pairs collected at the 
$\FourS$ resonance (``on-resonance'').
The \bkkkboth mode uses 426 \invfb or
$(467\pm 5)\times 10^6$ $\BB$ pairs collected on-resonance.
We also use approximately
 44\invfb\ collected 40~$\mev$ below the \FourS (``off-resonance'')
to study backgrounds.

A detailed description of the \babar\ detector is given in Ref.~\cite{babarNIM}. Charged-particle trajectories are measured with a five-layer, double-sided silicon vertex tracker (SVT) and a 40-layer drift chamber (DCH), both operating inside a 1.5-T magnetic field. Charged-particle identification (PID) is achieved by combining
information from a ring-imaging Cherenkov device and ionization energy loss (\itdedx) measurements from the DCH and SVT. Photons are detected and their energies measured in a CsI(Tl) electromagnetic calorimeter inside the magnet coil. Muon candidates are identified in the instrumented flux return of the solenoid.  

We use GEANT4-based~\cite{GEANT} software to simulate the detector response and account for the varying beam and environmental conditions. Using this software, we generate signal and background Monte Carlo (MC) event samples in order to estimate the efficiencies and expected backgrounds.

%% file: EventSelection.tex
\section{EVENT SELECTION AND BACKGROUNDS}
\label{sec:selection}

\subsection{\bkkkboth}

The \bkkkboth candidates are reconstructed
from three charged tracks that are each consistent with a kaon hypothesis.
The PID requirement is about 85\% efficient for kaons, with
a pion misidentification rate of around $2\%$.
The tracks are required to form a good-quality vertex. Also, the
total energy in the event must be less than 20 \gev.

Most backgrounds arise from random track combinations in 
$e^+e^-\to q\overline{q}~(q = u,d,s,c)$ events (hereafter referred to as
{\em continuum\/} events).  
These backgrounds peak at $\cosT = \pm 1$,  where 
$\theta_{T}$ is the angle in the \epem center-of-mass (CM) frame between the
thrust axis of the \B-candidate decay products and the thrust axis of the rest of the 
event. To reduce these backgrounds, we require
$|\cosT|<0.95$.
Additional continuum suppression is achieved by using
a neural network (NN) classifier with five input variables:
$|\cosT|$, $|\cosB|$, $|\dt/\dterr|$, $\Lratio$, and the output
of a \B-flavor tagging algorithm.  Here, $\theta_B$ is the
angle in the \epem CM frame between the \B-candidate momentum and the beam axis, $\dt$ is 
the difference between the decay times of the \Bp and \Bm candidates with $\dterr$
its uncertainty, and $\Lk = \sum_{j}|\pvec_j| P_{k}(\cos{\theta_j})$.
The sum includes every track and neutral cluster not used to form the \B 
candidate, and $\theta_j$ is the angle in the \epem CM frame between the momentum $\pvec_j$
and the \B-candidate thrust axis.
$P_k$ is the $k^{th}$ Legendre polynomial.  The NN is trained on signal 
MC events and \offpeak data.  We place a requirement on the NN output 
that removes 65\% 
of continuum events while removing only 6\% of signal events.

Further discrimination is achieved with the energy-substituted mass 
$\mes\equiv\sqrt{(s/2+{\mathbf {p}}_i\cdot{\mathbf{p}}_B)^2/E_i^2-p_B^2}$ 
and energy difference $\de \equiv E_B^*-\half\sqrt{s}$, where 
$(E_B,\pvec_B)$ and $(E_i, \pvec_i)$ are the four-vectors of the \B 
candidate and the initial electron-positron system measured in the laboratory
frame, respectively.  The asterisk denotes the \epem CM frame, and $s$
is the invariant mass squared of the electron-positron system.  
Signal events peak at the \B mass ($\approx 5.279 \gevcc$) for \mes, 
and at zero for \de.
We require $5.27<\mes<5.29 \gevcc$ and $|\de|<0.1 \gev$.  An \mes sideband 
region with $\mes<5.27 \gevcc$ is used for background characterization.
After the calculation of \mes and \de, we refit each \B candidate with the 
invariant mass of the candidate
constrained to agree with the nominal \B mass~\cite{Nakamura:2010zzi}, 
in order to improve the
resolution on the DP position and to ensure that 
Eq.~\eqref{eq:magicSum} is satisfied.  About 8\% of signal events have 
multiple \B candidates that pass the selection criteria.  If an
event has multiple \B candidates, we select the one with the best
vertex $\chi^2$.  
To avoid having events that
have candidates in both the \mes sideband and in the signal region,
the best-candidate selection is performed prior to
the $\mes$ and $\de$ selection.
The overall selection efficiency
for \bkkkboth is $33\%$.

We use MC simulation to study backgrounds from \B decays (\BB background).  
In this paper, we treat $\bkkkgeneric$ decays containing intermediate
charm decays as background, except for $\B\to\chiczero\kaon~(\chiczero\to\kaon\kaon)$, which we treat as signal.
Most of the \BB backgrounds come from $\B\to D^{(*)}X$ decays.
We study 20 of the most prominent \BpBm background modes using 
simulated exclusive samples, and split these modes into six classes,
summarized in Table~\ref{tab:BBbkg_KKKch}.  These classes have distinct
kinematic distributions, and so will be handled separately in the ML fit,
as described in Sec.~\ref{sec:ML}.
Class 1 contains various charmless
\Bp decays, the largest of which is $\Bp\to\Kp\Km\pip$.  Class
2 includes a number of decays containing $\Dz\to\KpKm$ in the
decay chain. Class 3 includes various decays containing
$\Dzb\to\Kp\pim$.  Class 4 consists of
$\Bp\to\Dzb\Kp~(\Dzb\to\Kp\pim)$ decays.
We also include classes
for signal-like $\bkkkboth$ decays coming from $\Bp\to~\Dzb\Kp$ (class 5) and 
$\Bp\to\jpsi\Kp$ (class 6).  These decays have the same 
\mes and \de distributions as signal, but can be distinguished from
charmless signal by their location on the DP.
We include a seventh \BB background class,  which contains the 
remaining inclusive
\BpBm and \BzBzb decays.

\begin{table*}[htbp]
\begin{center}
\caption{Summary of the \BB backgrounds in \bkkkboth.  The ``Expected yields'' column gives the
expected number of events for $467\times 10^6$ $\BB$ pairs, based on MC simulation. 
The ``Fitted yields'' column gives the fitted number of events from the best 
solution of the fit on the data (see Sec.~\ref{sec:fitResultsKKKch}).}
\label{tab:BBbkg_KKKch}
\begin{tabular}{llc@{\hspace{0.5cm}}c}
\hline\hline
Class  &   Decay       &  Expected yields &  Fitted yields \\
\hline
\noalign{\vskip1pt} 
1   &   $\Bp\to$ charmless    & $42 \pm 5$  & fixed \\   
2   &   $\Bp\to~\overline{D}^{(*)0}X, \Dzb\to\KpKm$   & $195 \pm 7$  & $170 \pm 21$ \\   
3   &   $\Bp\to~\overline{D}^{(*)0}X, \Dzb\to\Kp\pim$     & $117  \pm 5$  & $133 \pm 34$ \\   
4   &   $\Bp\to~\Dzb\Kp(\Dzb\to\Kp\pim)$     & $92 \pm 5$  & $23 \pm 9$ \\  
5   &   $\Bp\to~\Dzb\Kp(\Dzb\to\KpKm)$     & $233 \pm 13$ &  $238 \pm 22$ \\  
6   &   $\Bp\to\jpsi\Kp(\jpsi\to\KpKm)$ & $38 \pm 5$ & $45 \pm 10$ \\  
7   &   \BpBm/\BzBzb remaining   & $386 \pm 12$   &  $261 \pm 56$ \\  
\hline\hline
\end{tabular}
\end{center}
\end{table*}

\subsection{\bkksks}

The \bkksks candidates are reconstructed by combining a charged track with 
two $\KS\to\pip\pim$ 
candidates. The charged track is required to satisfy a kaon-PID requirement 
that is about 95\% efficient for kaons, with a pion misidentification rate of 
around $4\%$.
The \KS candidates are each required to have a mass within 12 \mevcc of 
the nominal
\KS mass and a lifetime significance exceeding 3 standard deviations.  
We also require that $\cos \alpha_{K_S} > 0.999$, where $\alpha_{K_S}$ is the 
angle between the momentum vector of the \KS candidate and the vector connecting 
the decay vertices of the \Bp and \KS candidates in the laboratory frame. 
The total energy in the event must be less than 20\gev.

To reduce continuum backgrounds, we require $|\cosT|<0.9$.  We also use 
the same NN
as for \bkkkboth, and place a requirement on the NN output
that removes 49\% of continuum events while removing 4\% of signal events.  
Finally, the \B candidates
are required to satisfy  $5.26<\mes<5.29 \gevcc$ and $|\de|<0.1 \gev$.  
An \mes sideband 
region with $\mes<5.26 \gevcc$ is used for background characterization.
After the calculation of \mes and \de, the \B candidates are refitted with a \B mass constraint.  
The overall selection efficiency for \bkksks (with both \Kspp) is $27\%$.

About 2\% of signal events have multiple \B candidates that pass the 
selection criteria.  In such cases, we choose the \B candidate 
whose \KS candidates have invariant masses closest
to the nominal \KS mass.  Because there can be multiple \B candidates 
that share one or more of the same kaon candidates, multiple \B 
candidates may still remain after
this step.   In this case, we select the \B candidate 
whose \Kp candidate has PID information 
most consistent with the kaon hypothesis. If multiple \B candidates 
still remain, we select the one with the
best vertex $\chi^2$.   The best candidate selection is performed prior to
the $\mes$ and $\de$ selection.

\BB backgrounds are studied with MC events.  We study 10 of the most prominent background
decay modes using simulated exclusive samples, and group them into 
three classes, summarized in Table~\ref{tab:BBbkg_KKsKs}.  Class 1 contains 
$\Bp\to~D^0\pip~(D^0\to\KS\KS)$ and $\Bp\to\KS~K^{*+}~(K^{*+}\to\KS\pi^+)$ decays.  
Class 2 contains various $\BpBm$
and $\BzBzb$ decays, dominated by the charmless decays $\bksksks$ and
$\Bz\to\kaon^{(*)+}\Km\KS$.  Signal-like 
$\bkksks$ decays coming from $\Bp\to~D^0\Kp$ make up class 3.
The remaining \BB backgrounds are grouped into a fourth class.

\begin{table*}[htbp]
\begin{center}
\caption{Summary of the \BB backgrounds in \bkksks.  The ``Expected yields'' column gives the
expected number of events for $471\times 10^6$ $\BB$ pairs, based on MC simulation. 
In the maximum-likelihood fit on the data (Sec.~\ref{sec:fitResultsKKsKs}), 
the yield of each class is fixed to its MC expectation. }
\label{tab:BBbkg_KKsKs}
\begin{tabular}{llc}
\hline\hline
Class   &   Decay       &  Expected yields  \\
\hline
\noalign{\vskip1pt} 
1   &   $\Bp\to\Dzb\pip~(\Dzb\to\KS\KS),$  & $6.1 \pm 1.2$  \\    
    &   $\Bp\to\KS~K^{*+}~(K^{*+}\to\KS\pi^+)$                          & \\
2   &   $\Bp/\Bz\to$ charmless    & $23 \pm 5$   \\   
3   &   $\Bp\to~\Dzb\Kp(\Dzb\to\KS\KS)$     & $8.1 \pm 1.6$   \\  
4   &   \BpBm/\BzBzb remaining   & $118 \pm 6$   \\  
\hline\hline
\end{tabular}
\end{center}
\end{table*}

\subsection{\bkkks}

\bkkks candidates are reconstructed by combining two charged tracks with a $\KS$ candidate.  The charged
tracks are required to be consistent with a kaon hypothesis.  For most events, we
apply tight kaon-PID requirements that are about $90\%$ efficient for kaons with a
 pion misidentification rate of around $1.5\%$.
Looser PID requirements ($\sim 95\%$ efficient, $\sim 6\%$ pion misidentification)
are applied in the $\mab<1.1\gevcc$ region, to increase the signal efficiency
for $\Bz\to\phi\KS$.  \KS candidates are reconstructed in both the $\Kspp$ and 
$\Kszz$ decay modes.  \Kspp candidates are required to have a mass within 20 \mevcc
of the nominal \KS mass, while \Kszz candidates are required to have a mass
$m_{\piz\piz}$ in the range $(m_{\KS}-20 \mevcc)<m_{\piz\piz}<(m_{\KS}+30 \mevcc)$,
where $m_{\KS}$ is the nominal \KS mass.   Both \Kspp and \Kszz candidates are 
required to have a lifetime significance of at least 3 standard deviations, 
and to satisfy $\cos \alpha_{K_S} > 0.999$.  The \piz candidates are formed from two photon
candidates, with each photon required to have a laboratory energy greater than 50 \mev and a transverse
shower profile consistent with an electromagnetic shower.

We reduce continuum backgrounds by requiring $|\cosT|<0.9$.  In addition, we use a NN
containing the variables $|\cosT|$, $|\cosB|$, and $\Lratio$.  
Since we are performing a time-dependent analysis of \bkkks, we omit $|\dt/\dterr|$
from the NN in order not to bias the fit.
We train the NN on signal MC events and \offpeak data.  We make a requirement on the NN output that removes
 26\% of continuum events in the \Kspp channel, and 24\% of continuum 
events in the \Kszz channel, with only a 2\% loss of signal events.
\B candidates must satisfy $5.26<\mes<5.29 \gevcc$ and 
$-0.06(-0.12)<\de<0.06\gev$ for \Kspp (\Kszz).   An \mes sideband region with 
$\mes<5.26 \gevcc$ is used for background characterization.
After the calculation of \mes and \de, the \B candidates are refitted with a \B mass constraint.  
The overall selection efficiency for \bkkks  is $31\%$  for \Kspp 
and $7\%$ for \Kszz.

The time difference $\deltat$ is obtained from the measured distance along 
the beam direction between 
the  positions of the $\Bz_{\rm sig}$ and 
$\Bz_{\rm tag}$ decay vertices, using the boost $\beta\gamma=0.56$ of 
the \epem\ system.
We require that \B candidates have $|\dt|<20 \ps$ and an uncertainty on $\dt$ less than $2.5 \ps$.  
To determine the flavor of $\Bz_{\rm tag}$ 
we use the $B$ flavor tagging algorithm of Ref.~\cite{Aubert:2009yr},
which produces six mutually exclusive tagging categories.
We also retain untagged events (about $23\%$ of signal events) in a 
seventh category, since these
events contribute to the measurements of the branching fractions,
although not to the \CP\ asymmetries.

Multiple \B candidates pass the selection criteria in
about 4\% of \Kspp signal events and 11\% of \Kszz signal events.
If an event has multiple candidates, we choose the \B candidate using 
criteria similar to those used for \bkksks.
The best candidate selection is performed prior to the $\mes$, $\de$, and 
$\dt$ selection.

\BB backgrounds are studied with MC events and grouped into 
five classes,
summarized in Table~\ref{tab:BBbkg_KKKs}.  We include classes for
signal-like $\bkkks$ decays coming from $\Bz\to~\Dm\Kp$ (class 1), 
$D_{s}^{-}\Kp$ (class 2), $\Dzb\KS$ (class 3), and 
$\jpsi\KS$ (class 4).
The remaining \BB backgrounds are grouped into a fifth class.

\begin{table*}[htbp]
\begin{center}
\caption{Summary of the \BB backgrounds in \bkkks.  The ``Expected yields'' column gives the
expected number of events for $471\times 10^6$ $\BB$ pairs, based on MC simulation. 
The ``Fitted yields'' column gives the fitted number of events from the best 
solution of the fit on the data (see Sec.~\ref{sec:fitResultsKKKs}). 
}
\label{tab:BBbkg_KKKs}
\begin{tabular}{llcccc}
\hline\hline
Class   &   Decay       & \multicolumn{2}{c}{Expected yields} & \multicolumn{2}{c}{Fitted yields} \\
        &               &  \Kspp       &  \Kszz                                      &   \Kspp    & \Kszz \\
\hline
\noalign{\vskip1pt}
1   &   $\Bz\to~\Dm\Kp(\Dm\to\Km\KS)$            & $42 \pm 13$ & $4 \pm 1$  & $36 \pm 7$ & $3.6 \pm 0.6$ \\    
2   &   $\Bz\to~D_{s}^-\Kp(D_{s}^-\to\Km\KS)$    & $33 \pm 6$ & $3 \pm 1$  & $11 \pm 4$ & $1.1 \pm 0.4$ \\   
3   &   $\Bz\to\Dzb\KS(\Dzb\to\Kp\Km)$ &  $10 \pm 1$ & $1.0 \pm 0.1$  & $16 \pm 5$ & $1.9 \pm 0.5$ \\  
4   &   $\Bz\to\jpsi\KS(\jpsi\to\KpKm)$          &  $10 \pm 1$ & $1.0 \pm 0.1$  & $4 \pm 4$ & $0.5 \pm 0.4$ \\ 
5   &   \BpBm/\BzBzb remaining                     & $141 \pm 7$ & $123 \pm 6$ &  $29 \pm 28$ & $48 \pm 18$ \\  
\hline\hline
\end{tabular}
\end{center}
\end{table*}

%% file: Likelihood.tex
\section{THE MAXIMUM-LIKELIHOOD FIT}
\label{sec:ML}

We perform an unbinned extended maximum-likelihood fit~\cite{Barlow:1990vc} to measure
the inclusive $\bkkkgeneric$ event yields and the resonant amplitudes
and \CP-violating parameters.  
The fit uses the variables $\mes$, $\de$, $\nn$, \mab, and \mbc 
 to discriminate signal from background.  
Events with both charges or tag flavors $\Q$ 
are simultaneously included in the fits in order to measure 
\CP violation.
For \bkkks, the additional variable $\deltat$ enables 
the determination of mixing-induced \CP violation.

The selected on-resonance data sample is assumed to consist of signal,
continuum background, and \B background components.

\subsection{The Likelihood Function}

\subsubsection{\bkkkboth and \bkksks}
\label{sec:ML_BpmModes}

The probability density function (PDF) ${\cal P}_i$ for an
event $i$ is the sum of the probability densities of all event components
 (signal, $q\overline{q}$ continuum background, $\BB$ background), namely
\begin{eqnarray}
\label{eq:theBpLikelihood}
        {\cal P}_i
        &\equiv& 
                N_{\rm sig}
                {\cal P}_{{\rm sig},i}
                + N_{q\overline q}\frac{1}{2}
                \left(1 - \Qi\Atagqq\right){\cal P}_{q\overline q,i}
                \nonumber \\[0.3cm]
        &&
		+\; \sum_{j=1}^{N^{\BB}_{\rm class}}
                N_{\BB j}\frac{1}{2}
		\left(1 - \Qi\Atagbbj\right)
                {\cal P}_{\BB j,i}~.
\end{eqnarray}
The parameters are defined in Table~\ref{tab:DefBpVarLik}.
\renewcommand{\arraystretch}{1.15}
\begin{table*}[htbp]  
\begin{center}
\caption{
Definition of parameters in the event PDF for \Bp decays shown in Eq.~\eqref{eq:theBpLikelihood}.
The \BB background classes are given in Tables~\ref{tab:BBbkg_KKKch} and \ref{tab:BBbkg_KKsKs}.
\label{tab:DefBpVarLik}}
\begin{tabular}{ll}
\hline\hline
Parameter           & Definition \\
\hline
$N_{\rm sig}$      & total fitted  $\bkkkgeneric$ signal yield in the data sample       \\
$N_{q\overline q}$ & fitted continuum yield \\
$\Qi$           &   charge of the \B candidate, $+1$ or $-1$ \\
$\Atagqq$          & charge asymmetry in continuum events \\
$N^{\BB}_{\rm class}$              & number of $\BB$-related background classes considered in the fit \\
$N_{\BB j}$         & fitted or fixed yield in  $\BB$ background class $j$ \\
$\Atagbbj$          & charge asymmetry in  $\BB$ background class $j$ \\
\hline\hline
\end{tabular}
\end{center}
\end{table*}
\renewcommand{\arraystretch}{1.}

The PDFs ${\cal P}_{X,i}$ 
have the general form
\beqn
\label{eq:pdfVarsBpModes}
        {\cal P}_{X,i}  & \equiv & 
        P_{X,i}(\mes) P_{X,i}(\de) P_{X,i}(\nn,\sab,\sbc) \times \nonumber \\
    & &	P_{X,i}(\sab,\sbc,\Q)\,.
\eeqn
This form neglects some small correlations between observables.  Biases due to
correlations in the signal PDF are assessed using MC events passed through
a GEANT4-based detector simulation (see Sec.~\ref{sec:Systematics}).

The extended likelihood function is given by
\begin{equation}
        {\cal L} \;\propto\;  
        e^{-\bar N}\,
        \prod_{i}^{N} {\cal P}_{i}~,
\end{equation}
where $N$ is the number of events entering into the fit, 
and $\bar N \equiv N_{\rm sig} + N_{q\overline{q}} + \sum_{j=1}^{N^{\BB}_{\rm class}}
 N_{\BB j}$ is the total fitted number of events. 

A total of 43 parameters are allowed to  vary in the \bkkkboth fit. 
They include eight yields 
(signal, continuum, and six \BB background yields) and $30$ parameters for the 
complex amplitudes $a_j$ from Eq.~\eqref{eq:aj_Bpm} (see Table~\ref{tab:isobarSummary_KKKch_SolnIandII} in Sec.~\ref{sec:fitResults}).
The last five parameters are $\Atagqq$, one parameter each for the continuum \mes and \de
PDFs, and the means of the signal \mes and \de PDFs (see Sec.~\ref{sec:pdfsOtherVars}).
The $\Atagbbj$ are fixed to their MC expectations, except for classes 5 and 6,
in which they are fixed to the world average~\cite{Nakamura:2010zzi} and 0, 
respectively.

A total of 41 parameters are allowed to vary in the \bkksks fit. They 
include two yields 
(signal and continuum) and 16 parameters for the 
complex amplitudes $a_j$ (see Table~\ref{tab:isobarSummary_KKsKs_SolnI} 
in Sec.~\ref{sec:fitResults}).
The last 23 parameters are $\Atagqq$, one parameter each for the shapes 
of the continuum \mes and \de
PDFs, the means of the signal \mes and \de PDFs, and 18 parameters for the
continuum \nn PDFs (nine $h_{0i}$ and nine $g_i$; see Sec.~\ref{sec:pdfsOtherVars}).
The $\Atagbbj$ are fixed to their MC expectations, except for  class 3,
which is fixed to the world average~\cite{Nakamura:2010zzi}.

\subsubsection{\bkkks}

For this decay we use a similar unbinned maximum likelihood fit to that 
described in
Sec.~\ref{sec:ML_BpmModes}, but there are some significant differences.
The components in the fit may be separated by the flavor and tagging category of 
the tag-side \B decay.

The probability density function ${\cal P}_i^\cat$ for an
event $i$ in tagging category $\cat$~\cite{Aubert:2009yr} is the sum of the probability densities 
of all components, namely
\begin{eqnarray}
\label{eq:theBzLikelihood}
        {\cal P}_i^\cat
        &\equiv& 
                N_{\rm sig} f^\cat
                {\cal P}_{{\rm sig},i}^\cat 
                +\; N^\cat_{q\overline q}{\cal P}_{q\overline q,i}^\cat
                +\; N_{\BB} f^\cat {\cal P}_{\BB,i}^\cat \,.
\end{eqnarray}

The parameters are defined in Table~\ref{tab:DefBzVarLik}.  The signal PDF 
includes components for the \BB background classes 1-4 listed in 
Table~\ref{tab:BBbkg_KKKs}, since they lead to the same \kkks final state.
\renewcommand{\arraystretch}{1.15}
\begin{table*}[htbp]
\begin{center}
\caption{
Definition of parameters in the event PDF for \Bz decays shown in Eq.~\eqref{eq:theBzLikelihood}.
The \BB background classes are described in Table~\ref{tab:BBbkg_KKKs}.
\label{tab:DefBzVarLik}}
\begin{tabular}{ll}
\hline\hline
Parameter           & Definition \\
\hline
\noalign{\vskip1pt} 
$N_{\rm sig}$      & total fitted $\bkkkgeneric$ signal yield in the data sample,  including \BB background classes 1-4 \\
$f^\cat$ & fraction of events that are tagged in category $\cat$, with $\sum_{\cat} f^\cat = 1$  \\
         & This fraction is assumed to be the same for signal and \BB background events    \\
$N^\cat_{q\overline q}$ & fitted continuum yield in tagging category $\cat$ \\
$\Qtagi$           & tag flavor of the event, defined to be $+1$ for a $\Bz_{\rm tag}$ and $-1$ for a $\Bzb_{\rm tag}$ \\
$N_{\BB}$            & fitted yield in  $\BB$ background class 5  \\
\hline\hline
\end{tabular}
\end{center}
\end{table*}
\renewcommand{\arraystretch}{1.}
The PDFs ${\cal P}_{X,i}^{\cat}$ 
are the products of PDFs for one or more variables,
\beqn
\label{eq:pdfVarsBzMode}
        {\cal P}_{X,i}^{\cat}  & \equiv & 
        P_{X,i}^\cat(\mes) P_{X,i}^\cat(\de) P_{X,i}^\cat(\nn,\sab,\sbc) \times \nonumber \\
    & &	P_{X,i}^\cat(\sab,\sbc,\dt,\dterr,\Qtag)~,
\eeqn
where $i$ is the event index.
Not all the PDFs depend on the tagging category;
the general notations $P_{X,i}^\cat$ and ${\cal P}_{X,i}^{\cat}$ are used for simplicity.

The extended likelihood function evaluated for events in all tagging categories is given by
\begin{equation}
        {\cal L} \;\equiv\;  
        \prod_{\cat=1}^{7} e^{-\bar N^\cat}\,
        \prod_{i}^{N^\cat} {\cal P}_{i}^\cat~,
\end{equation}
where $N^\cat$ is the number of events entering into the fit in category $\cat$, 
and $\bar N^\cat \equiv N_{\rm sig} f^\cat + N^\cat_{q\overline q} + N_{\BB}  f^\cat$ is the 
total fitted number of events in category $\cat$. 

The maximum-likelihood fit is performed simultaneously over both the \Kspp and \Kszz modes. 
The signal isobar model parameters are constrained to be equal for both modes, but otherwise
the PDFs may differ.  

A total of 90 parameters are allowed to vary in the fit. They include the 18 inclusive yields 
(for both \Kspp and \Kszz, there are nine yields: signal, \BB, and seven continuum yields, 
one per tagging category). We also allow 32 parameters for the 
complex amplitudes $a_j$ to vary (22 are shown in Table~\ref{tab:isobarSummary_KKKs_SolnI},
six are $b$ and $\delta$ parameters corresponding to the parameters in Table~\ref{tab:CPV_KKKs_SolnI}, and four 
describe the background classes 1-4
in Table~\ref{tab:BBbkg_KKKs}, which we model as non-interfering isobars).
The remaining 40 parameters include 38 parameters that describe the 
continuum PDF shapes (one \de parameter and 18 \nn parameters, for both \Kspp and \Kszz),
as well as the means of the signal \mes and \de
PDFs for \Kspp only (see Sec.~\ref{sec:pdfsOtherVars}).

\subsection{The Dalitz Plot and $\dt$ PDFs}
\label{sec:DPdeltaT}

For \bkkkboth and \bkksks, the signal DP PDFs  are given by
\beq
\label{eq:DPpdfBp}
P_{{\rm sig}}(\sab,\sbc,\Q) = d\Gamma(\sab,\sbc,\Q) \varepsilon(\sab,\sbc)\,,
\eeq
where $d\Gamma$ is defined in Eq.~\eqref{eq:partialWidth}, and 
$\varepsilon$ is the DP-dependent selection efficiency,
determined from MC simulation.  We assume equal efficiencies for
$\Bp$ and $\Bm$ events, and consider a possible asymmetry as a 
systematic uncertainty.

For \bkkks, the time- and DP-dependent signal PDF is given by
\beqn
\label{eq:DPdtpdfBz}
P_{{\rm sig}}^{\cat}(\sab,\sbc,\dt,\dterr,\Qtag) & = & \nonumber \\
   d\Gamma(\sab,\sbc,\dt,\Qtag) \varepsilon(\sab,\sbc) & \otimes & {\cal R}(\dt, \dterr)\,, 
\eeqn
where $d\Gamma$ is defined in Eq.~\eqref{eq:dalitz_plot_rate}
and the $\dt$ resolution function is a sum of three Gaussian distributions.
The parameters of the \dt resolution function and the tagging-category-dependent
mistag rate are determined by a fit to fully 
reconstructed $\Bz$ decays~\cite{Aubert:2009yr}.

To account for finite resolution on DP location, the
signal PDFs are convolved with a              
$2\times 2$-dimensional resolution function
\begin{equation}
  \label{eq:resolutionDP}
  \Res(\sab^r,\sbc^r,\sab^t,\sbc^t)~,
\end{equation}
which represents the probability for an event with true
DP coordinates $(\sab^t,\sbc^t)$
to be reconstructed with coordinates $(\sab^r,\sbc^r)$.
It obeys the unitarity condition
\begin{equation}
  \int\int 
  \Res(\sab^r,\sbc^r,\sab^t,\sbc^t)
  \,d\sab^r d\sbc^r = 1~\forall~\sab^t,\sbc^t.
\end{equation}
The $\Res$ function is obtained from MC simulation.

For \bkkkboth and \bkksks, the \BB background DP PDFs are histograms obtained 
from MC samples.  The histograms have variable bin sizes calculated using an adaptive 
binning method to ensure that fine binning is used where the DP distributions 
have narrow structures.  

For \bkkks, the generic \BB background DP PDFs are likewise histograms 
obtained from MC samples.  The background classes 1-4 given in 
Table~\ref{tab:BBbkg_KKKs},
however, are modeled as non-interfering isobars, and so
their DP- and time-dependence is included in Eq.~\eqref{eq:DPdtpdfBz}.   

The DP PDFs for continuum events are described by histograms similar to 
those for \BB backgrounds.  The PDFs are modeled with data taken from 
\mes sidebands, with a correction applied for \BB backgrounds present in the
sidebands.  

For \bkkks, the \dt distribution of the continuum events is
modeled with the sum of an exponential decay and prompt component, 
convolved with a double-Gaussian resolution function. The 
parameters are taken from a fit to data in the \mes sideband. The 
\dt distribution of the generic \BB backgrounds is 
modeled in the same way, but the parameters are taken from a fit
to MC samples. In the nominal fit, we assume zero \CP violation in the 
backgrounds, but as a systematic we include \CP violation in the 
\BB exponential decay component.

\subsection{PDFs of Other Fit Variables}
\label{sec:pdfsOtherVars}

The \mes and \de distributions of signal events are described by 
modified Gaussians of the form
\beq
P(x) \propto \exp \Big[-\frac{(x - x_0)^2}{2 \sigma_{\pm}^2 + \alpha_{\pm} (x - x_0)^2}\Big],
\eeq
where $\sigma_{+}$ and $\alpha_{+}$ are used when $x>x_0$, and 
$\sigma_{-}$ and $\alpha_{-}$ when $x<x_0$.  Most parameters are 
taken from fits to signal MC events.  The means $x_0$ are allowed to vary
 in the nominal fits to data, except for the \bkkks, \Kszz channel.

The \mes distributions for continuum events are modeled with a threshold
function~\cite{argus},
while the \de distributions are modeled with first-order polynomials.  
The \mes and
\de shape parameters are allowed to vary in the nominal fits.

A variety of PDFs are used to describe the \mes and \de distributions of the
various \BB background categories.  
The PDF shapes for each category are taken from MC simulation.  
Those \BB backgrounds that have the same true final state as signal events
are modeled with the same \mes, \de, and \nn PDFs as signal events.

The output of the \nn does not have an easily parameterized shape,
so we split the distribution into ten bins, with the bin size chosen so 
that approximately equal numbers of signal events are expected in each bin;
continuum events peak at larger values of the bin number.  
The binned \nn is then easily described using histogram PDFs.  The
PDFs for signal and \BB background events are taken from fits to 
MC events.  In the case of \bkkkboth, due to the large number of signal events,
 we obtain the histogram bin heights for signal from a 
separate fit to data, and then fix these parameters in the nominal fit.
For continuum events, the \nn output is correlated with the distance 
from the center of the DP.
To account for this correlation, the continuum \nn PDF is given by
a histogram with bin heights $h_i$ equal to $h_{0i} + g_i\dDP$. Here,
\dDP is the smallest of (\mab,\mbc,\mac). 

\subsection{Fitting Method}
\label{sec:fitDetails}
The ML fits are performed with {\tt MINUIT}~\cite{James:1975dr}.  Proper
normalization of the DP PDFs poses a technical challenge in these fits, 
because some of the resonance amplitudes vary rapidly as functions of the DP.
The normalization of these PDFs is performed using a numerical 2-dimensional
integration algorithm that makes use of adaptive binning~\cite{Genz1980295}.
The speed of this algorithm allows the masses and
decay widths of resonances to be varied in the fit. The normalization of the DP
PDFs is recalculated at each step in the fit for which these parameters 
are varied.

%% file: ModelDetermination.tex
\section{Determination of Dalitz Model}
\label{sec:modelDetermination}

\begin{figure}[htbp]
\includegraphics[width=7.0cm,keepaspectratio]{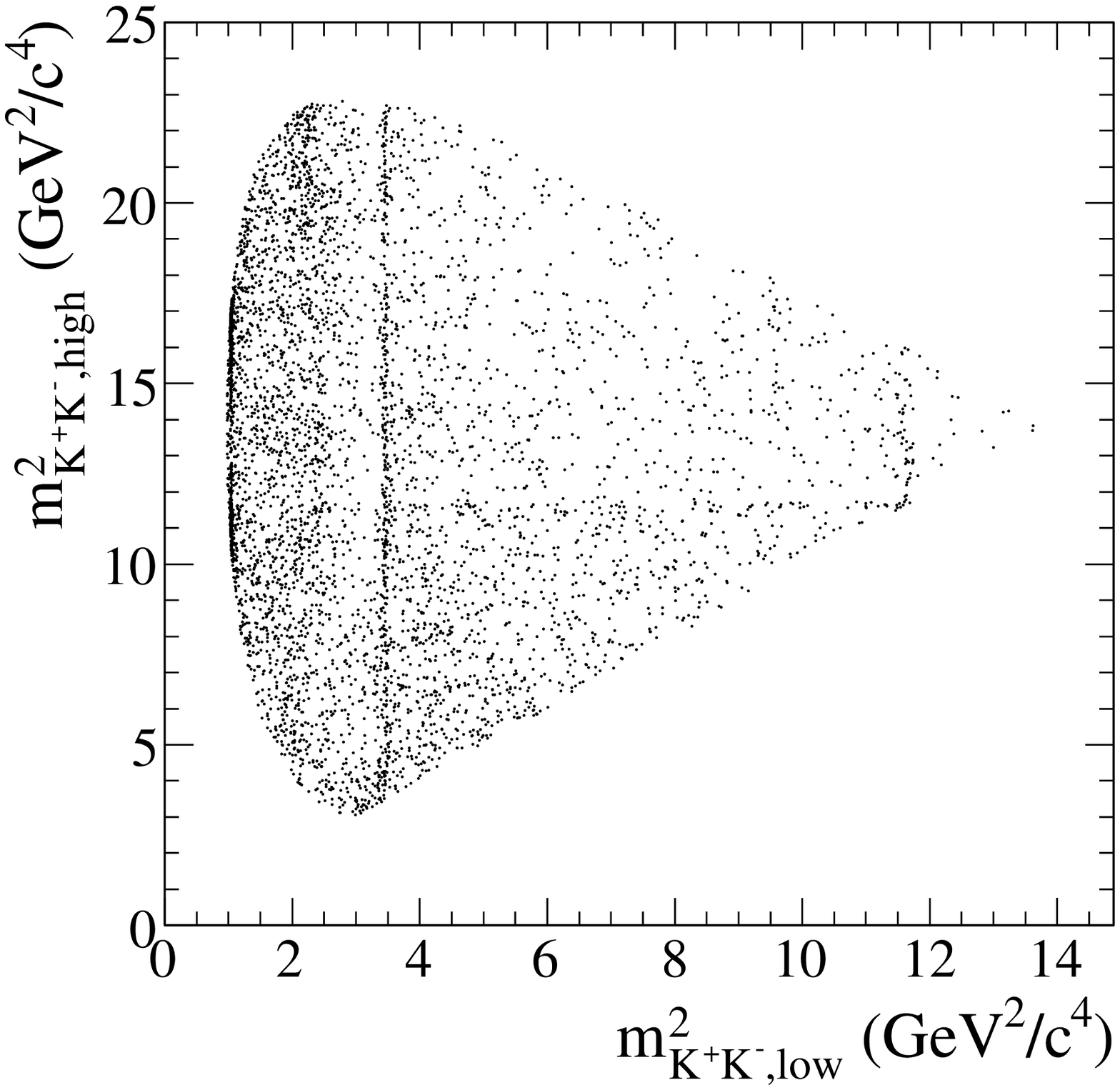}
\includegraphics[width=7.0cm,keepaspectratio]{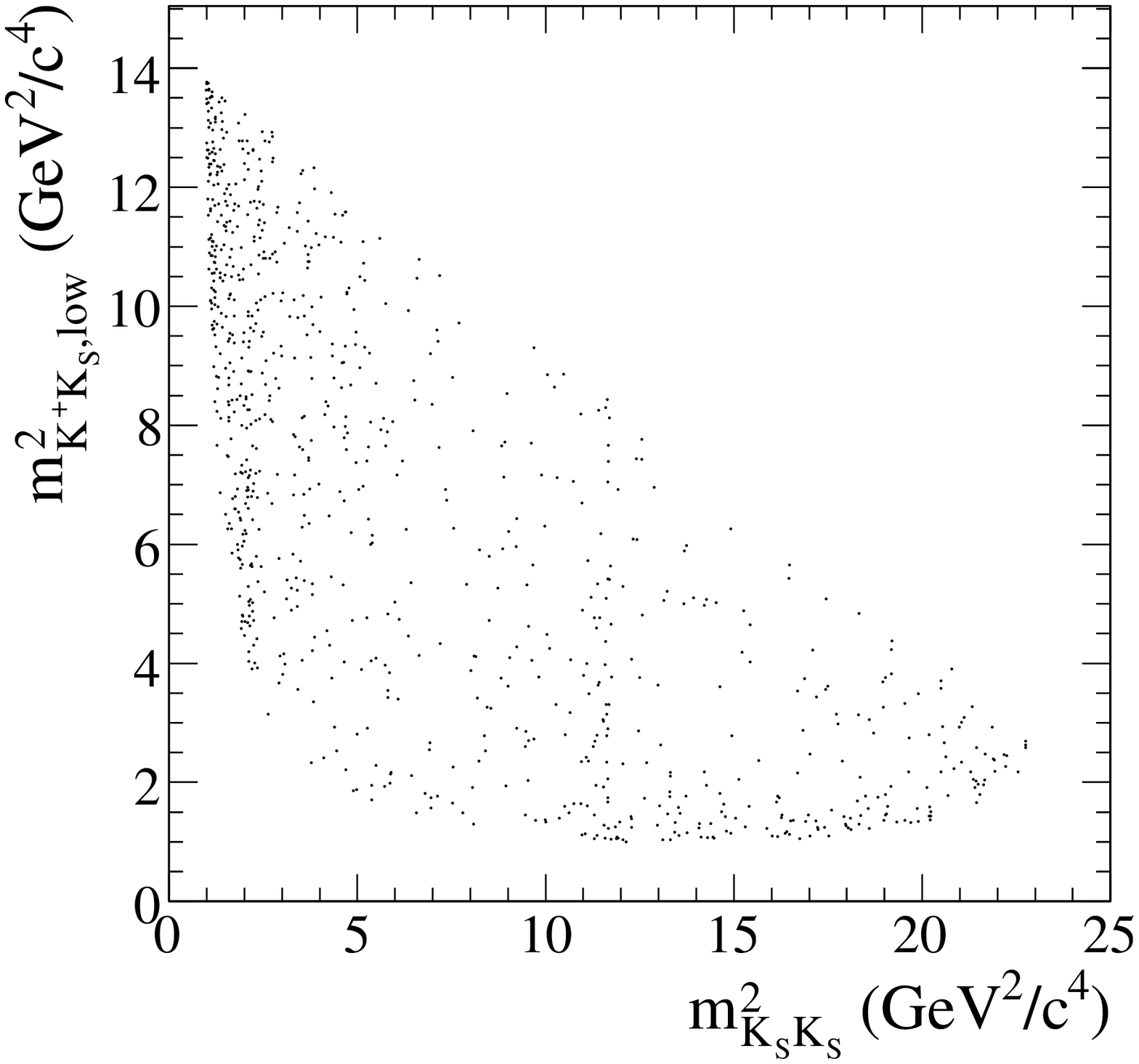}
\includegraphics[width=7.0cm,keepaspectratio]{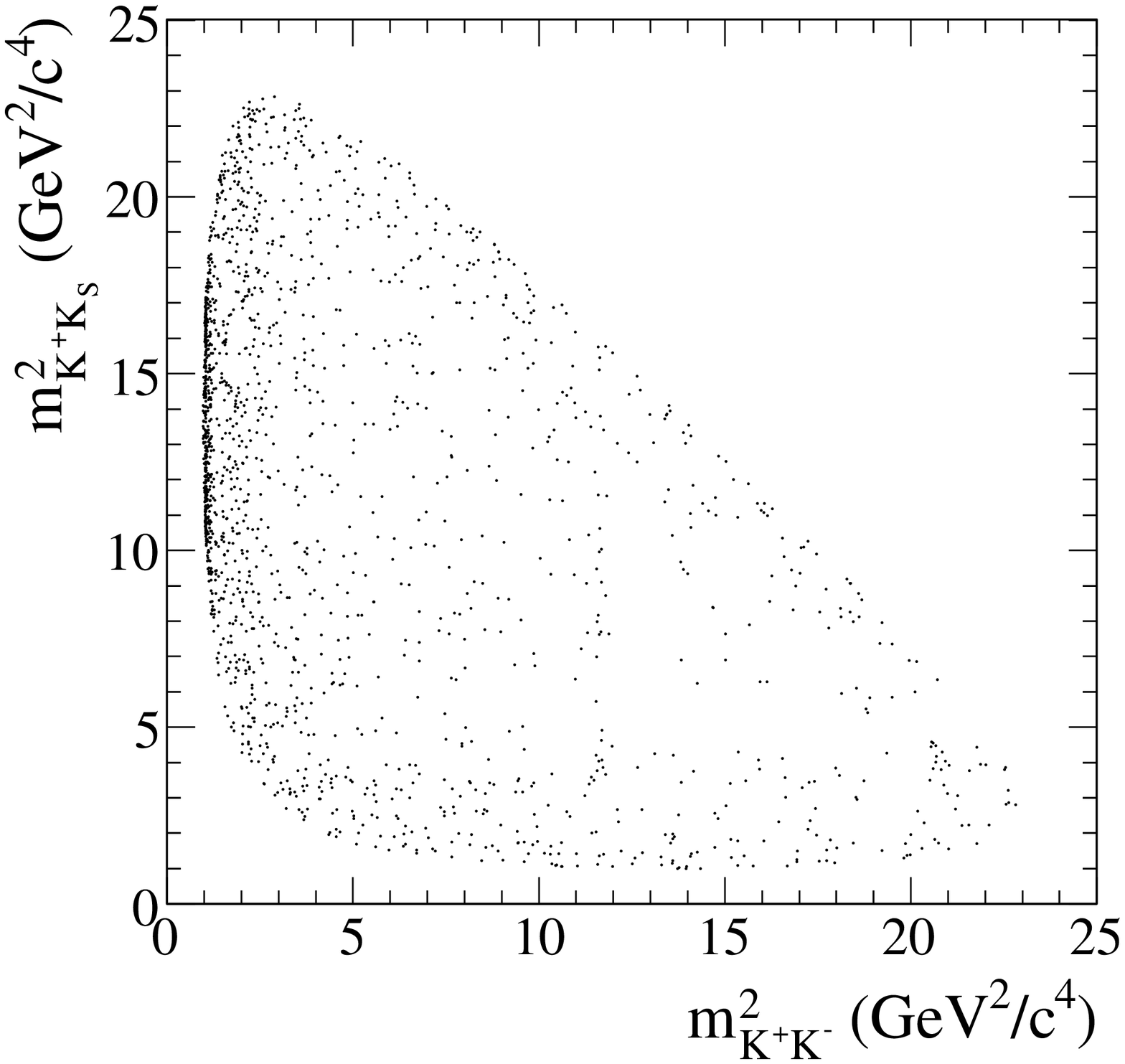}
\caption{\label{fig:scatterDP}  Dalitz plots for \bkkkboth (top), \bkksks (middle),
and \bkkks (bottom).  Points correspond to candidates in data that pass the full event selection, with
an additional requirement that the \nn output be 7 or less, in order to enhance
the signal.
}
\end{figure}

The Dalitz plots for the three \bkkkgeneric modes 
are shown in Fig.~\ref{fig:scatterDP}.
Before fitting \Acp parameters, we first decide which
resonances and NR terms to include in the DP model
for each of the \bkkkgeneric modes.  Because the 
\bkkkboth mode has the largest number of events, we primarily
use it to guide our decision-making, but the other
modes are useful as well.  The studies in this section
are performed in a ``\CP-blind'' fashion, which means that
we constrain the \CP-violating parameters of the signal and
background components to zero, except in the case of \bkkks, where we
constrain  $\betaeff$ to $\beta$ for all isobars.

One important goal is to understand the nature of the 
so-called $f_X(1500)$ resonance seen in several previous 
analyses.
Both \babar ~\cite{Aubert:2006nu,Aubert:2007sd} and 
Belle~\cite{Garmash:2004wa,Nakahama:2010nj}
have modeled this resonance
as a scalar particle, but while \babar\ has found its mass and width 
to be inconsistent with any established resonance, Belle has found a 
mass and width consistent with the $f_0(1500)$.  There is also confusion
surrounding the branching fraction to $f_X(1500)\kaon$.
Belle's \bkkkboth and \bkkks analyses both find multiple solutions
for the fit fraction for $f_X$.  Some solutions favor a small fit
fraction, less than 10\%, while others favor a large fit fraction, greater
than 50\%.  \babar\ obtained a small fit fraction in \bkkks, but a large fit 
fraction in \bkkkboth. A large, broad structure around 
$m_{\Kp\Km} = 1500\mevcc$ is also seen by \babar\ in 
$\bkkpi$~\cite{Aubert:2007xb} but not in $\bkskspi$~\cite{Aubert:2008aw}.
\babar's analysis of \bksksks~\cite{Lees:2011nf} does not provide
evidence for the $f_X(1500)$.

\subsection{\bkkkboth}
We initially perform a fit to \bkkkboth using the
same DP model as \babar's previous analysis of this mode,
which includes the resonances $\phiI$, $\fI$,
$f_{X}(1500)$, $\fIII$, and $\chi_{c0}$, and an
exponential NR model [Eq.~\eqref{eq:NRexp}].  We allow the NR 
parameter $\alpha$, as well as the mass and width of the 
$f_{X}(1500)$, to vary in the fit. The $f_{X}(1500)$ is taken 
to have a spin of zero.  We refer to this hereafter as \bkkkboth Model A. 
We find 
fit parameters consistent 
with \babar's previous analysis.

To see how well the fit model describes the DP
distribution, we calculate angular moments, defined as
\begin{equation}
\plmoment \equiv \int_{-1}^{1}{ d\Gamma P_{\ell}(\cosHth) d\cosHth},
\label{eq:plmoment}
\end{equation}
where \thetaTh is the helicity angle between $K_{3}$ and $K_{1}$, measured 
in the rest frame of $K_{1}K_{2}$, $P_\ell$ is the $\ell$-th Legendre polynomial,
and the differential decay rate $d\Gamma$ is given 
in Eq.~\eqref{eq:partialWidth}.
Note that the angular moments are functions of $\mab$ but we suppress 
this dependence in our notation.  Angular moments plotted as a
function of $\mab$ are an excellent tool
for  visualizing the agreement between the fit model and data, as they 
provide more information than ordinary DP projections, in particular
spin information.

\begin{figure*}[htbp]
\includegraphics[width=8.9cm,keepaspectratio]{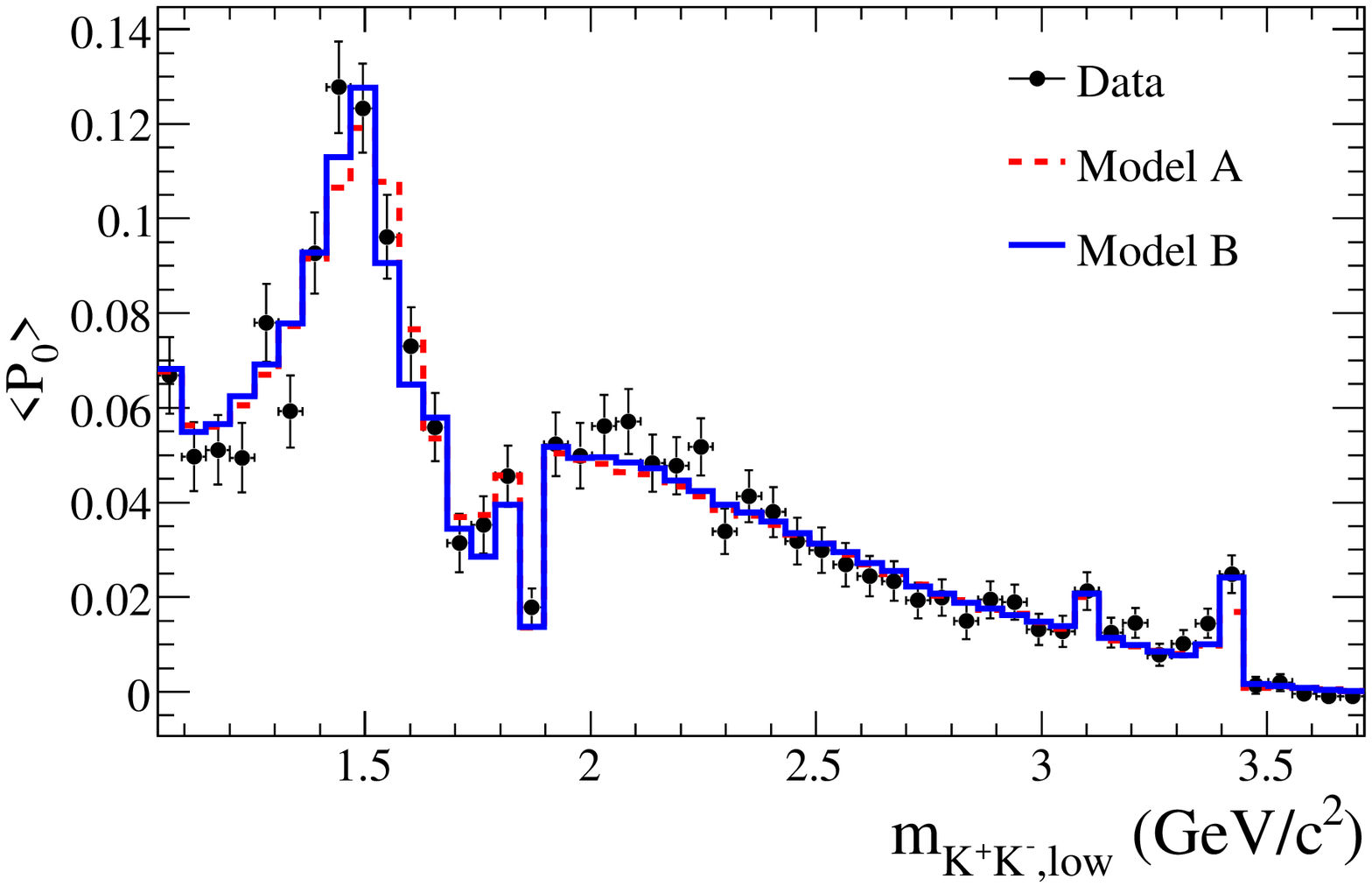}
\includegraphics[width=8.9cm,keepaspectratio]{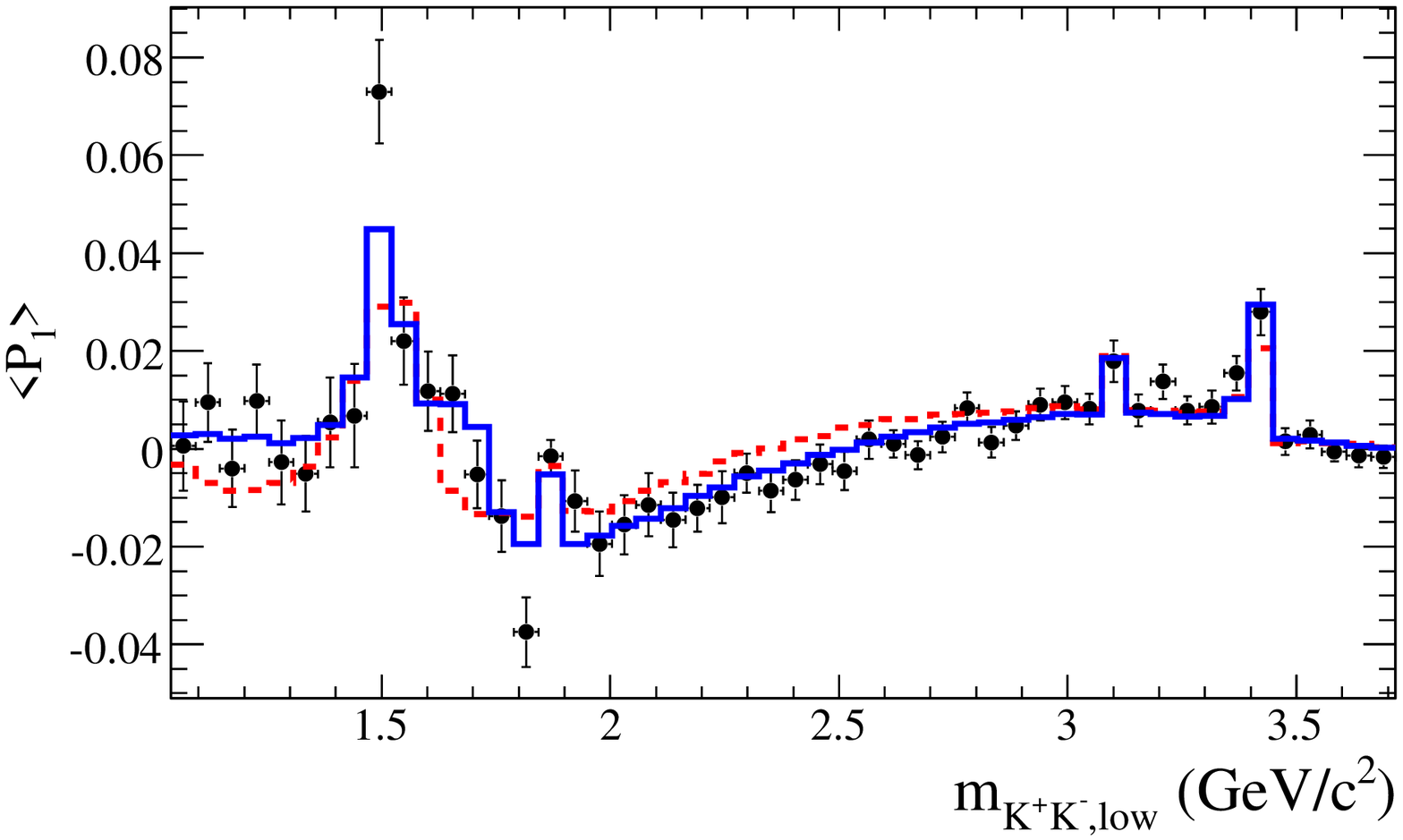}
\includegraphics[width=8.9cm,keepaspectratio]{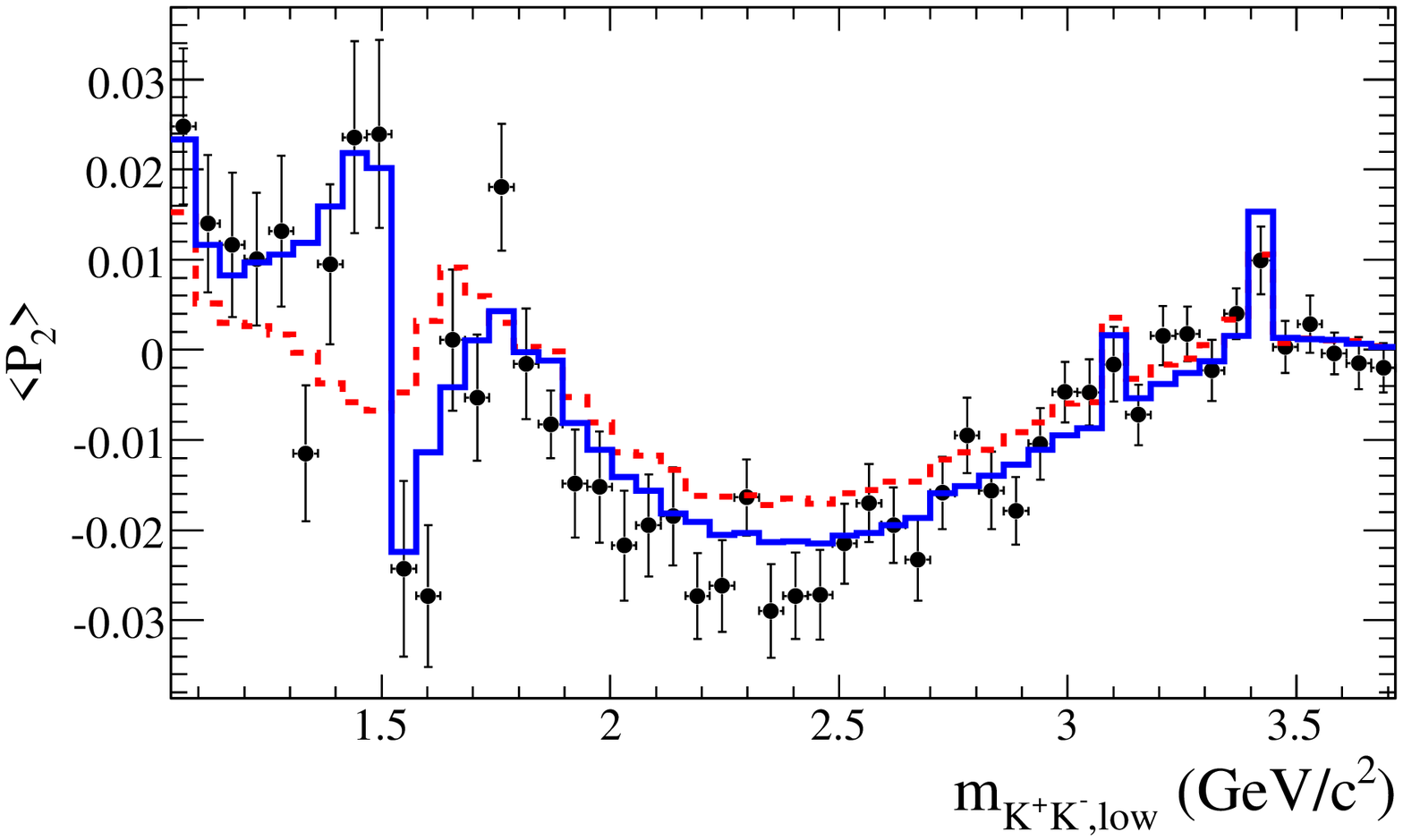}
\includegraphics[width=8.9cm,keepaspectratio]{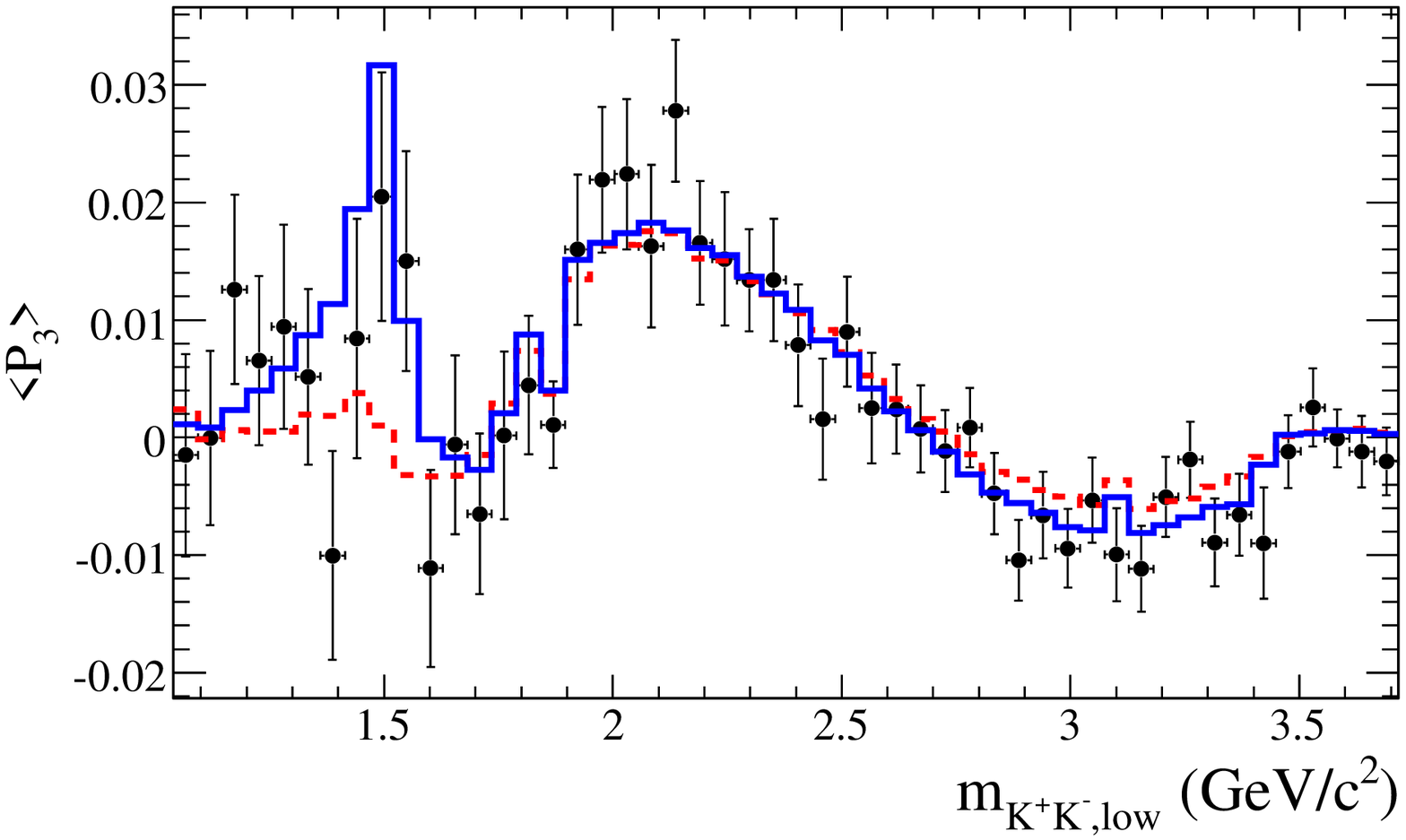}
\includegraphics[width=8.9cm,keepaspectratio]{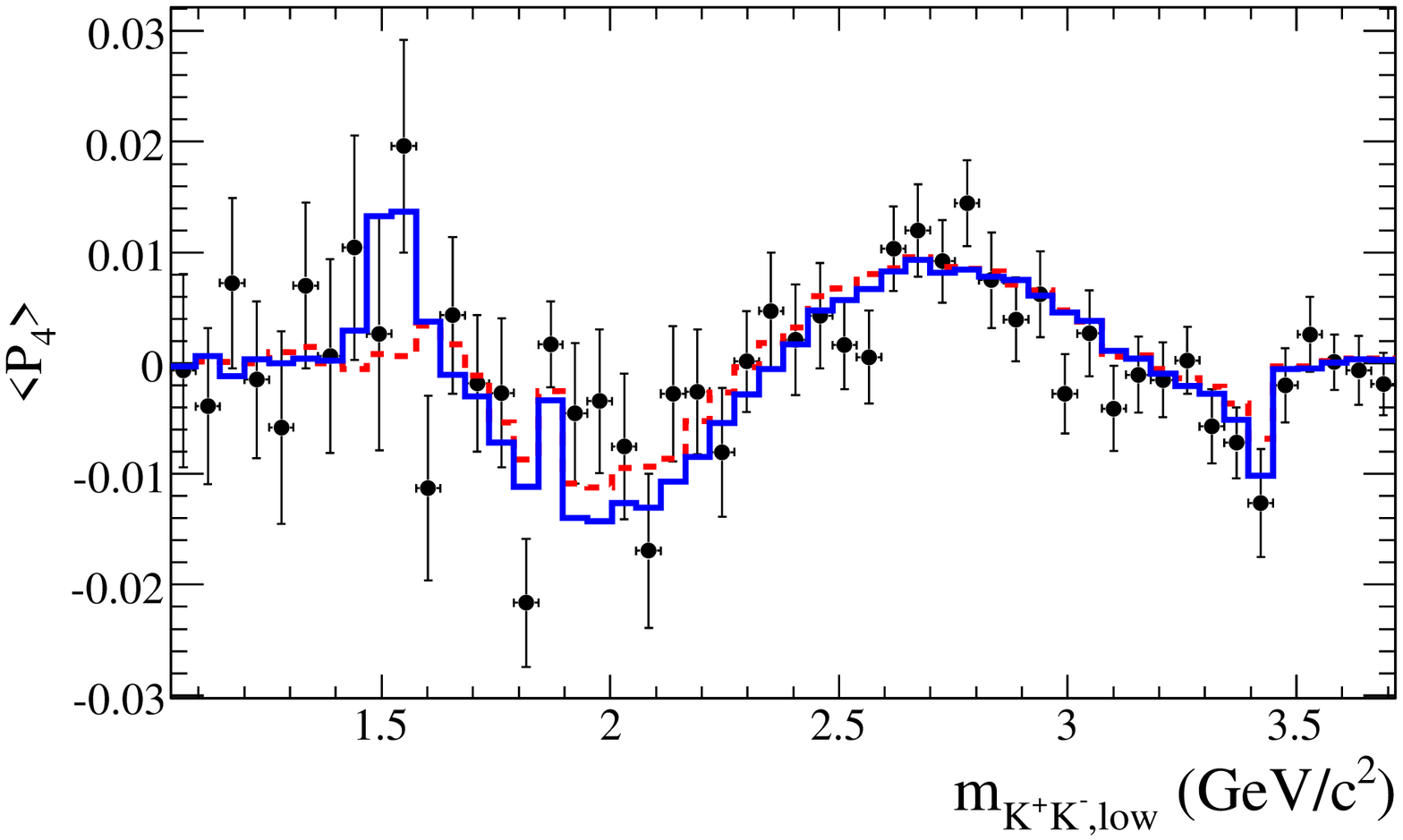}
\caption{\label{fig:angmoments_KKKch}  \bkkkboth angular moments in the region $\mab>1.04 \gevcc$, 
computed for signal-weighted data, compared to Model A (dashed line) and Model B (solid line).    
The signal weighting is performed using the \splot method.  Events with $\mkk$ near 
the $D^0$ mass are vetoed.
}
\end{figure*}

\begin{figure}[htbp]
\includegraphics[width=8.9cm,keepaspectratio]{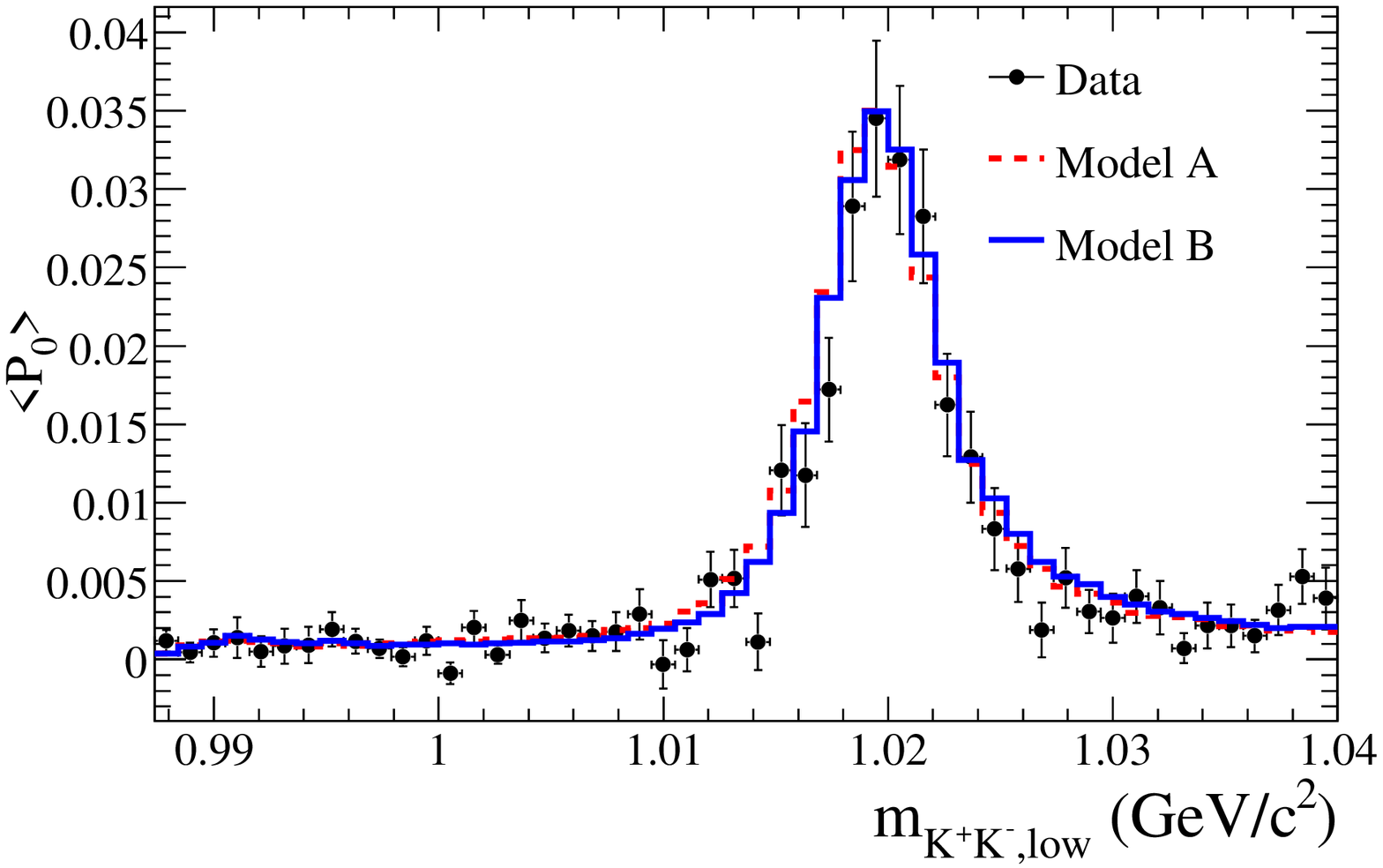}
\includegraphics[width=8.9cm,keepaspectratio]{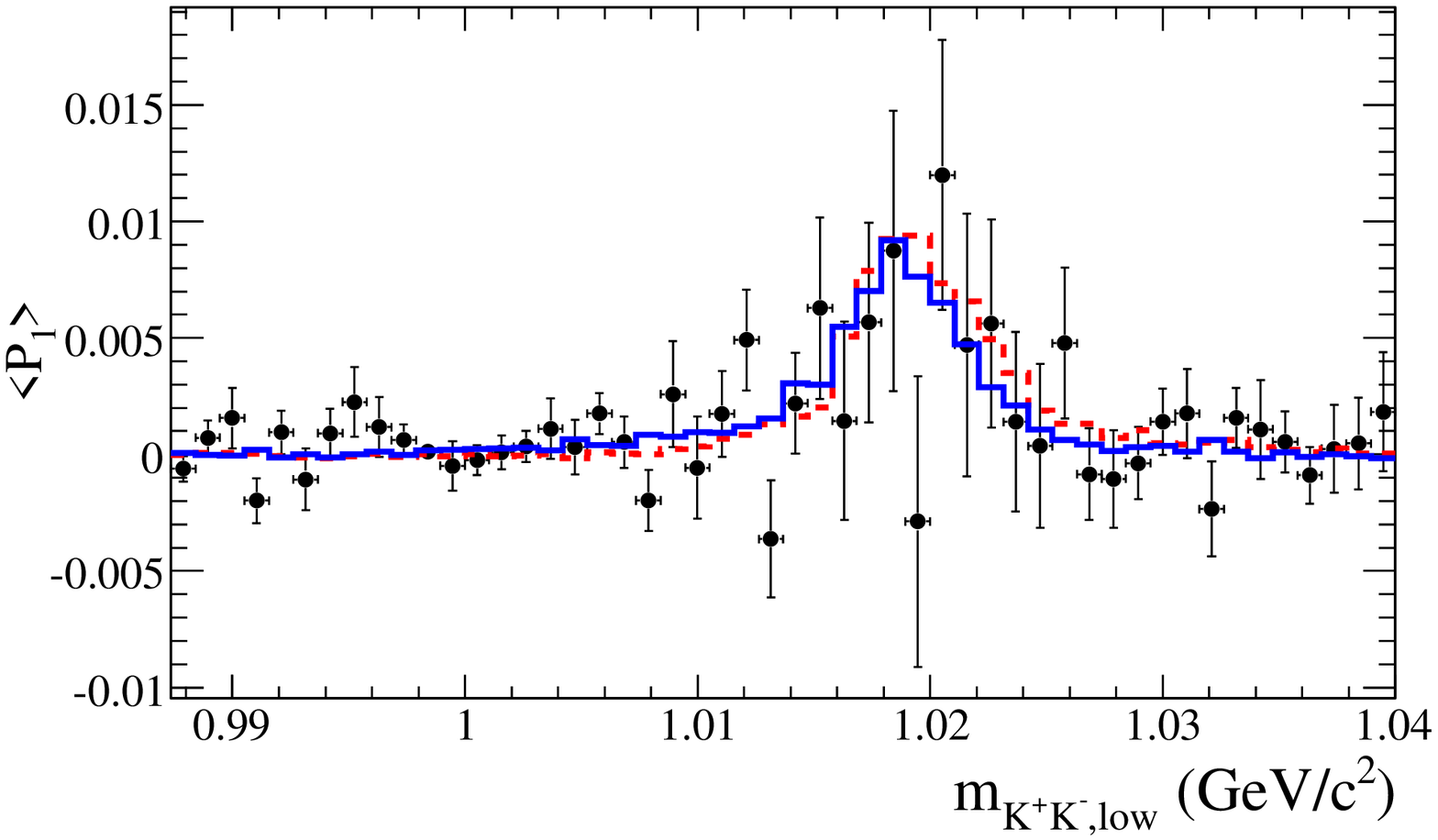}
\includegraphics[width=8.9cm,keepaspectratio]{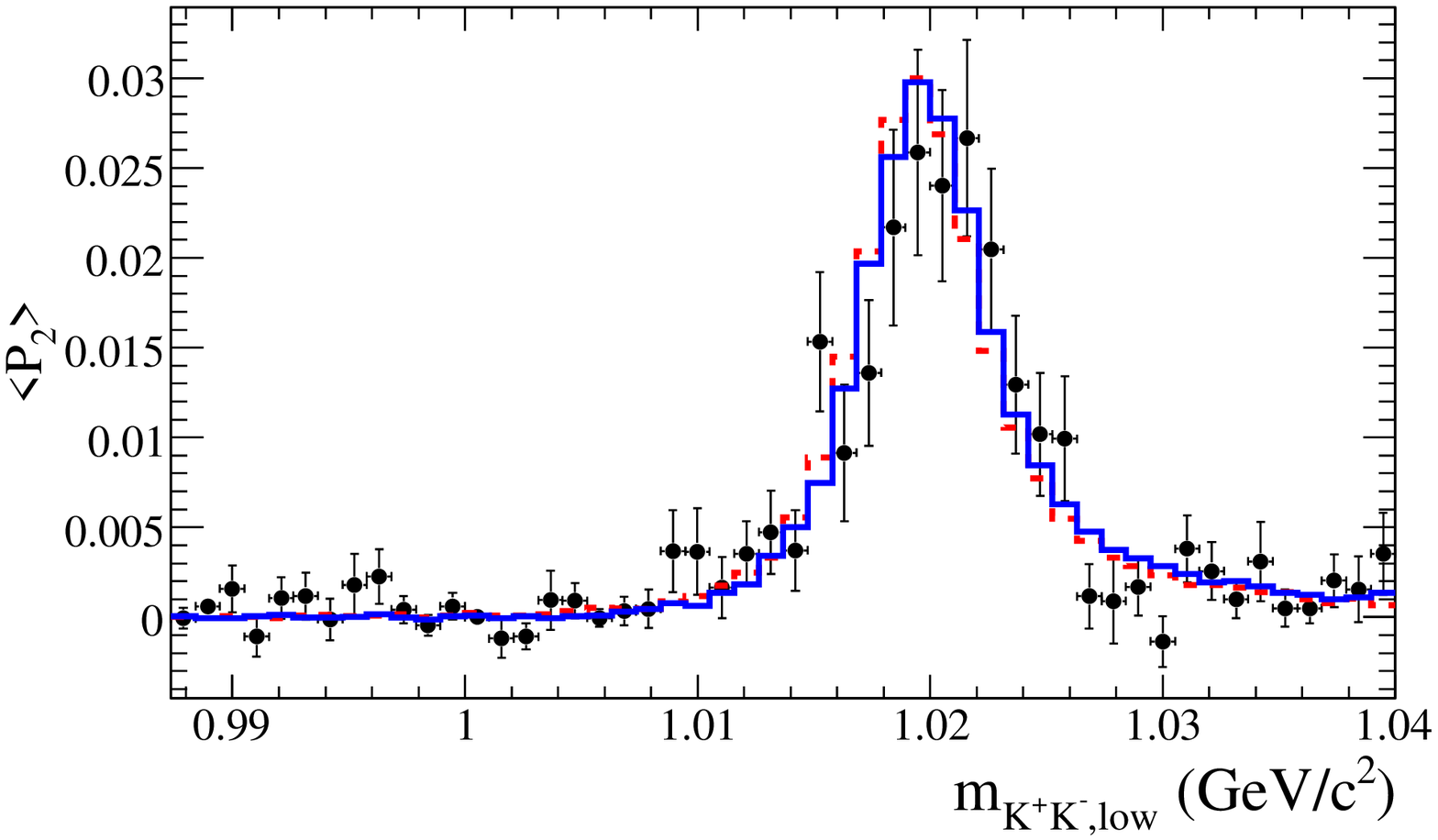}
\caption{\label{fig:angmoments_KKKch_lowmass}  \bkkkboth angular moments in the region $\mab<1.04 \gevcc$, computed for signal-weighted data, compared to Model A (dashed line) and Model B (solid line).  The signal weighting is performed using the \splot method.
}
\end{figure}

If we assume that no $K_{1}K_{2}$ partial-waves of a higher order 
than D-wave contribute, and we temporarily ignore the effects of 
symmetrization, then we can express the overall decay amplitude as a sum
of S-wave, P-wave, and D-wave terms:
\begin{eqnarray}
\Amp(m_{12},\cosHth) & = & \Amp_{S} P_0(\cosHth) + \Amp_P e^{i\phi_P} P_{1}(\cosHth) \nonumber \\
 &  &  + \Amp_D e^{i\phi_D} P_{2}(\cosHth),
\end{eqnarray}
where $\Amp_{k}$ and $\phi_{k}$ are real-valued functions of \mab, and 
we have factored out the S-wave phase.  
We can then  calculate the angular moments:
\begin{eqnarray}
 \langle P_0\rangle & = & \frac{{\cal A}_S^2 + {\cal A}_P^2 + {\cal A}_D^2}{\sqrt{2}} \nonumber\,, \\
 \langle P_1\rangle & = & \sqrt{2} {\cal A}_S {\cal A}_P \cos{\phi_P} + \frac{2\sqrt{10}}{5} {\cal A}_P {\cal A}_D \cos{(\phi_P-\phi_D)} \nonumber\,, \\
 \langle P_2\rangle & = & \sqrt{\frac{2}{5}} {\cal A}_P^2 + \frac{\sqrt{10}}{7}{\cal A}_D^2 + \sqrt{2} {\cal A}_S {\cal A}_D \cos{\phi_D} \nonumber\,, \\
 \langle P_3\rangle & = & \frac{3}{5}\sqrt{\frac{30}{7}} {\cal A}_P {\cal A}_D \cos{(\phi_P-\phi_D)} \nonumber\,, \\
 \langle P_4\rangle & = & {\frac{\sqrt{18}}{7}}{\cal A}_D^2. 
\label{eq:plmomentlist}
\end{eqnarray}
The symmetrization of the \bkkkboth amplitude spoils the validity of 
Eq.~\eqref{eq:plmomentlist}.  Nevertheless, the angular moments can be calculated 
both for signal-weighted data and for the
fit model, providing a useful tool for checking how well the isobar model describes the data.
In Fig.~\ref{fig:angmoments_KKKch}, we show angular moments for data compared to 
the fit model, in the region of the DP above the $\phiI$.  The data is signal-weighted using the 
\splot~\cite{Pivk:2004ty} technique.  The fit model histograms are made by simulating large numbers
of events based on the fit results.
In Fig.~\ref{fig:angmoments_KKKch_lowmass}, we show
the angular moments in the $\phiI$ region.

The angular moments, in particular $\langle P_2\rangle$, show that Model A 
does not describe the data well 
in the $f_X(1500)$ region.  If we replace the $f_X(1500)$ with
the $\fII$ and the $\ftwop$, there is an improvement in $2\logL$ of
17 units.   As we will discuss shortly, this replacement is also motivated by a
peak in $\langle P_2\rangle$ seen in $\bkksks$.

We also vary the NR model.  The exponential NR model 
is not very flexible; it 
assumes no phase motion and only an S-wave term.  We fit with a
polynomial model [Eq.~\eqref{eq:NRpoly}] instead, which contains S-wave and 
P-wave terms and allows for phase motion.  There is an improvement 
in $2\logL$ of 233 units.  However, the polynomial model has nine more degrees
of freedom than the exponential model.  We refer to this model [which
replaces the $f_X(1500)$ with the $\fII$ and the $\ftwop$, and which uses
the polynomial NR model] hereafter as Model B for \bkkkboth.  
We compare the angular 
moments for Model B to data in Figs.~\ref{fig:angmoments_KKKch} and
\ref{fig:angmoments_KKKch_lowmass}.  Model B
matches the data significantly better than Model A, especially
for $\langle P_1\rangle$ and $\langle P_2\rangle$.

\subsection{\bkksks}
Next we examine \bkksks, initially including the resonances
$\fI$, $f_{X}(1500)$, $\fIII$, and $\chi_{c0}$.  We take the 
$f_{X}(1500)$ mass and width from the \bkkkboth Model A result.  We also
include a polynomial NR model, but without the P-wave term, which
is forbidden.  We call this Model A for \bkksks.  

\begin{figure}[htbp]
\includegraphics[width=8.9cm,keepaspectratio]{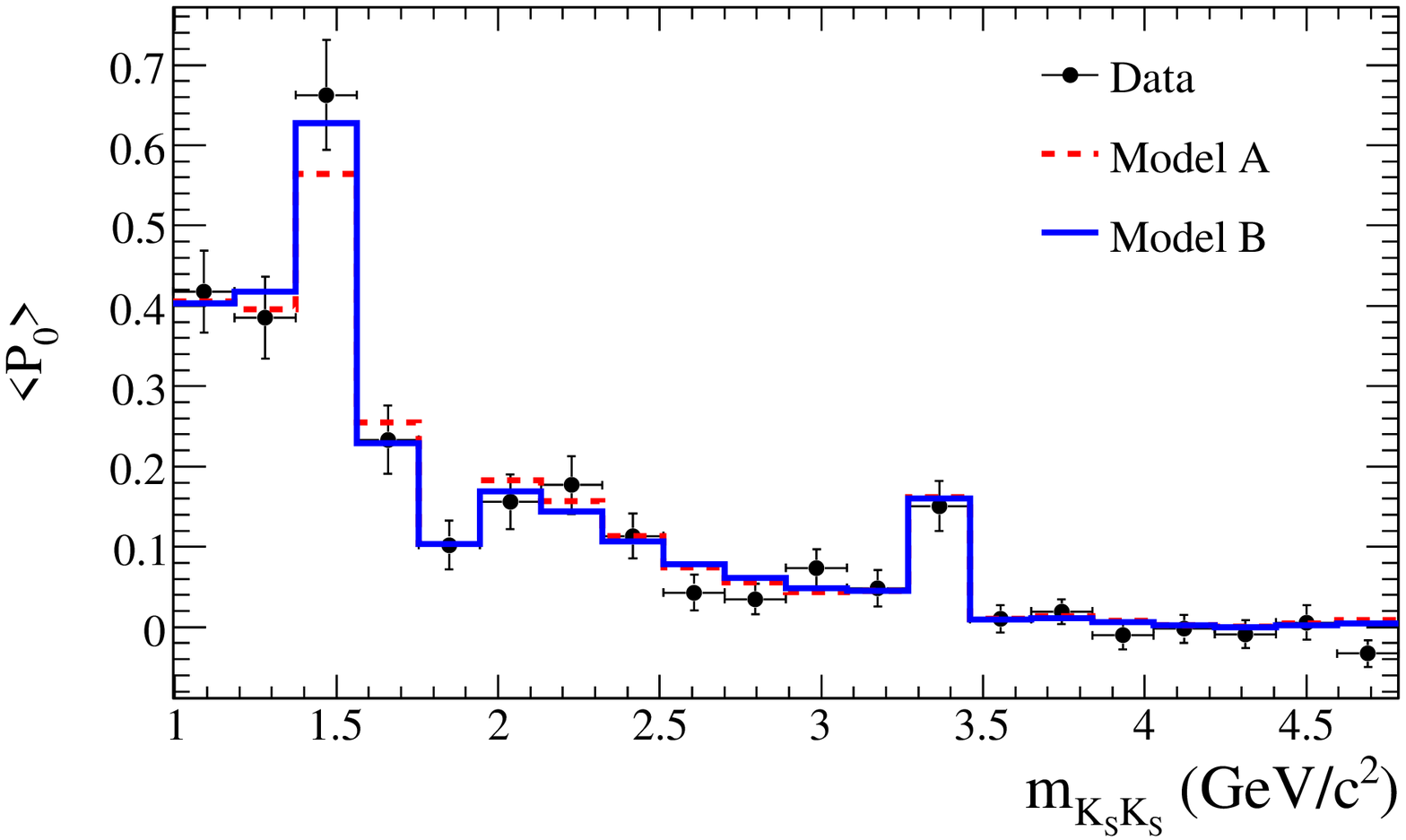}
\includegraphics[width=8.9cm,keepaspectratio]{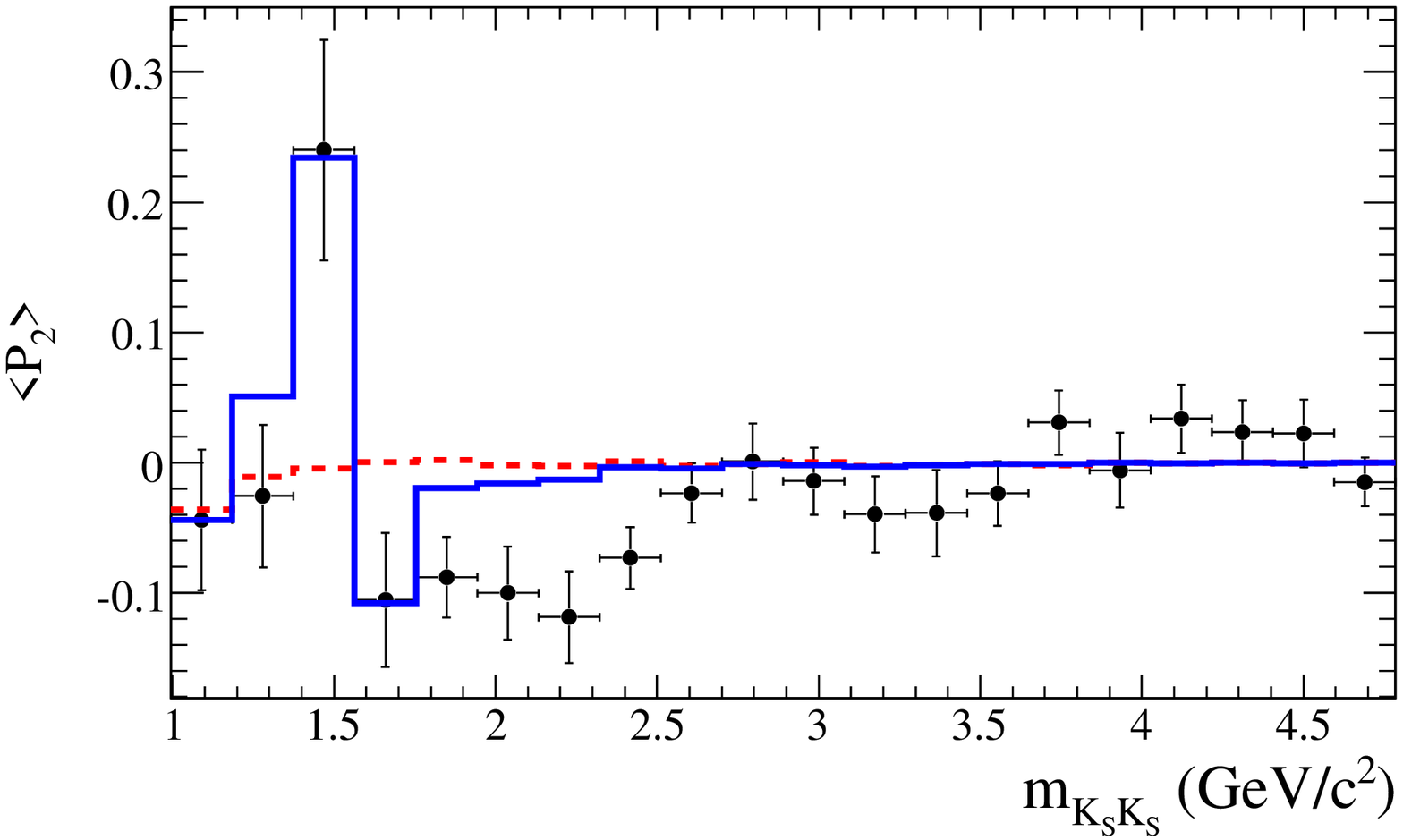}
\includegraphics[width=8.9cm,keepaspectratio]{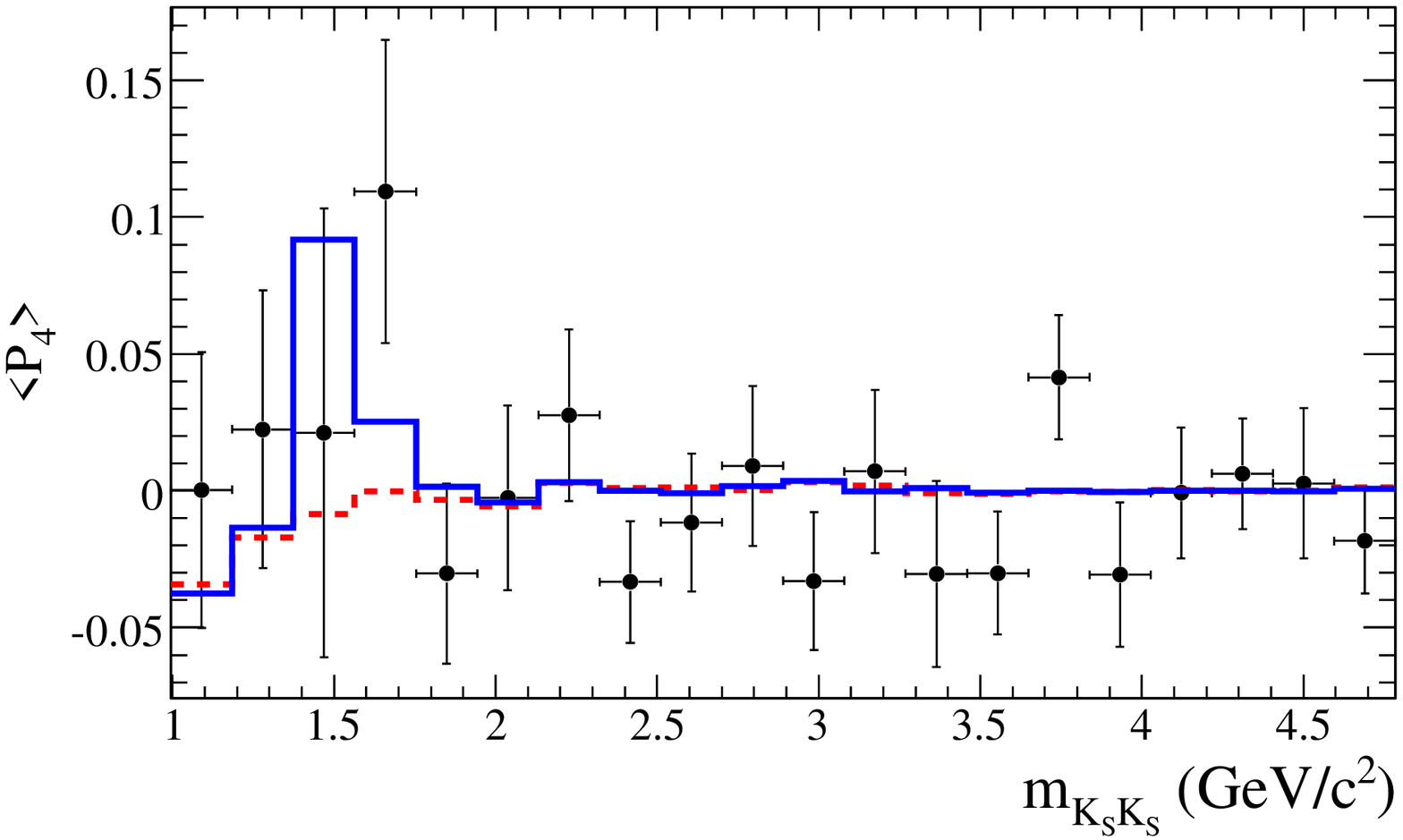}
\caption{\label{fig:angmoments_KKsKs}  \bkksks angular moments computed for 
signal-weighted data, compared to Model A (dashed line) and Model B (solid line).  
The signal weighting is performed using the \splot method.
Events with $\mksks$ near the $D^0$ mass are vetoed.
}
\end{figure}

In Fig.~\ref{fig:angmoments_KKsKs}, we show the angular 
moments for this model, compared to signal-weighted data. 
Assuming there are no higher-order $\KS\KS$ partial waves 
than D-wave, Eq.~\eqref{eq:plmomentlist} is valid for \bkksks.  
However, because odd partial waves are forbidden in this 
channel, the odd angular moments are automatically zero.
The peak in $\langle P_2\rangle$ around $1.5\gevcc$ in 
Fig.~\ref{fig:angmoments_KKsKs} suggests the presence of a 
tensor resonance. We replace the $f_{X}(1500)$ with the
$\fII$ and $\ftwop$, and call this Model B for \bkksks.  
Model B improves $2\logL$ by 37 units over Model A.
The angular moments for Model B are shown in 
Fig.~\ref{fig:angmoments_KKsKs}.  Neither model does a good
job of describing $\langle P_2\rangle$ in the region 
$1.8 <\mab < 2.5 \gevcc$.  As an alternative, we use 
the model from \babar's \bksksks analysis~\cite{Lees:2011nf}, which 
includes \fI, \fIII,
$f_2(2010)$, \chiczero, and an exponential
NR model like in Eq.~\eqref{eq:NRexp}, except without 
the second term.  For this model,
$2\logL$ is 52 units worse than for Model B.  We then
add the \ftwop to this model, but its $2\logL$ is 
still 19 units worse than Model B.
Adding the $f_2(2010)$ or
$f_2(2300)$ resonance to Model B significantly improves 
$2\logL$ and improves the modeling of the $\langle P_2\rangle$
distribution, but no evidence for these resonances is seen
in \bkkkboth, which has a much higher signal yield. Therefore, we do
not include either of these resonances in our model.  We
will, however, include these resonances as part of our 
evaluation of systematic uncertainties.

\subsection{\bkkks}
Lastly, we examine \bkkks.  We initially fit with the same
model used in \babar's previous analysis, which includes
the resonances $\phiI$, $\fI$,
$f_{X}(1500)$, $\fIII$, and $\chi_{c0}$, and the extended 
exponential NR model given in Eq.~\eqref{eq:NRexpext}. We
take the mass and width of the $f_{X}(1500)$ from the
\bkkkboth Model A result. We hereafter refer to this model as 
Model A for \bkkks.
Belle's most recent \bkkks analysis uses this same
model, although with a different mass and width for the
$f_{X}(1500)$.  

\begin{figure*}[htbp]
\includegraphics[width=8.9cm,keepaspectratio]{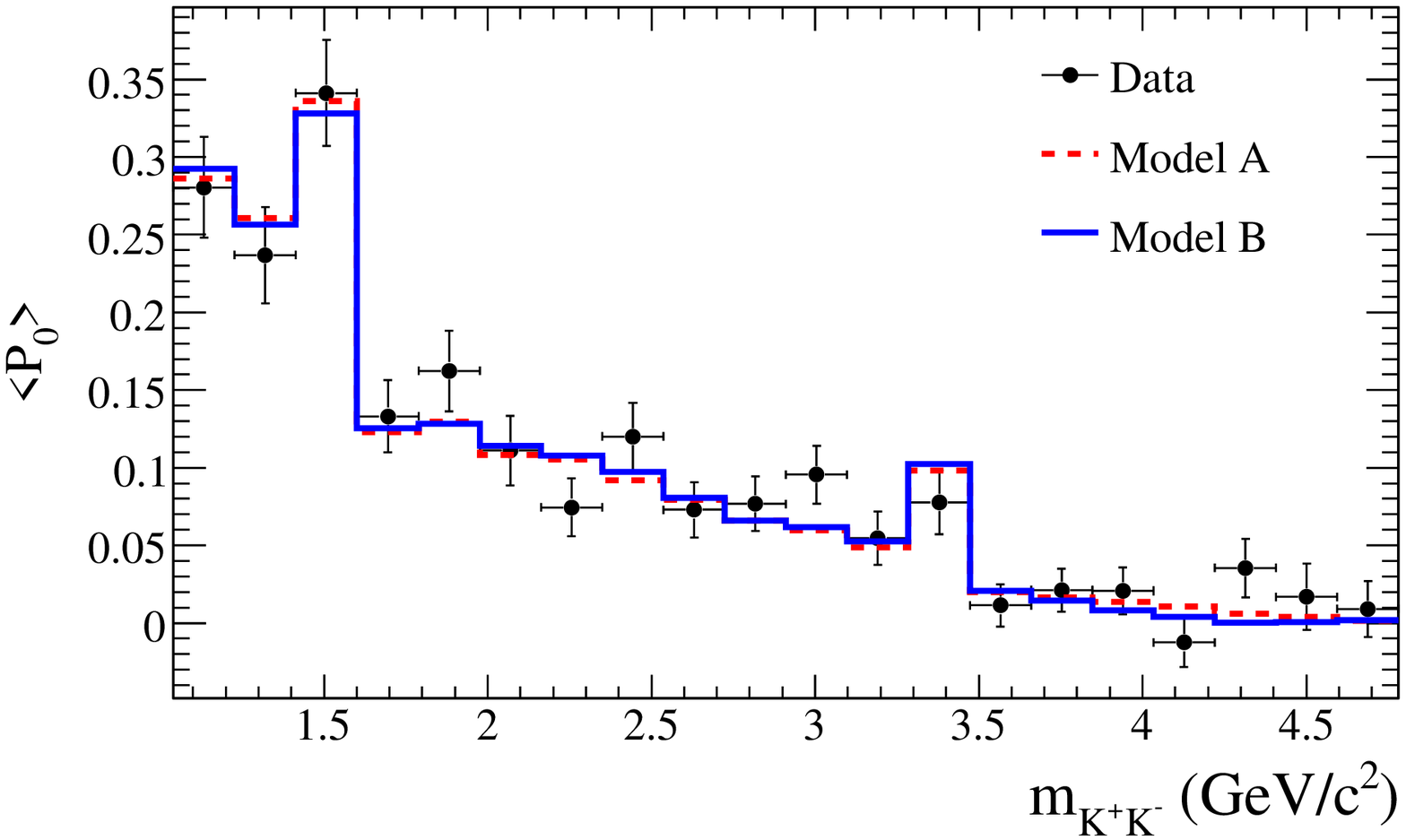}
\includegraphics[width=8.9cm,keepaspectratio]{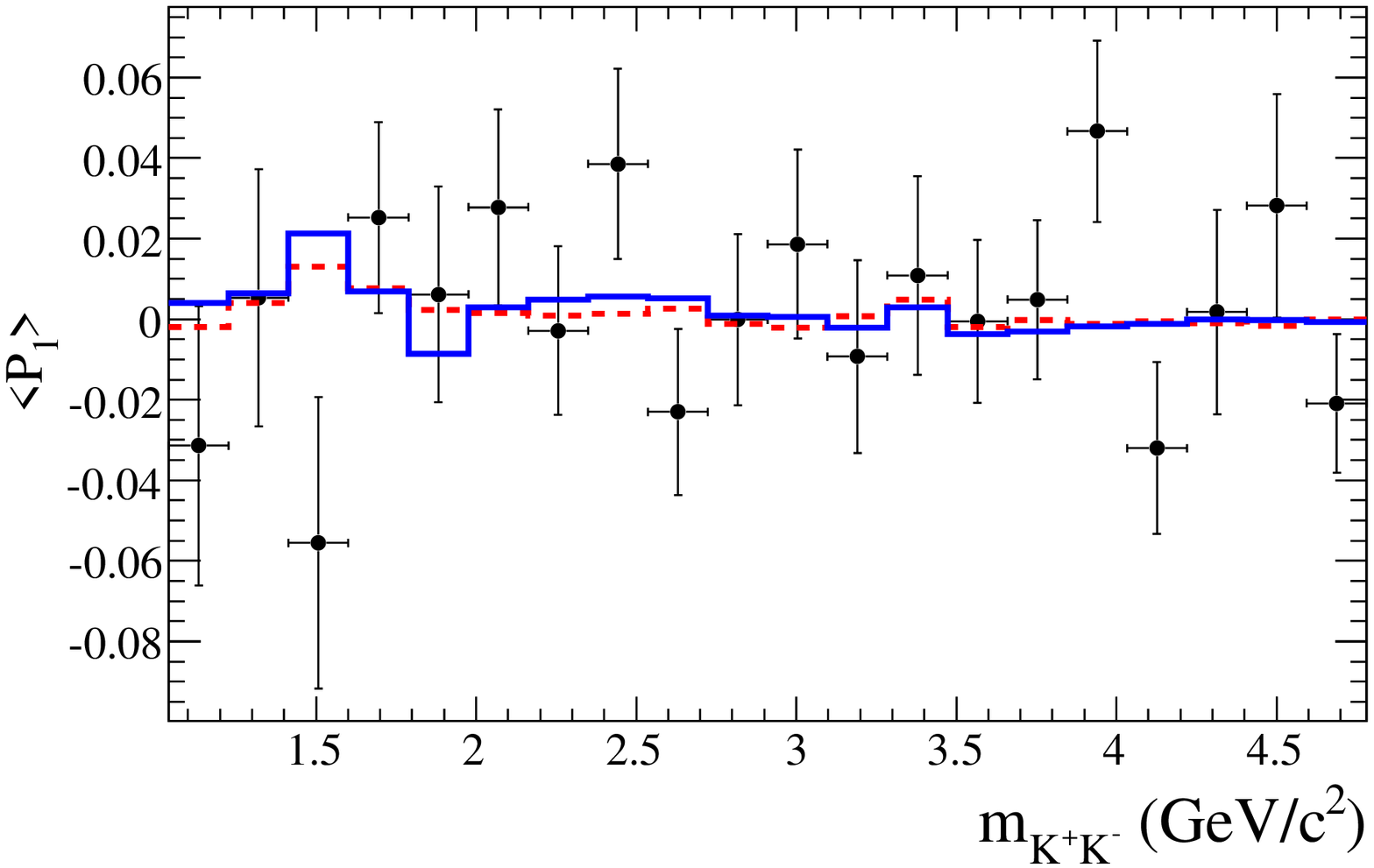}
\includegraphics[width=8.9cm,keepaspectratio]{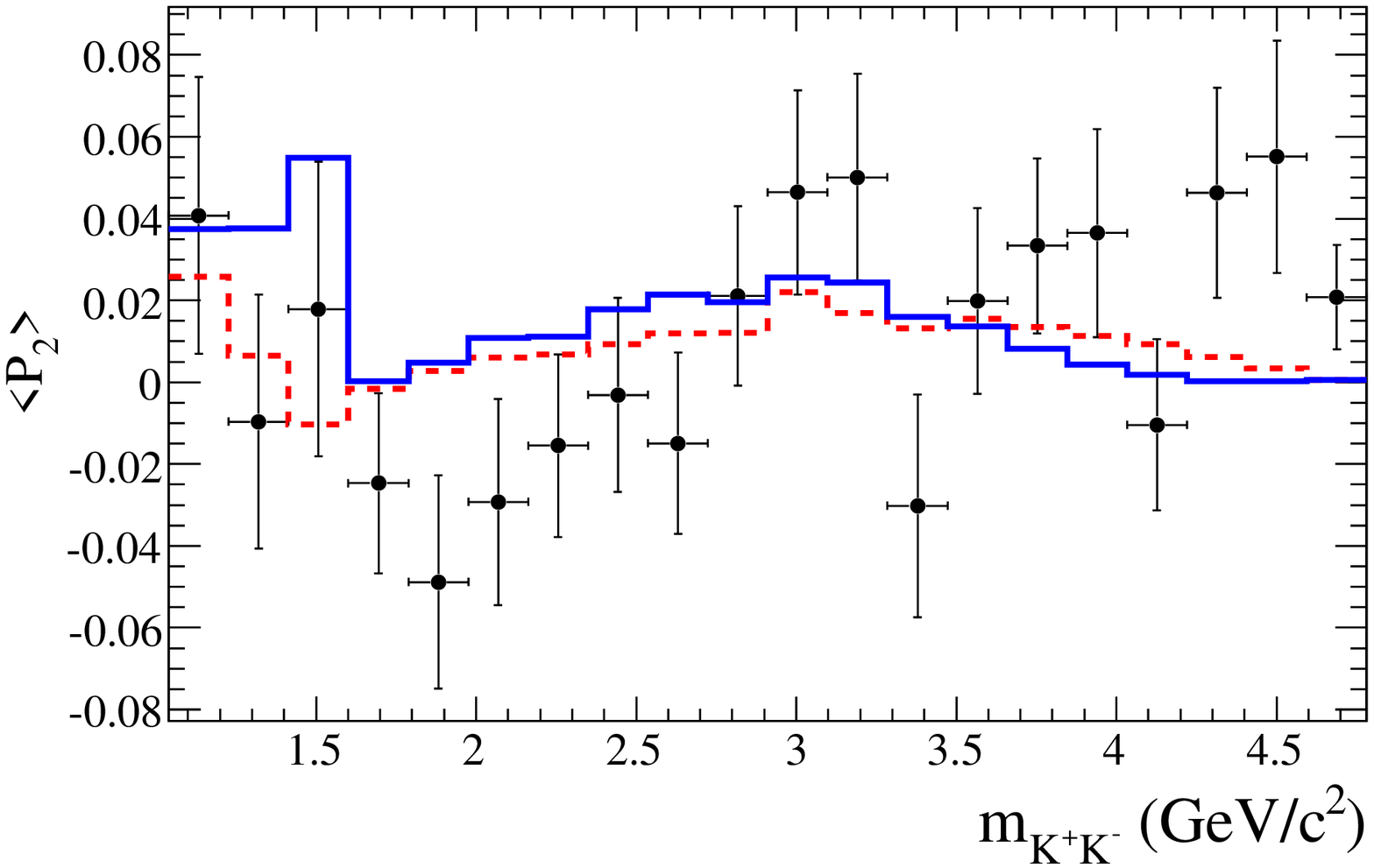}
\includegraphics[width=8.9cm,keepaspectratio]{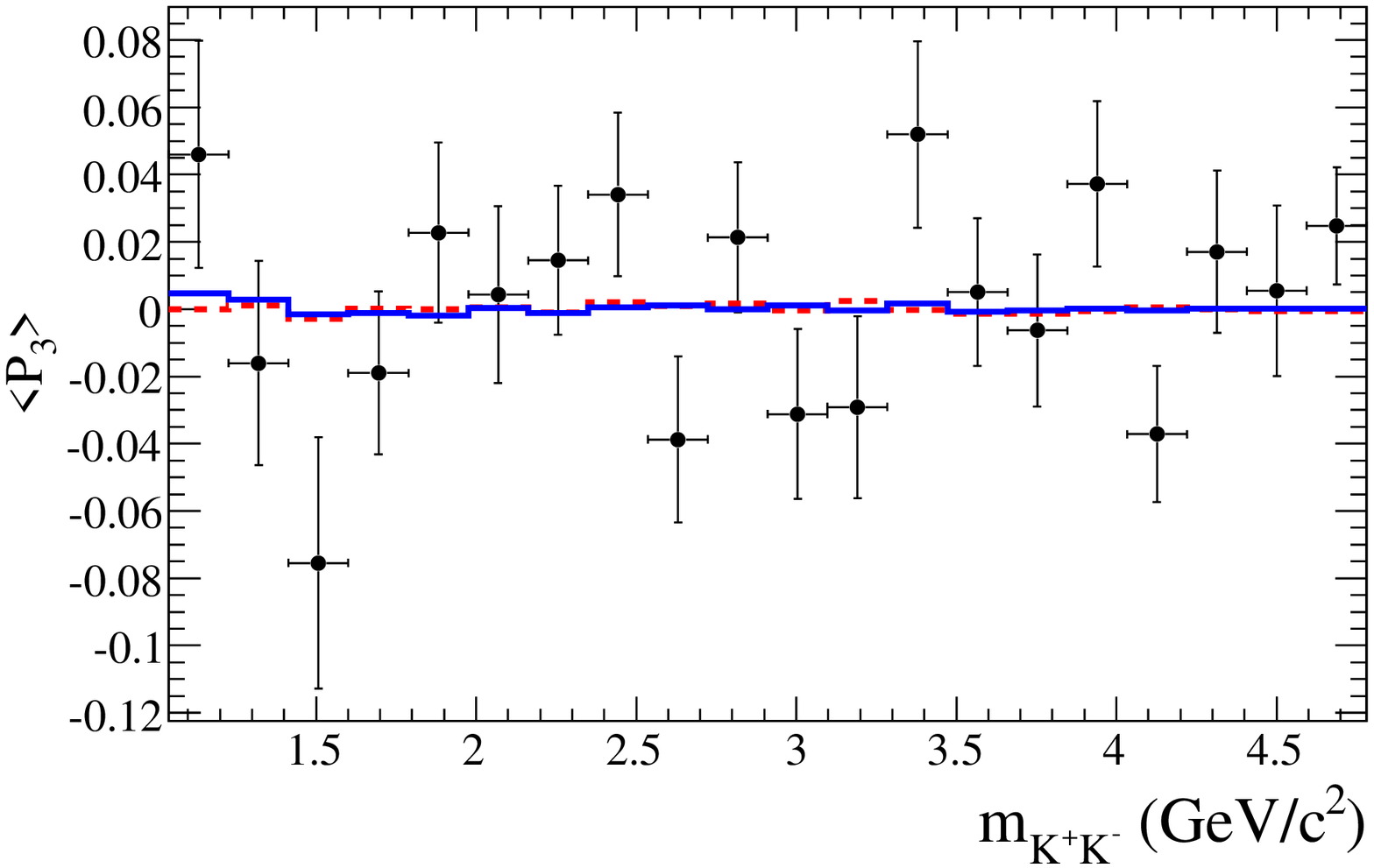}
\includegraphics[width=8.9cm,keepaspectratio]{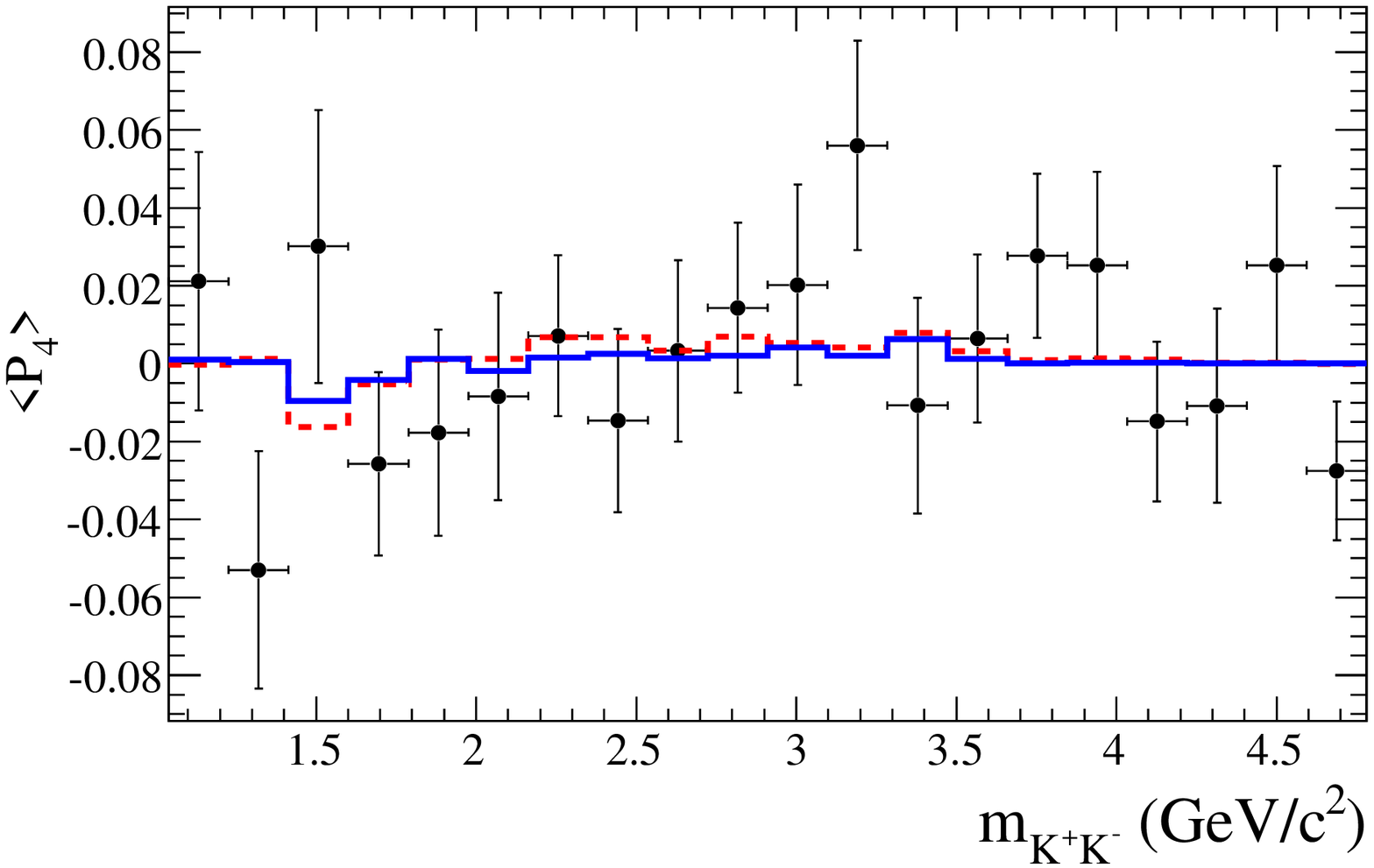}
\caption{\label{fig:angmoments_KKKs}  \bkkks angular moments in the region $\mab>1.04 \gevcc$, computed for 
signal-weighted data, compared to Model A (dashed line) and Model B (solid line).  
The signal weighting is performed using the \splot method.  The plots are made using 
the \Kspp mode only.  Events with $m_{\Kpm\KS}$ near the $\Dp$ or $\Ds$ mass
are vetoed.
}
\end{figure*}

Using Model A, we obtain fit results consistent with \babar's
previous measurement.  In Fig.~\ref{fig:angmoments_KKKs}, 
we show the angular moments for this model compared to data.
The angular moments in \bkkks are complicated due to the relative
minus sign between \Bz and \Bzb amplitudes for odd-$L$ resonances
[Eq.~\eqref{eq:signflip}].  
To account for this, when computing the odd angular 
moments, we weight the events
by $-\Qtag$, where $\Qtag$ is the flavor of the
$\Bz_{\rm tag}$.  Then, Eq.~\eqref{eq:plmomentlist} is valid for \bkkks, except
that for the odd angular moments, the right-hand side must be multiplied by 
\mbox{$(1-2w)/((\deltamd\tau_{\Bz})^2+1)$}, 
which is a dilution factor caused by mistagging and \Bz-\Bzb mixing.

We replace the $f_X(1500)$ by the $\fII$, $\ftwop$, and $\fIII$, and
this improves $2\logL$ by $18$ units.  We then replace the NR model
with a polynomial NR model containing S-wave and P-wave terms.  
This improves $2\logL$ by an additional $13$ units.  We refer to this model
as Model
B for $\bkkks$;  its angular moments are shown in 
Fig.~\ref{fig:angmoments_KKKs}.  The improvement of Model B over Model A
is not evident by examining the angular moments by eye, but Model B 
provides a considerably better likelihood.

\subsection{Conclusion}
For each of the three decay modes, Model B produces a better fit to the data
than Model A, at the cost of more free parameters. Model B also eliminates
the need for the hypothetical $f_X(1500)$ state.  The NR parameterization used
in Model B greatly improves the fit likelihood in \bkkkboth, and its large
number of parameters make it very flexible.  A benefit of this flexibility
is that the fit results are then less dependent on the particular 
choice of NR parameterization.
Model B also has a similar form in 
all three modes (the only difference is the absence of P-wave states in 
\bkksks), aiding comparison of results between the modes.  
In addition to the studies already mentioned, we tested for the presence
of the $f_0(1370)$, $f_2(1270)$, $f_2(2010)$, and $f_2(2300)$
in each mode, and in \bkkkboth and \bkkks, we tested for the $\phi(1680)$.
We did not find evidence for any of these resonances. We also tested
for the following isospin-1 resonances: $a_0^0(1450)$ in each of the
three modes, and $a_0^\pm(980)$ and $a_0^\pm(1450)$ in 
\bkksks and \bkkks only.  We did not find evidence for any of 
these resonances.
We henceforth use Model B as the nominal fit model for each mode,
and only include these additional resonances to evaluate
systematic uncertainties.

\clearpage

%% file: FitResults.tex
\section{RESULTS}
\label{sec:fitResults}

\subsection{\bkkkboth}
\label{sec:fitResultsKKKch}
The maximum-likelihood fit of $12240$ candidates results in 
yields of $5269\pm 84$ signal events, $6016\pm 91$ continuum 
events, and  $912\pm 54$ $\BB$ events, 
where the uncertainties are statistical only.    

In order to limit the number of fit parameters and improve
fit stability, we constrain the \Acp and $\Delta\phi$ of the 
$\fII$, $\ftwop$, and $\fIII$ to be equal in the fit 
({\ie, the $b$ and $\delta$ parameters, defined in Eq.~\eqref{eq:isobarPars},
are constrained to be the same for these isobars).  We also 
constrain the \Acp and $\Delta\phi$ of the S-wave and 
P-wave NR terms to be equal. Since the \Acp in 
$\Bp\to~c\overline{c}\Kp$ decays is known to be small~\cite{Nakamura:2010zzi},  
we fix the \Acp of the $\chiczero$ to 0 in the fit.  
Only {\em relative\/} values of $c$, $\phi$, and $\Delta\phi$ 
are measurable, so as references we fix $c=1$ and $\phi=0$ for the 
NR term $a_{S0}$ and $\Delta\phi=0$ for all NR terms.

When the fit is repeated starting from input parameter values 
randomly chosen within wide ranges above and below the nominal 
values for the magnitudes and within the $[0-360^{\circ}]$ interval 
for the phases, we observe convergence toward two solutions with 
minimum values of the negative log likelihood function $-2\logL$
that are separated by $5.6$ units.  We will refer to them as
Solution I (the global minimum) and Solution II (a local
minimum).  The two solutions have nearly identical values for
most parameters, but differ greatly for some of the isobar
parameters.  The isobar parameters for both solutions are given in 
Table~\ref{tab:isobarSummary_KKKch_SolnIandII}.
The correlation matrices of the isobar parameters are given 
in Ref.~\cite{epaps}.

\begin{table}[htbp]
\center
\caption{Isobar parameters (defined in Eq.~\eqref{eq:isobarPars}) for \bkkkboth, Solutions I and II.  
The same $b$ and $\delta$ parameters are 
used for the $\fII$, $\ftwop$, and $\fIII$, and we choose to quote their
fitted values in the $\ftwop$ rows.  The NR coefficients are defined in Eq.~\eqref{eq:NRpoly}.  
Phases are given in degrees.
Only statistical uncertainties are given.  
}
\begin{tabular}{ll|cc}
\hline \hline
\multicolumn{2}{l|}{Parameter }   &  Solution I &  Solution II   \\
\hline
\noalign{\vskip1pt} 
$\phiI$$\Kpm$        &    $c$   &    $0.0311\pm 0.0043 $   &     $0.043\pm 0.009 $   \\
     &    $\phi$   &    $177\pm 13 $   &     $-53\pm 13 $   \\
     &    $b$   &    $-0.064\pm 0.022 $   &     $-0.037\pm 0.022 $   \\
     &    $\delta$   &    $11\pm 7 $   &     $-10\pm 6 $   \\
\hline
\noalign{\vskip1pt} 
$\fI$$\Kpm$        &    $c$   &    $1.64\pm 0.23 $   &     $1.5\pm 0.5 $   \\
     &    $\phi$   &    $118\pm 12 $   &     $-34\pm 11 $   \\
     &    $b$   &    $0.040\pm 0.041 $   &     $-0.32\pm 0.11 $   \\
     &    $\delta$   &    $4.5\pm 3.3 $   &     $-12\pm 7 $   \\
\hline
\noalign{\vskip1pt} 
$\fII$$\Kpm$        &    $c$   &    $0.179\pm 0.031 $   &     $0.28\pm 0.07 $   \\
     &    $\phi$   &    $-45\pm 11 $   &     $-41\pm 15 $   \\
\hline
\noalign{\vskip1pt} 
$\ftwop$$\Kpm$        &    $c$   &    $0.00130\pm 0.00022 $   &     $0.00160\pm 0.00038 $   \\
     &    $\phi$   &    $34\pm 10 $   &     $43\pm 16 $   \\
     &    $b$   &    $-0.07\pm 0.05 $   &     $-0.09\pm 0.05 $   \\
     &    $\delta$   &    $-0.8\pm 2.8 $   &     $0.5\pm 2.6 $   \\
\hline
\noalign{\vskip1pt} 
$\fIII$$\Kpm$        &    $c$   &    $0.254\pm 0.044 $   &     $0.32\pm 0.08 $   \\
     &    $\phi$   &    $44\pm 9 $   &     $45\pm 16 $   \\
\hline
\noalign{\vskip1pt} 
$\chiczero$$\Kpm$        &    $c$   &    $0.114\pm 0.017 $   &     $0.170\pm 0.038 $   \\
     &    $\phi$   &    $9\pm 12 $   &     $31\pm 15 $   \\
     &    $\delta$   &    $-2\pm 6 $   &     $-2\pm 6 $   \\
\hline
NR   &  &  &  \\
         &    $b$   &    $-0.030\pm 0.022 $   &     $-0.062\pm 0.024 $   \\
$a_{S0}$  &   $c$   &   $  1.0 $ (fixed) &   $  1.0 $ (fixed)  \\
    &    $\phi$    &     $  0 $ (fixed)  &    $ 0   $ (fixed) \\
$a_{S1}$        &    $c$   &    $2.09\pm 0.38 $   &     $0.4\pm 1.2 $   \\
     &    $\phi$   &    $160\pm 14 $   &     $1\pm 162 $   \\
$a_{S2}$        &    $c$   &    $0.33\pm 0.08 $   &     $0.45\pm 0.35 $   \\
     &    $\phi$   &    $157\pm 12 $   &     $-65\pm 19 $   \\
$a_{P0}$        &    $c$   &    $1.6\pm 0.5 $   &     $2.3\pm 1.9 $   \\
     &    $\phi$   &    $7\pm 20 $   &     $130\pm 25 $   \\
$a_{P1}$        &    $c$   &    $0.80\pm 0.07 $   &     $0.85\pm 0.30 $   \\
     &    $\phi$   &    $-159\pm 6 $   &     $-114\pm 12 $   \\
$a_{P2}$        &    $c$   &    $0.49\pm 0.15 $   &     $0.77\pm 0.38 $   \\
     &    $\phi$   &    $-110\pm 17 $   &     $-60\pm 18 $   \\
\hline  \hline
\end{tabular}
\label{tab:isobarSummary_KKKch_SolnIandII}. 
\end{table}

Figure~\ref{fig:kkkch_projections} shows distributions of
\mes, \de and the \nn output.  Figure~\ref{fig:kkkch_DP}
shows the \mab, \mbc, and \mac
distributions for signal- and background-weighted events, using the 
\splot~\cite{Pivk:2004ty} technique.

\begin{figure*}[htbp]
\includegraphics[width=5.9cm,keepaspectratio]{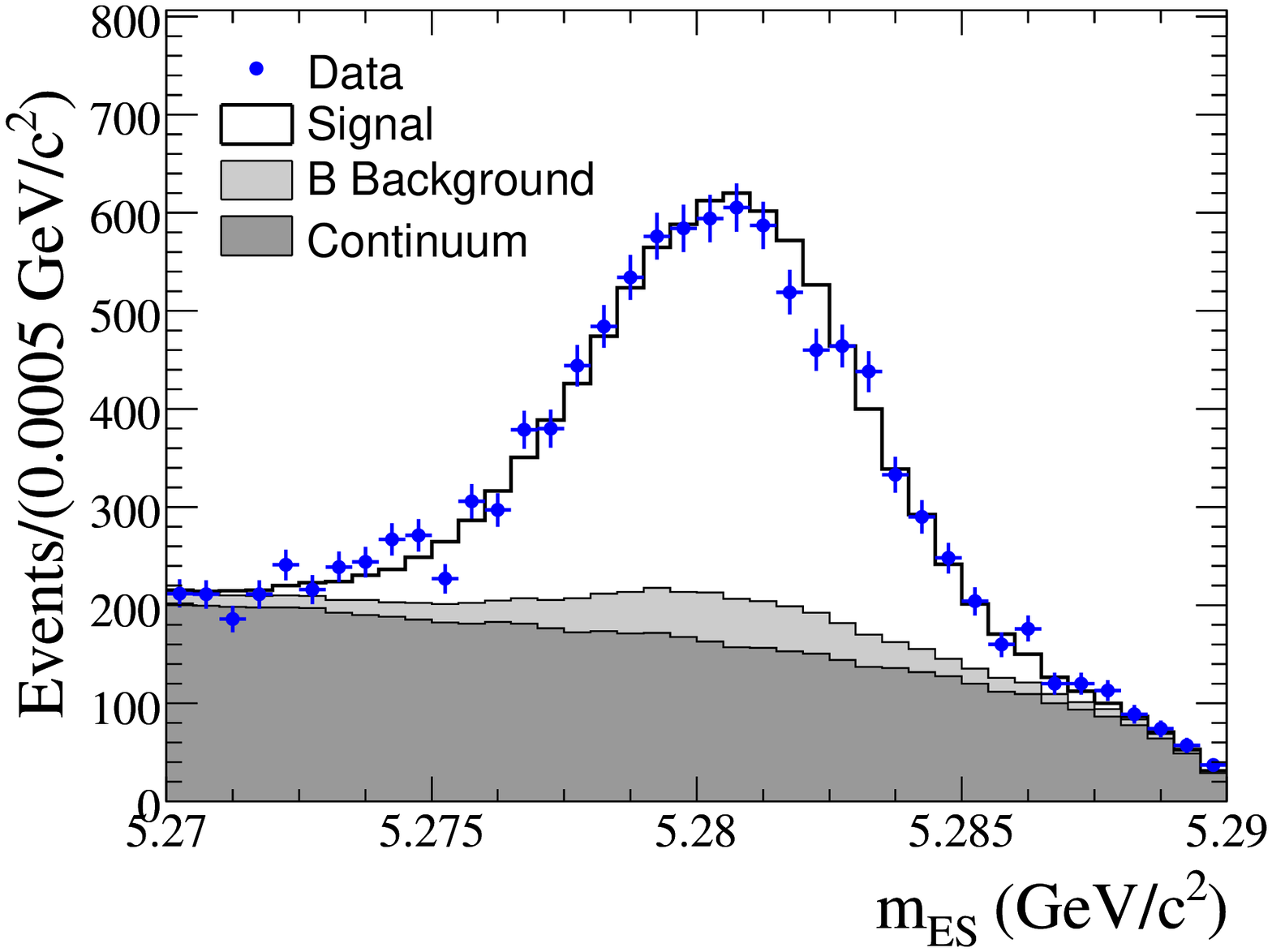}
\includegraphics[width=5.9cm,keepaspectratio]{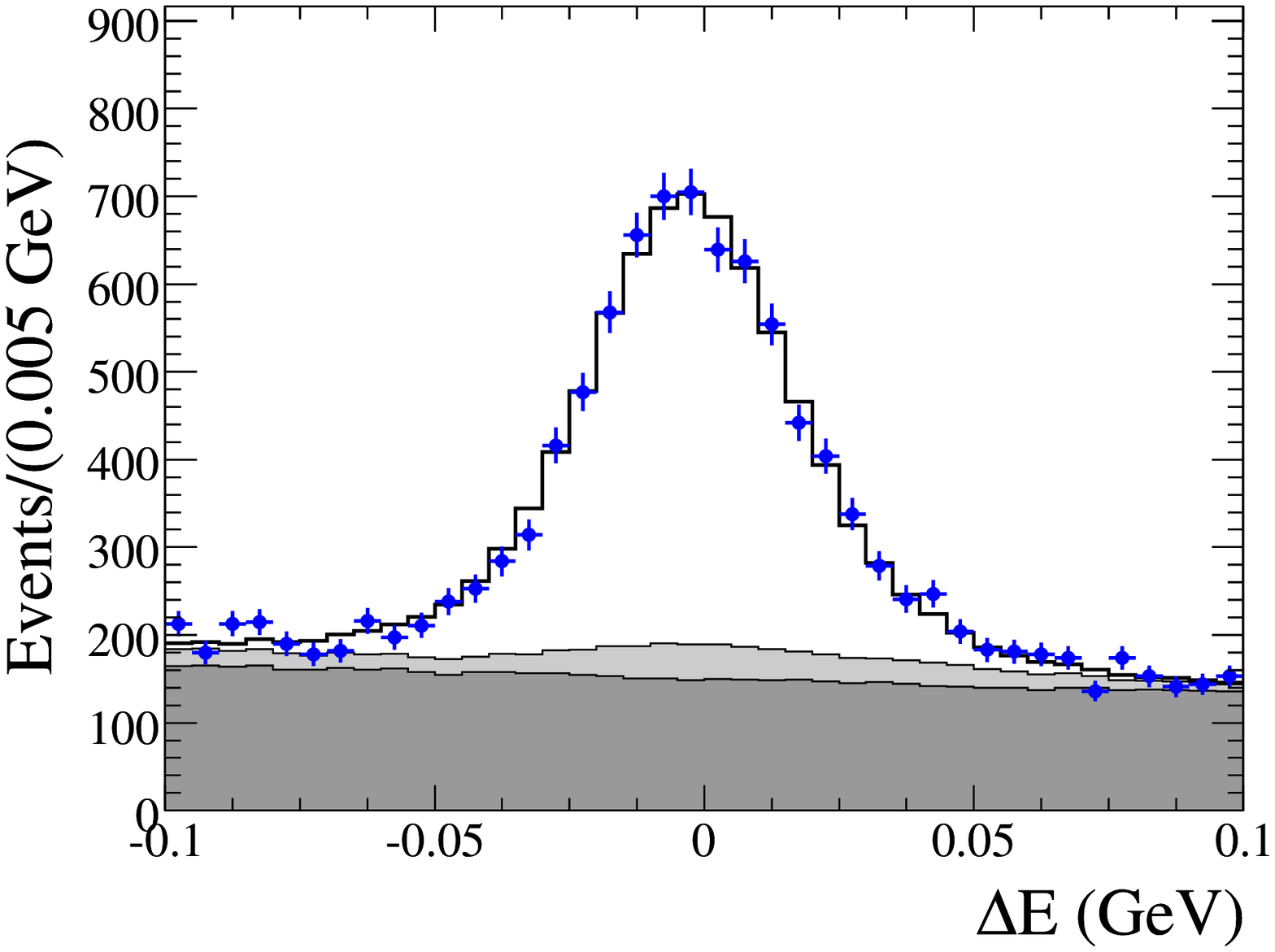}
\includegraphics[width=5.9cm,keepaspectratio]{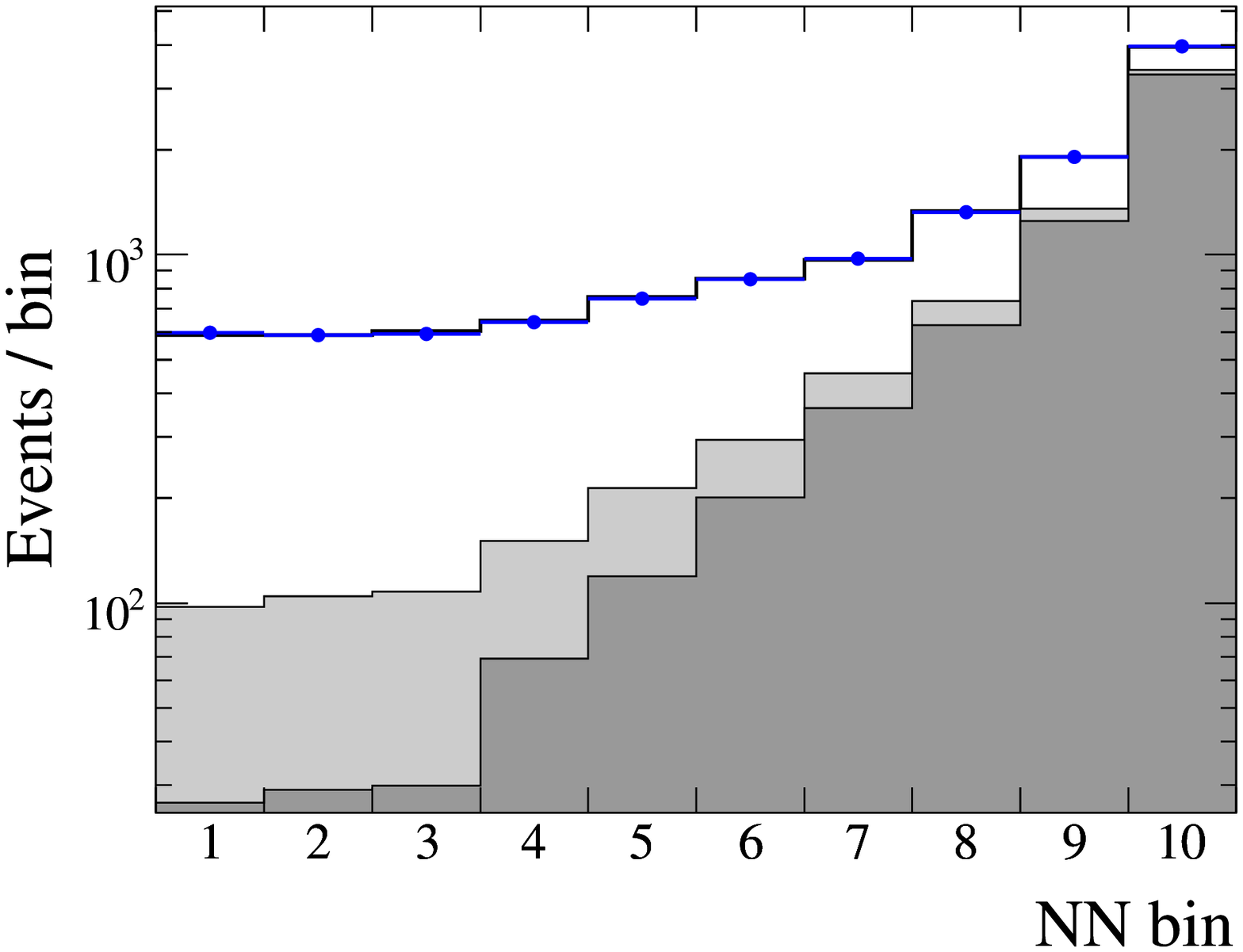}
  \caption{\label{fig:kkkch_projections}Distributions of $\mes$ (left), 
	$\de$ (center), and $\nn$ output (right) for \bkkkboth.
    The NN output is shown in vertical log scale.}
\end{figure*}

\begin{figure*}[htbp]
\includegraphics[width=7.9cm,keepaspectratio]{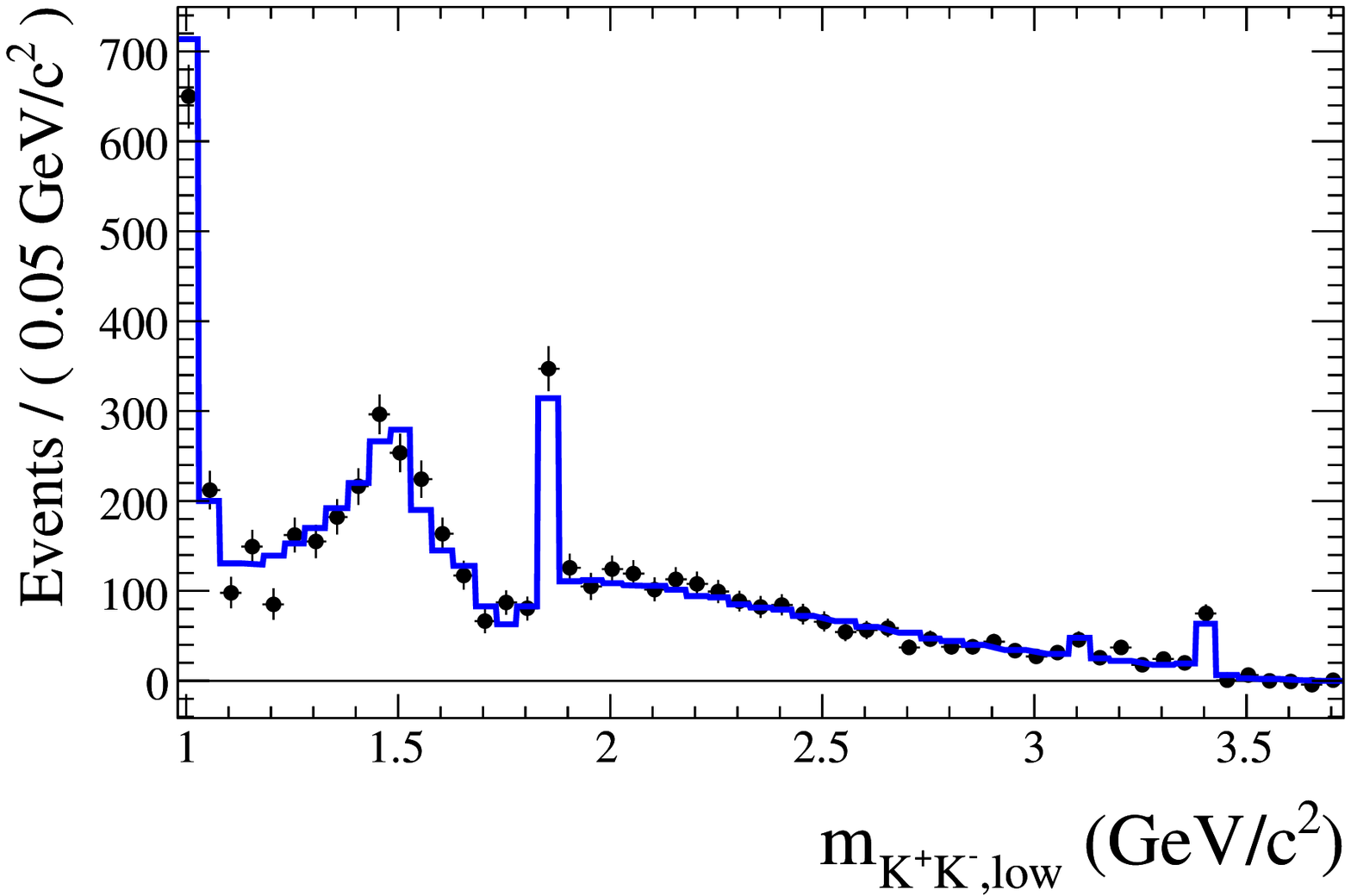}
\includegraphics[width=7.9cm,keepaspectratio]{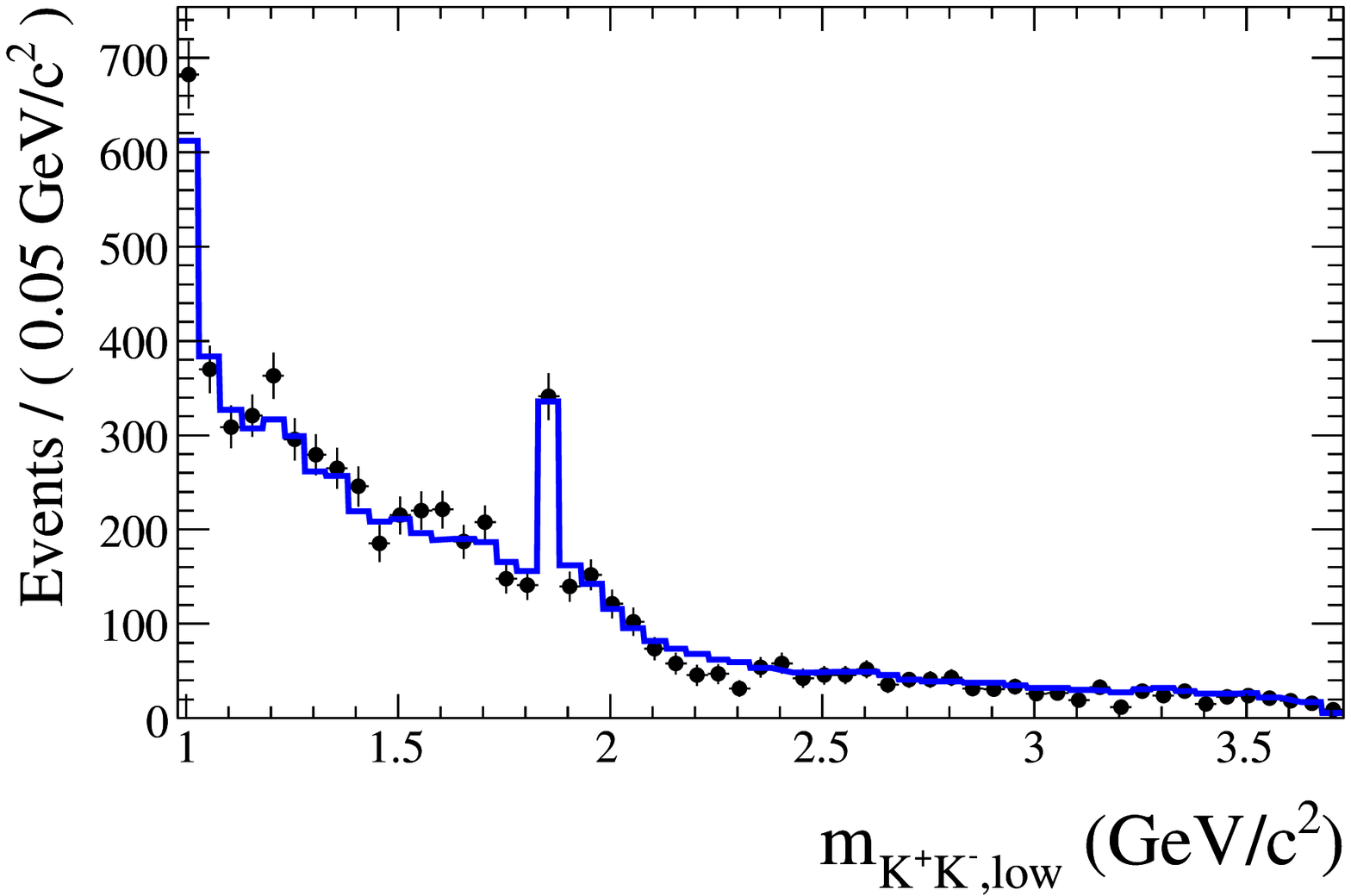}
\includegraphics[width=7.9cm,keepaspectratio]{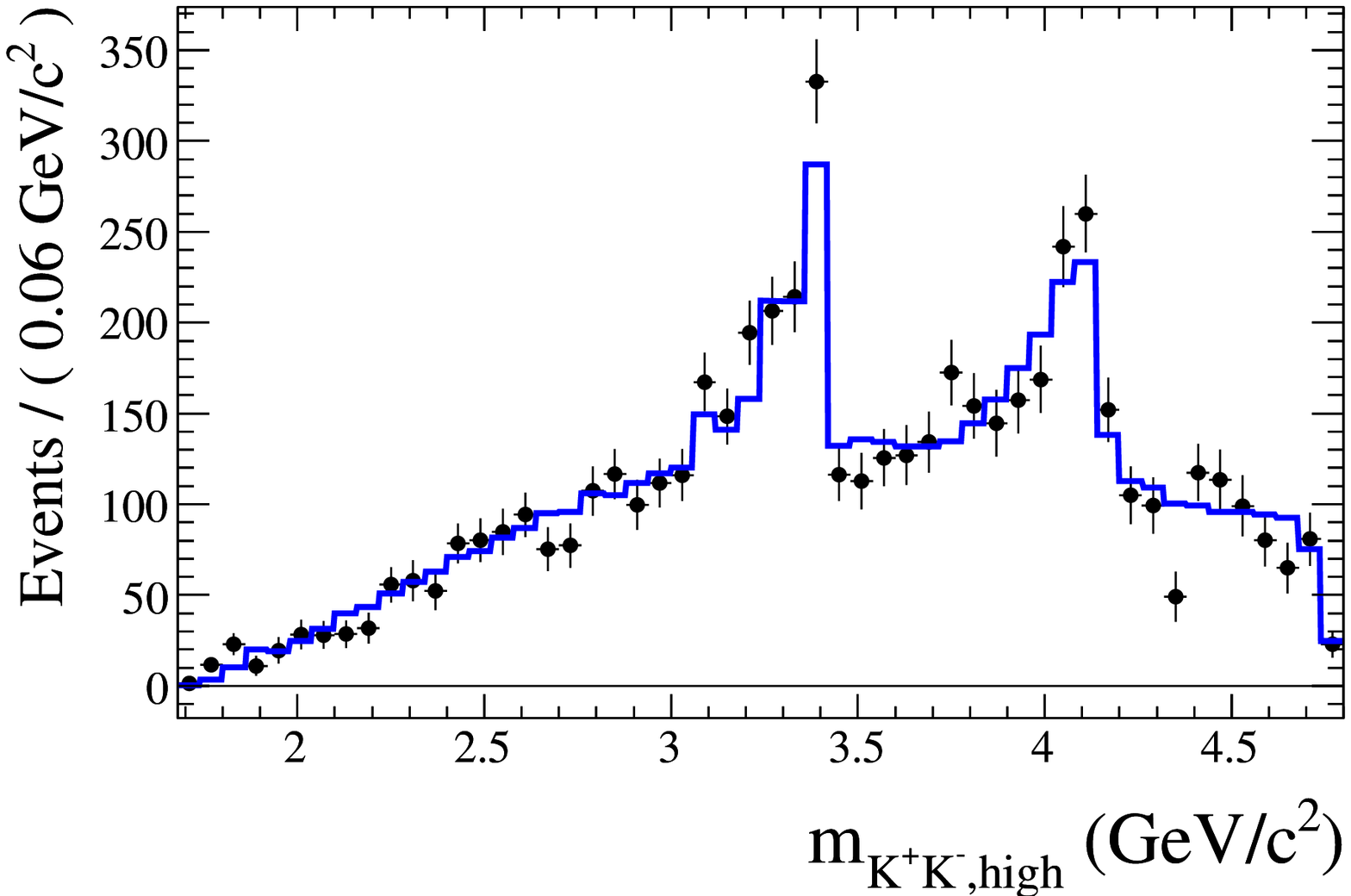}
\includegraphics[width=7.9cm,keepaspectratio]{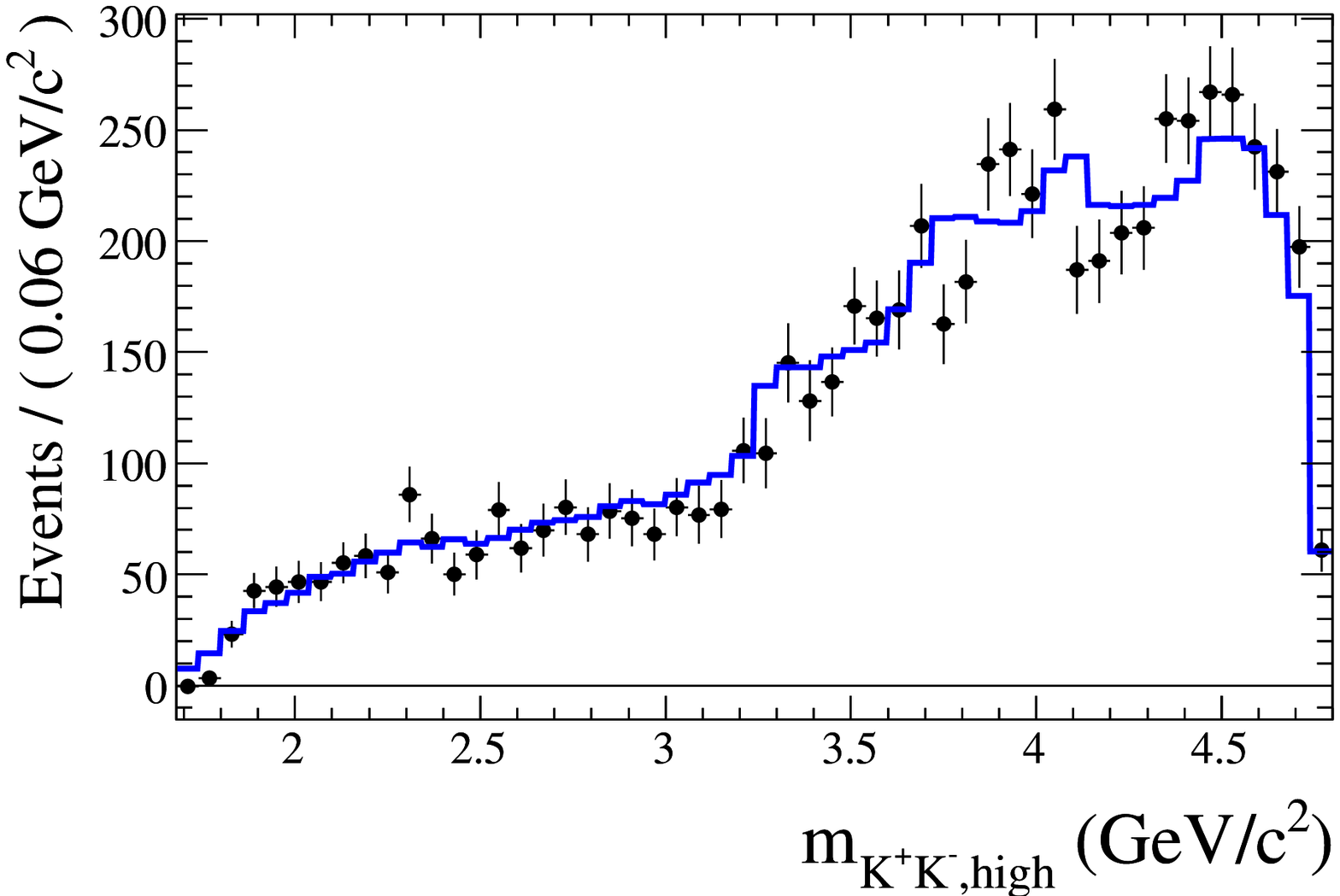}
\includegraphics[width=7.9cm,keepaspectratio]{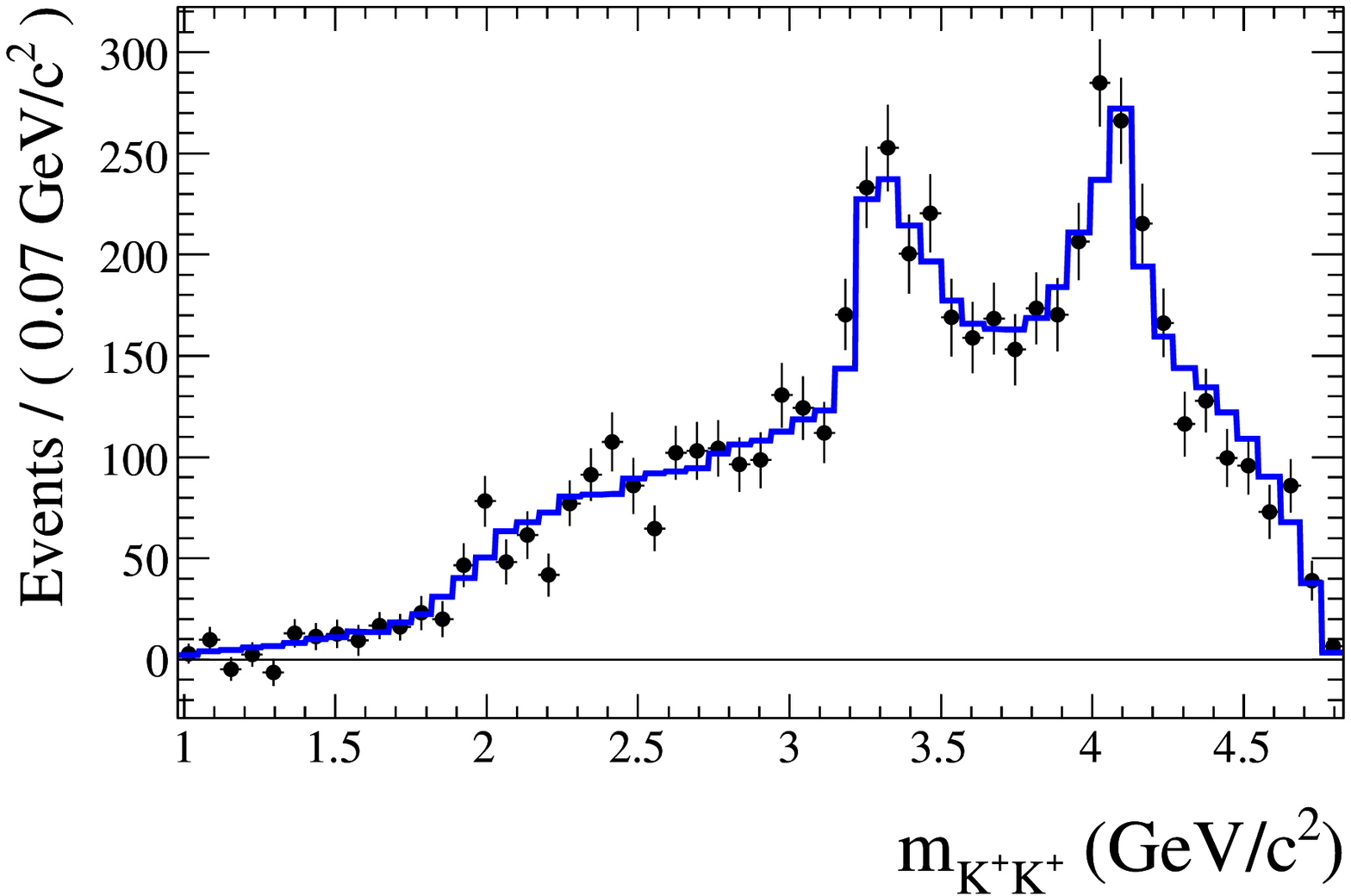}
\includegraphics[width=7.9cm,keepaspectratio]{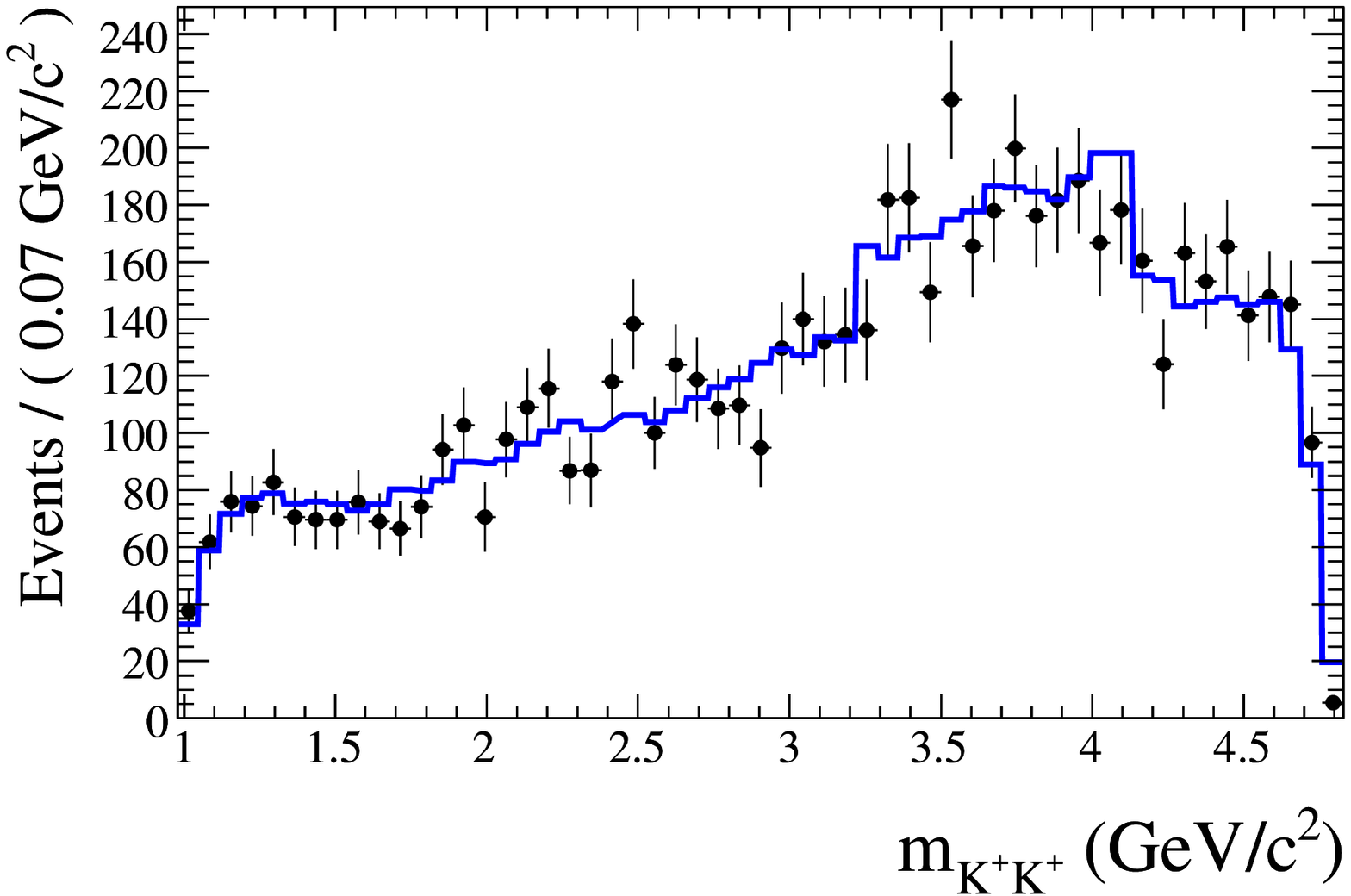}
\caption{\label{fig:kkkch_DP}Distributions of $\mab=\mkkloL$, $\mbc=\mkkhiL$, and $\mac=\mkpkpL$,
 for signal-weighted (left) and background-weighted (right) \bkkkboth candidates in data.  The event weighting is performed using the \splot method.   
The fit model (histograms)
is shown superimposed over data (points). The signal includes the signal-like \BB 
backgrounds (classes 5 and 6 in Table~\ref{tab:BBbkg_KKKch}).  
The four main peaks in the upper signal
plot are, from left to right: the \phiI, \fII/\ftwop, $D^0$ (background), and \chiczero.
The horn-like peaks in the middle and lower signal plots are reflections from the \phiI. 
The \chiczero is also visible around $3.4\gevcc$ in the middle signal plot. 
The upper background plot has a \phiI peak (mainly due to continuum) and $D^0$ peak
(mainly due to \BB).}
\end{figure*}

The fit result for Solution I is summarized in Table~\ref{tab:fitresult_KKKch_SolnI}.
The systematic uncertainties are described in Sec.~\ref{sec:Systematics}.
We report branching fractions for individual decay
channels by multiplying the inclusive branching fraction by the fit fractions.
This neglects interference between decay channels.
The inclusive branching fraction is computed as
\beq
{\cal B}(\bkkkboth) = \frac{N_{\rm sig}}{\bar{\varepsilon}\NBB}\,,
\eeq
where \NBB is the total number of \BB pairs and $\bar{\varepsilon}$
is the average efficiency, estimated by weighting MC events by the
measured DP distribution, $|\Amp|^2 + |\Ampbar|^2$.  We assume 
equal number of $\BpBm$ and $\BzBzb$ pairs from the $\FourS$.
We find ${\cal B}(\bkkkboth)= (34.6\pm 0.6 \pm 0.9) \times 10^{-6}$, including
the $\chi_{c0}\Kp$ channel. 
We find an inclusive charmless branching
fraction (excluding  $\chi_{c0}\Kp$) of 
${\cal B}(\bkkkboth)= (33.4\pm 0.5 \pm 0.9) \times 10^{-6}$.

\begin{table*}[htbp]
\center
\caption{
Branching fractions (neglecting interference), \CP asymmetries, and
\CP-violating phases (see Eq.~\eqref{eq:deltaphi}) for \bkkkboth.  
The ${\cal B}(\Bp\to R\Kp)$ column gives the branching fractions to intermediate
resonant states, corrected for secondary branching fractions obtained from
Ref.~\cite{Nakamura:2010zzi}.
Central
values and uncertainties are obtained from Solution I. In addition to quoting
the overall NR branching fraction, we quote the S-wave and P-wave NR branching
fractions separately.  
}
\begin{tabular}{l|c@{\hspace{0.3cm}}ccc}
\hline \hline
\noalign{\vskip1pt} 
Decay mode &  ${\cal B}(\bkkkboth)\times \fitfrac_j~(10^{-6})$ &  ${\cal B}(\Bp\to R\Kp)~(10^{-6})$   &    \Acp (\%)&  $\Delta\phi_j$ (deg) \\
\hline
\noalign{\vskip1pt} 
$\phiI$$\Kp$   &    $4.48\pm 0.22 ^{+0.33}_{-0.24} $ &    $9.2 \pm 0.4 ^{+0.7}_{-0.5}$  &    $12.8\pm 4.4 \pm 1.3 $ &    $23\pm 13 ^{+4}_{-5} $    \\
$\fI$$\Kp$   &    $9.4\pm 1.6 \pm 2.8 $ &                                               &    $-8\pm 8 \pm 4 $ &    $9\pm 7 \pm 6 $    \\
$\fII$$\Kp$   &    $0.74\pm 0.18 \pm 0.52 $  &  $17 \pm 4 \pm 12$                      &    &       \\
$\ftwop$$\Kp$   &    $0.69\pm 0.16 \pm 0.13 $ &   $1.56 \pm 0.36 \pm 0.30$            &  $14\pm 10 \pm 4 $ &    $-2\pm 6 \pm 3 $    \\
$\fIII$$\Kp$   &    $1.12\pm 0.25 \pm 0.50 $ &                                        &            &       \\
$\chiczero$$\Kp$   &    $1.12\pm 0.15 \pm 0.06 $ &    $184 \pm 25 \pm 14$             &     &    $-4\pm 13 \pm 2 $    \\
NR            &    $22.8\pm 2.7 \pm 7.6 $   &                                         &   $6.0\pm 4.4 \pm 1.9 $    &     $0$ (fixed)     \\
NR (S-wave)   &    $52^{+23}_{-14} \pm 27 $ &                                         &    &      \\
NR (P-wave)   &    $24^{+22}_{-12} \pm 27 $ &                                         &     &       \\
\hline  \hline
\end{tabular}
\label{tab:fitresult_KKKch_SolnI}
\end{table*}

Fit fraction matrices giving the values of $\fitfrac_{jk}$ for Solutions I and II are 
shown in the Appendix.  
Solution I has large
destructive interference between the S-wave and P-wave NR decays.
Solution II has a smaller $\fI$ fit fraction and large destructive
interference between the $\fI$ and nonresonant decays.

We also calculate an overall charmless \Acp by integrating the charmless $|\Amp|^2$
and $|\Ampbar|^2$ over the DP. We find the charmless 
$\Acp(\bkkkboth)= (-1.7^{+1.9}_{-1.4} \pm 1.4) \%$.  There is negligible difference between Solutions 
I and II for this quantity.

We plot the signal-weighted \mab distribution separately for \Bp and \Bm
events in Fig.~\ref{fig:kkkch_mkklo_BpBm}.  
Solutions I and II yield $\Acp(\phi(1020)) = (12.8 \pm 4.4)\%$ and
$(7.4 \pm 4.5)\%$, respectively, where the uncertainties
are statistical only.
We perform a likelihood scan in $\Acp(\phi(1020))$, shown in 
Fig.~\ref{fig:kkkch_scan_phi_acp}.  At each scan point, many fits are performed
with random initial parameters, and the fit with the largest likelihood is
chosen.  Thus, the likelihood scan properly accounts for any local minima.
The \Acp is found to differ from 0 at the $2.8$ standard deviation
level ($2.9\sigma$ if one uses only the statistical uncertainties).

\begin{figure*}[htbp]
\includegraphics[width=8.9cm,keepaspectratio]{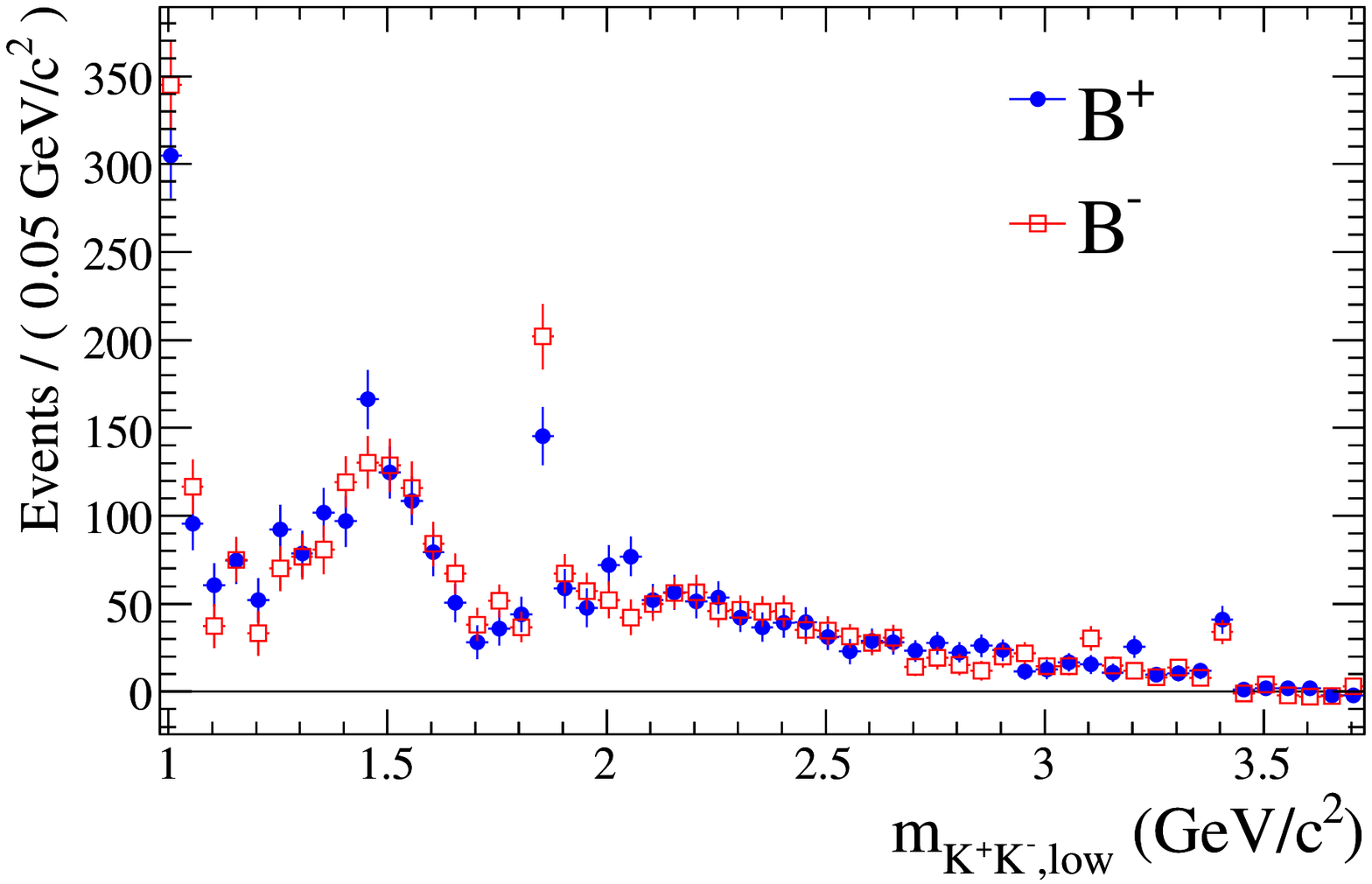}
\includegraphics[width=8.9cm,keepaspectratio]{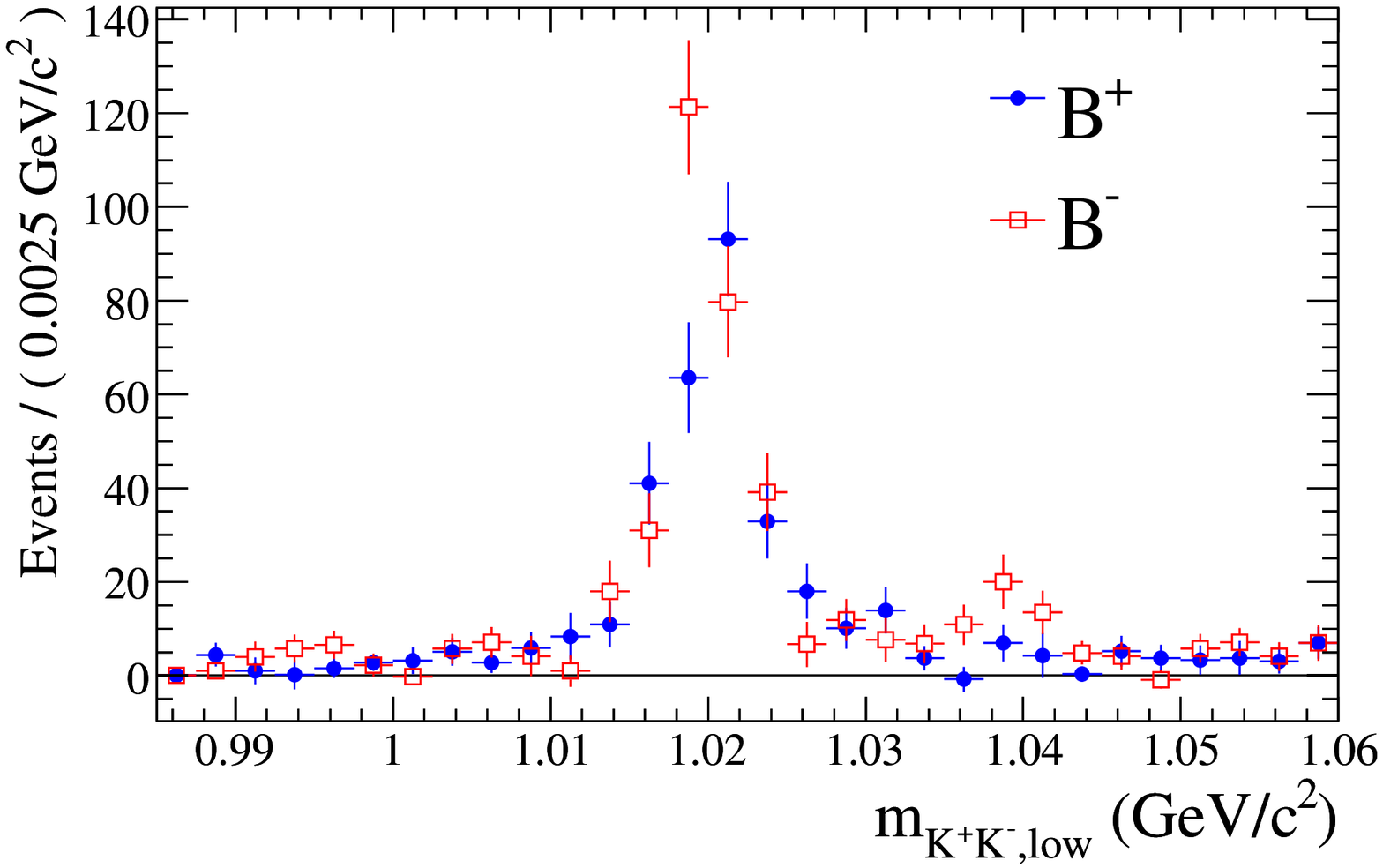}
\caption{\label{fig:kkkch_mkklo_BpBm}   
Signal-weighted \mab distribution for \bkkkboth candidates in data, plotted separately for \Bp and \Bm 
events, for the entire DP range (left), and the $\phiI$-region only (right).
The event weighting is performed using the \splot method.  
Signal includes irreducible \BB backgrounds (classes 5 and 6 in Table~\ref{tab:BBbkg_KKKch}). 
}
\end{figure*}

\begin{figure}[htbp]
\includegraphics[width=8.9cm,keepaspectratio]{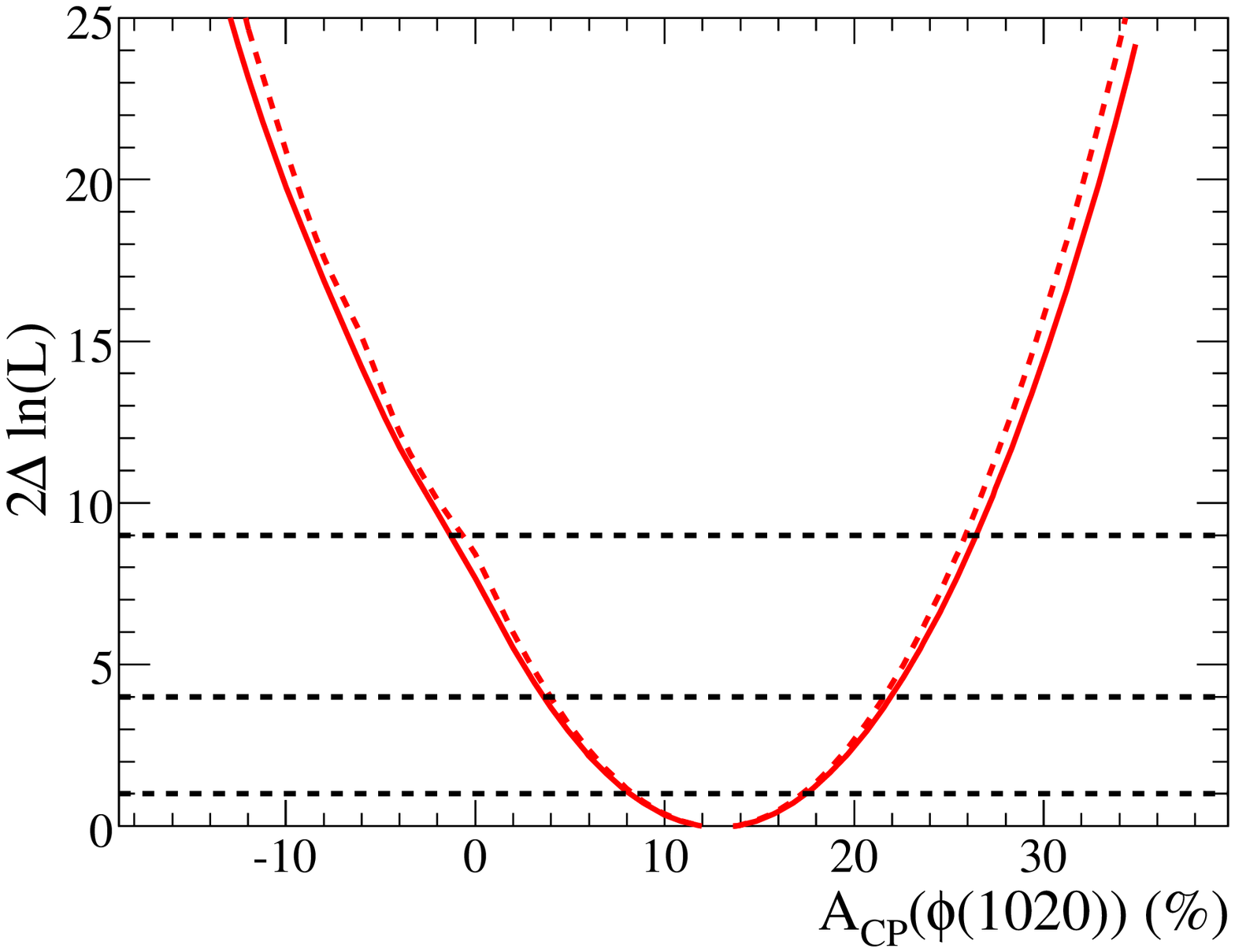}
\caption{\label{fig:kkkch_scan_phi_acp}   
Scan of $2\Delta \logL$, with (solid line) and without (dashed line)
systematic uncertainties, as a function of $\Acp(\phi(1020))$ in \bkkkboth.  
}
\end{figure}

Solution II exhibits a very large \Acp for the $\fI\Kp$ channel, but 
in Solution I this \Acp is consistent with 0.   A likelihood scan
in $\Acp(\fI)$ is shown in Fig.~\ref{fig:kkkch_scan_fI_acp}, in 
which the two solutions are clearly visible.

\begin{figure}[htbp]
\includegraphics[width=8.9cm,keepaspectratio]{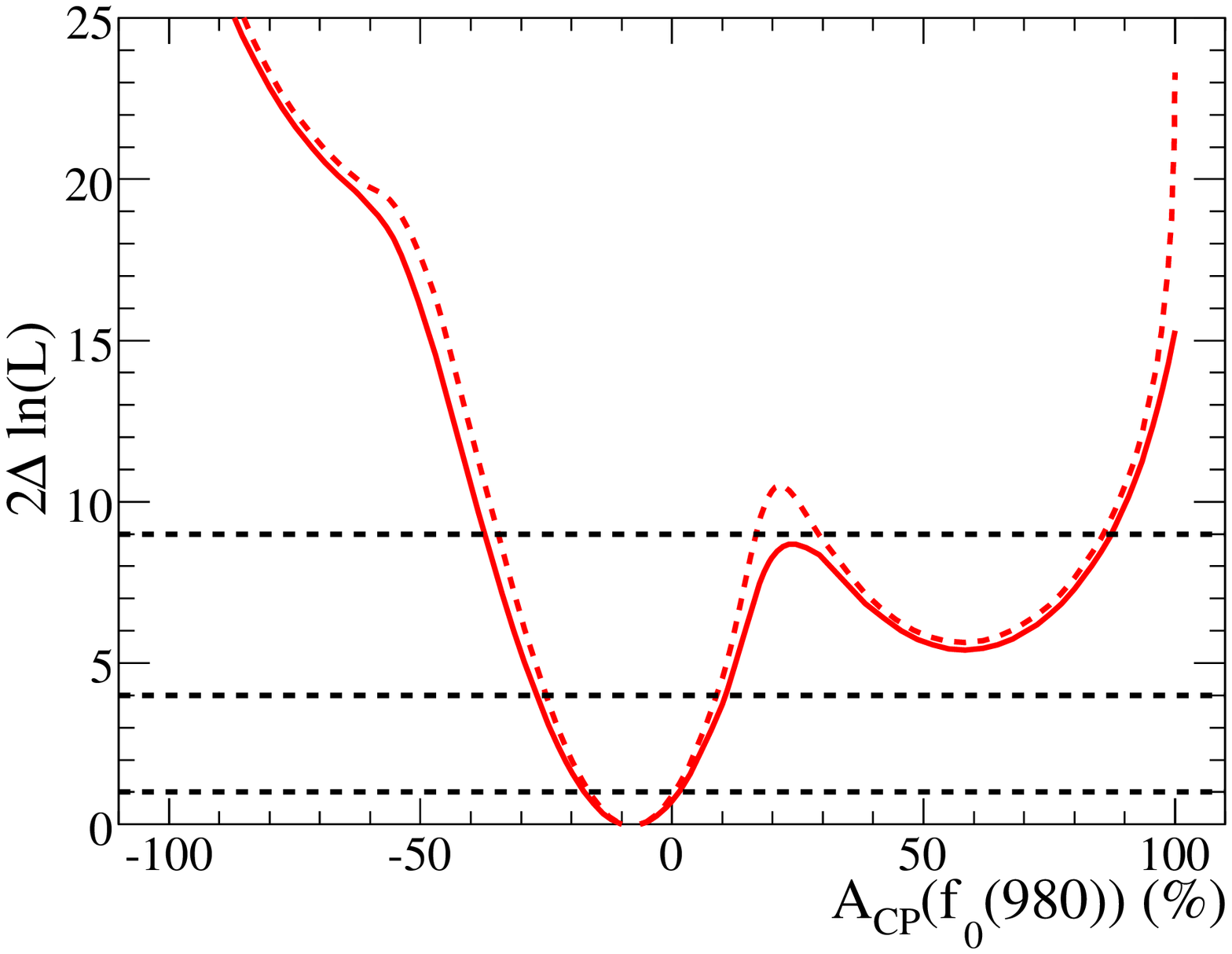}
\caption{\label{fig:kkkch_scan_fI_acp}   
Scan of $2\Delta \logL$, with (solid line) and without (dashed line)
systematic uncertainties,
as a function of $\Acp(\fI)$ in \bkkkboth.}
\end{figure}

\subsection{\bkksks}
\label{sec:fitResultsKKsKs}

The maximum-likelihood fit of $3012$ candidates results in 
yields of $636\pm 28$ signal events and  $2234\pm 50$ continuum 
events, where the uncertainties are statistical only.  
The \BB yields are fixed to the expected number of events 
(Table~\ref{tab:BBbkg_KKsKs}), 
for a total of $155$ events.

In order to limit the number of fit parameters, we 
constrain the $\Acp$ and $\Delta\phi$ of every 
charmless isobar to be equal in the fit.
We fix $\Acp$ for $\chiczero\Kp$ to 0, but
leave the corresponding $\Delta\phi$ parameter free to vary in 
the fit. 
Recalling that only {\em relative\/} values of $\Delta\phi$ 
are measurable, our choice is therefore to measure the difference
between $\Delta\phi$ for the $\chiczero$ and the reference
$\Delta\phi$ shared by all the other isobars.

Many fits are performed with randomly chosen starting values for
the isobar parameters.  In addition to the global minimum, 14
other local minima are found with values of $-2\logL$ within
9 units (3$\sigma$) of the global minimum.  These different 
solutions vary greatly in their isobar parameters, but have
consistent signal yields and values of \Acp.

Figure~\ref{fig:bkksks_projections} shows the distributions of
\mes, \de, and the \nn output, compared to the fit model.  
Figure~\ref{fig:kksks_DP}  shows the \mab, \mbc, and \mac
distributions for signal- and background-weighted events, using the 
\splot technique.
We plot the signal-weighted \mab distribution separately for \Bp and \Bm
events in Fig.~\ref{fig:kksks_mvar_BpBm}.

\begin{figure*}[htbp]
\includegraphics[width=5.9cm,keepaspectratio]{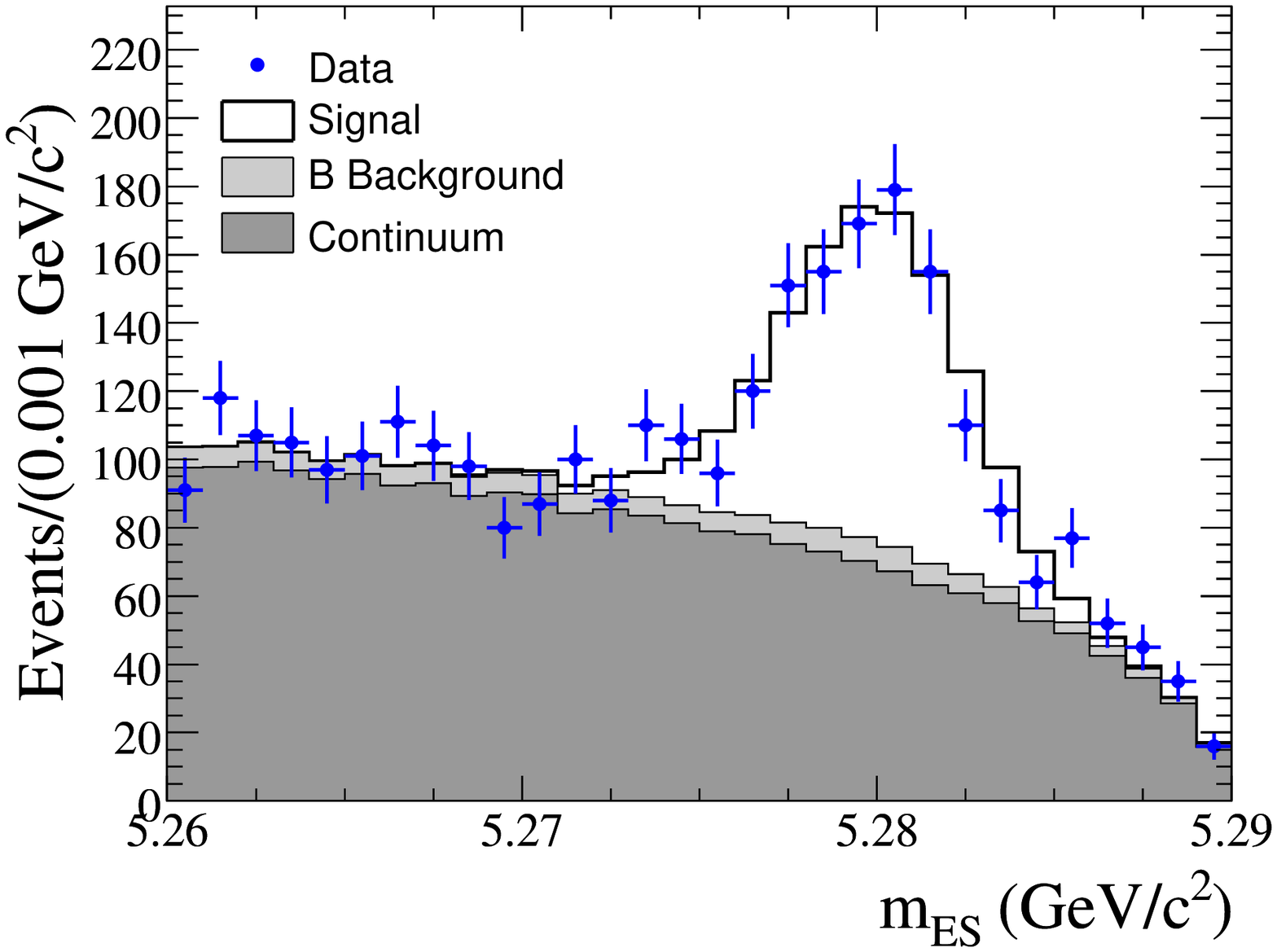}
\includegraphics[width=5.9cm,keepaspectratio]{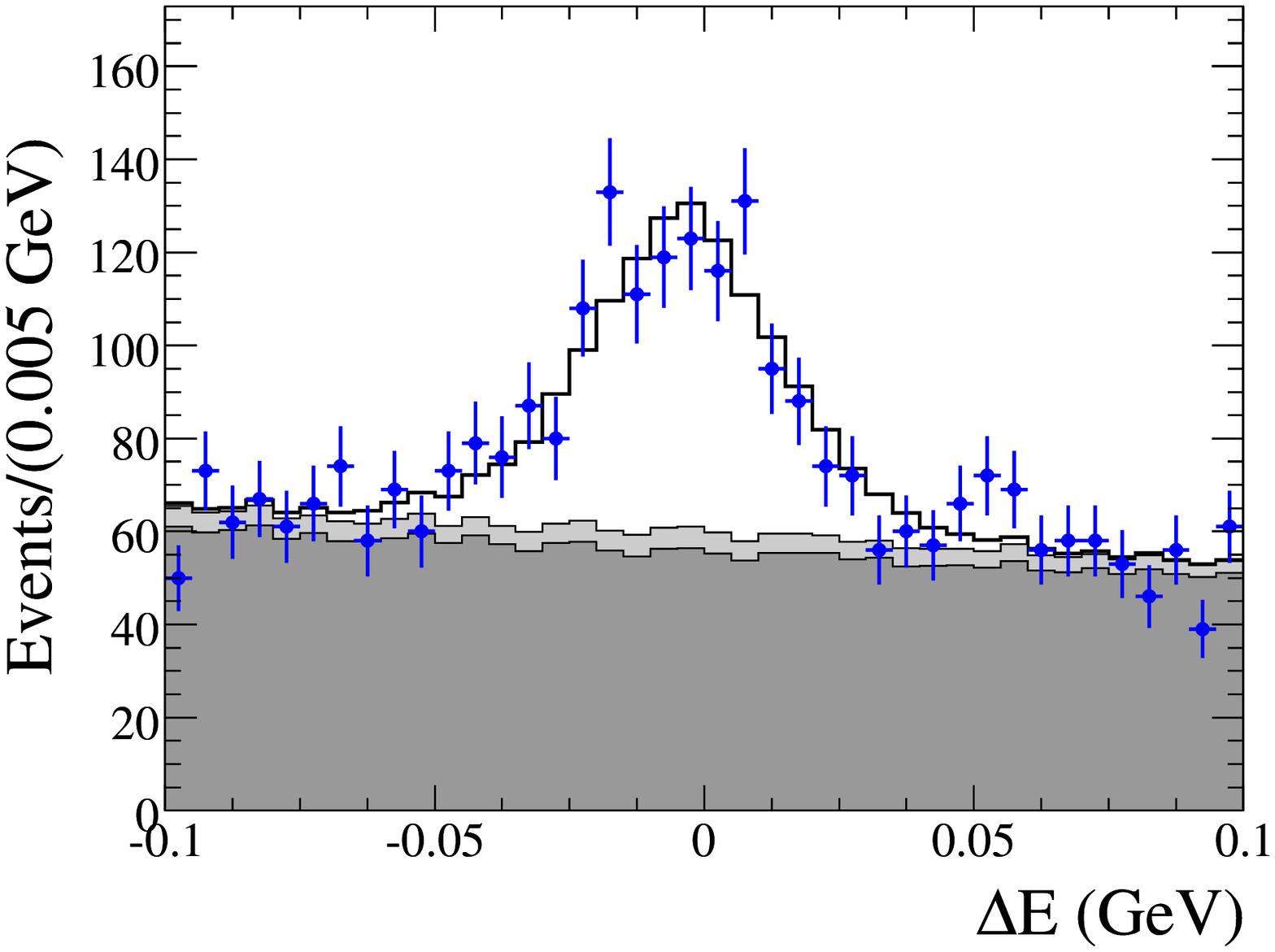}
\includegraphics[width=5.9cm,keepaspectratio]{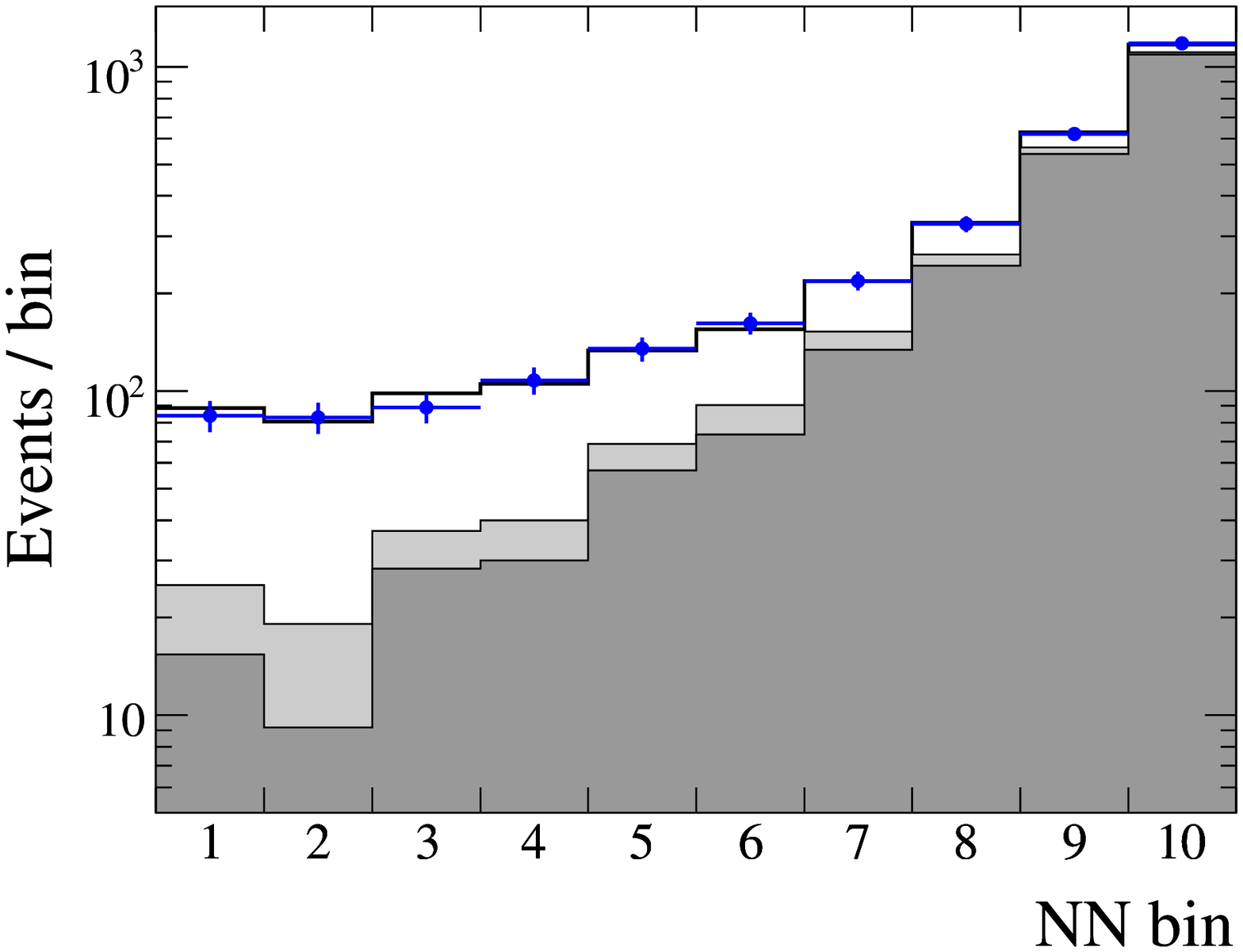}
  \caption{\label{fig:bkksks_projections}   Distributions of $\mes$ (left), 
	$\de$ (center), and $\nn$ output (right) for \bkksks.
    The NN output is shown in vertical log scale.}
\end{figure*}

\begin{figure*}[htbp]
\includegraphics[width=7.9cm,keepaspectratio]{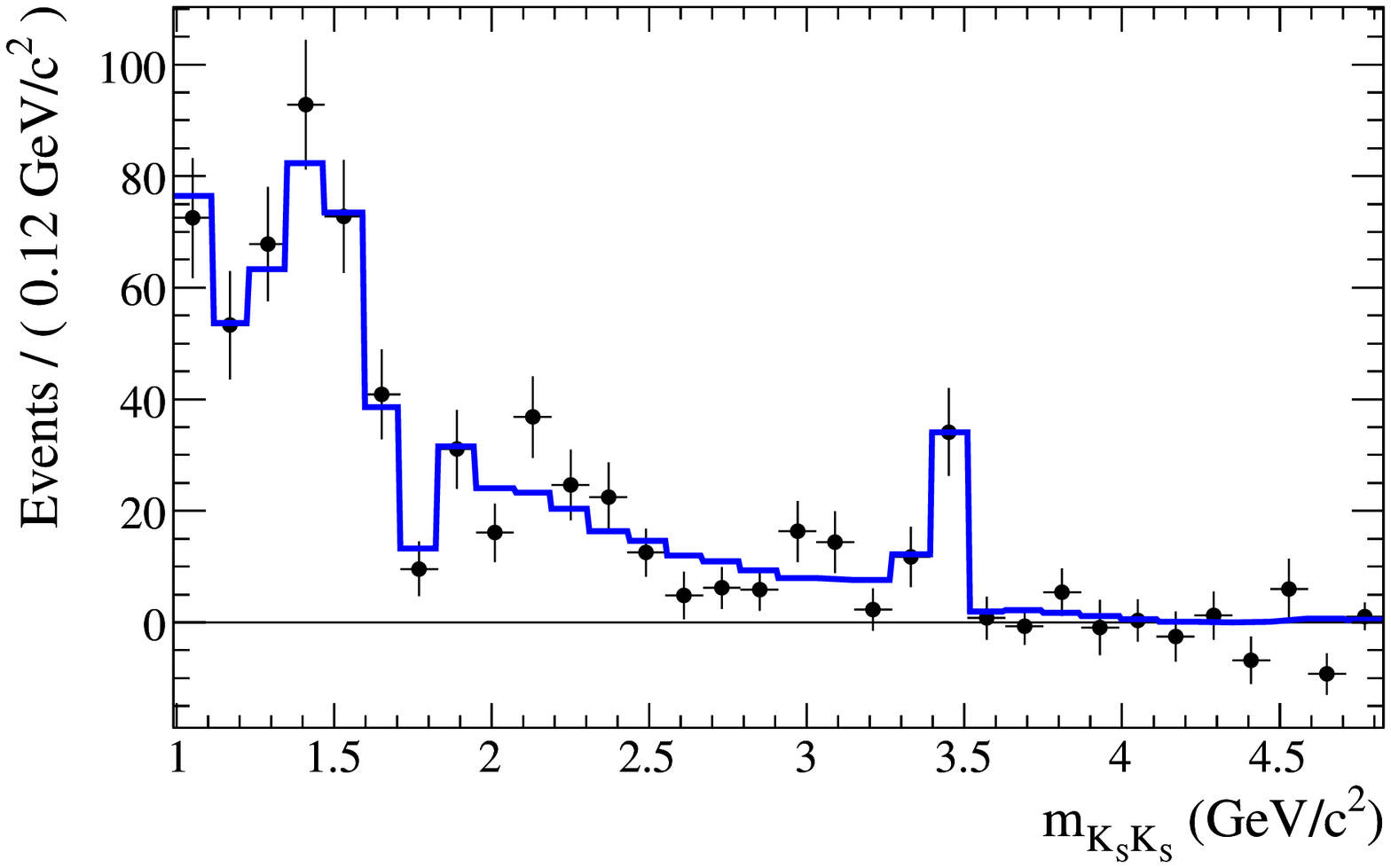}
\includegraphics[width=7.9cm,keepaspectratio]{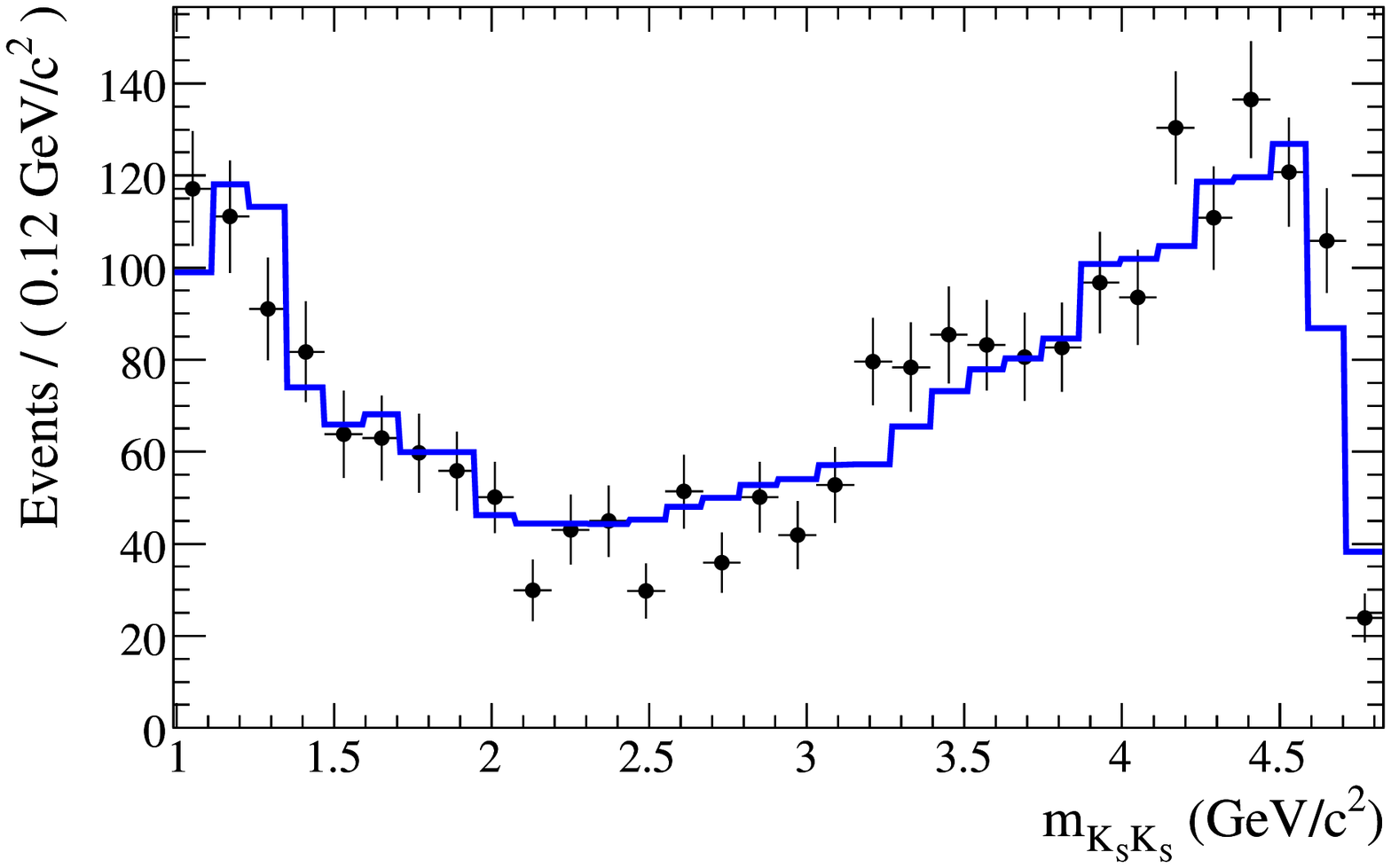}
\includegraphics[width=7.9cm,keepaspectratio]{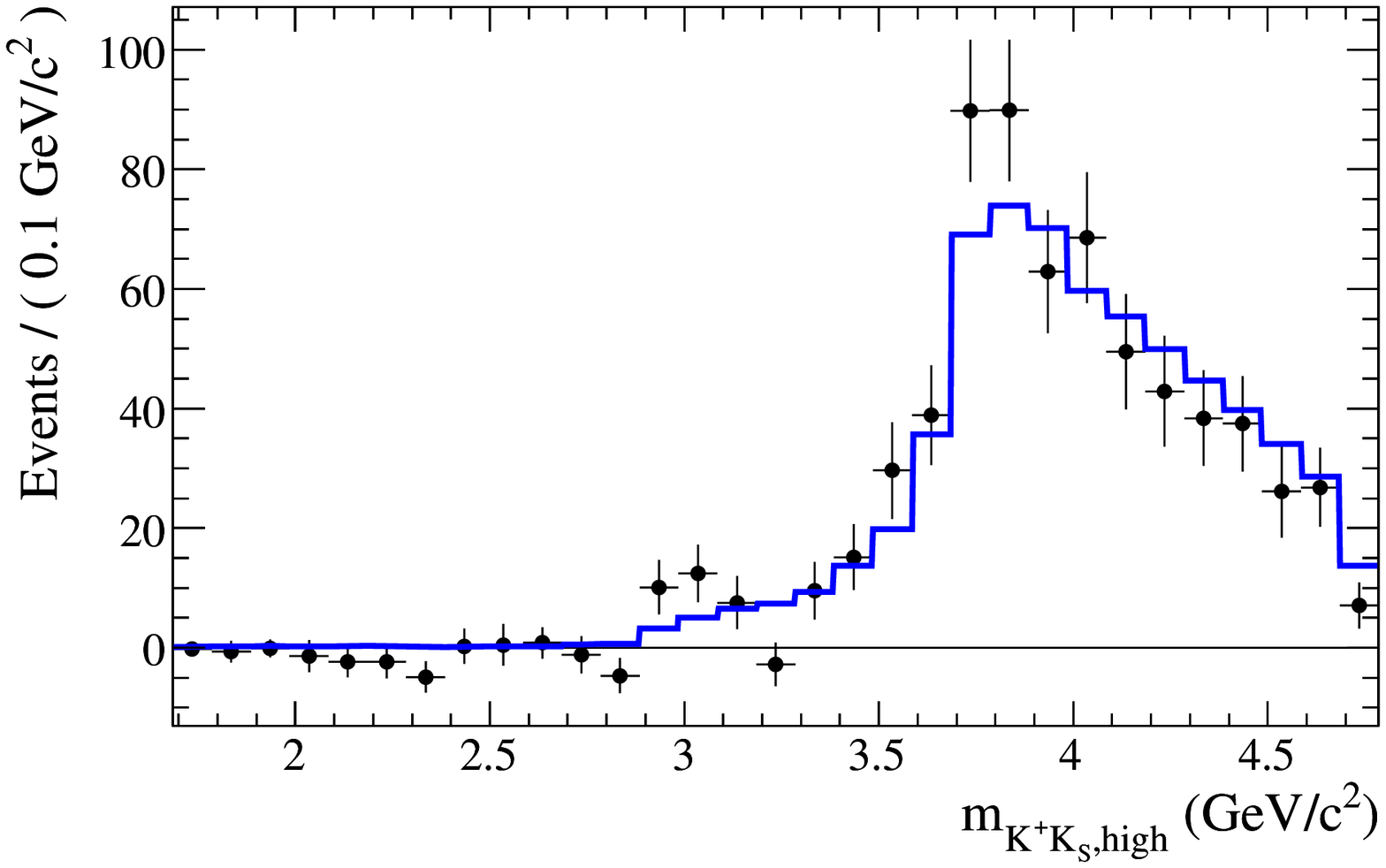}
\includegraphics[width=7.9cm,keepaspectratio]{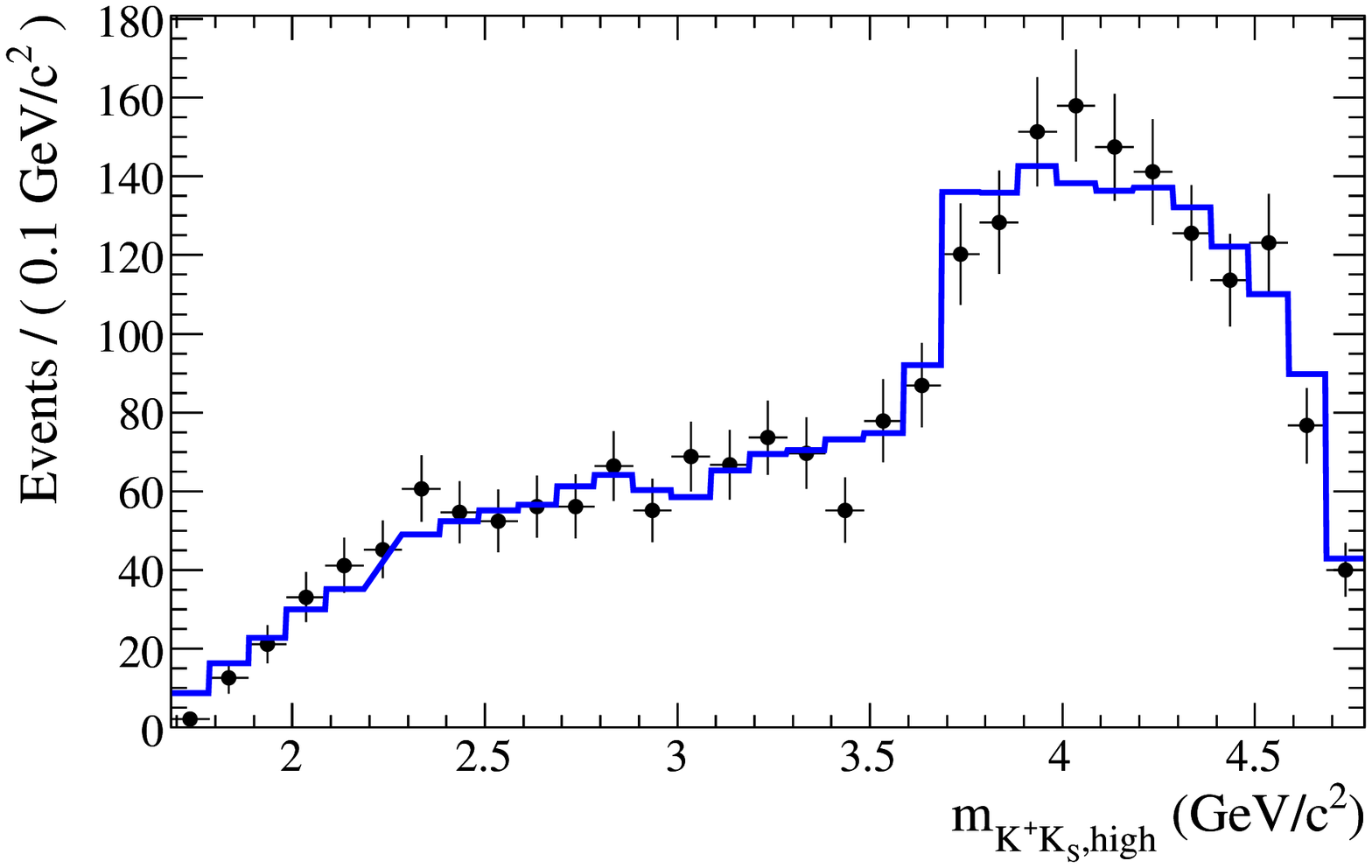}
\includegraphics[width=7.9cm,keepaspectratio]{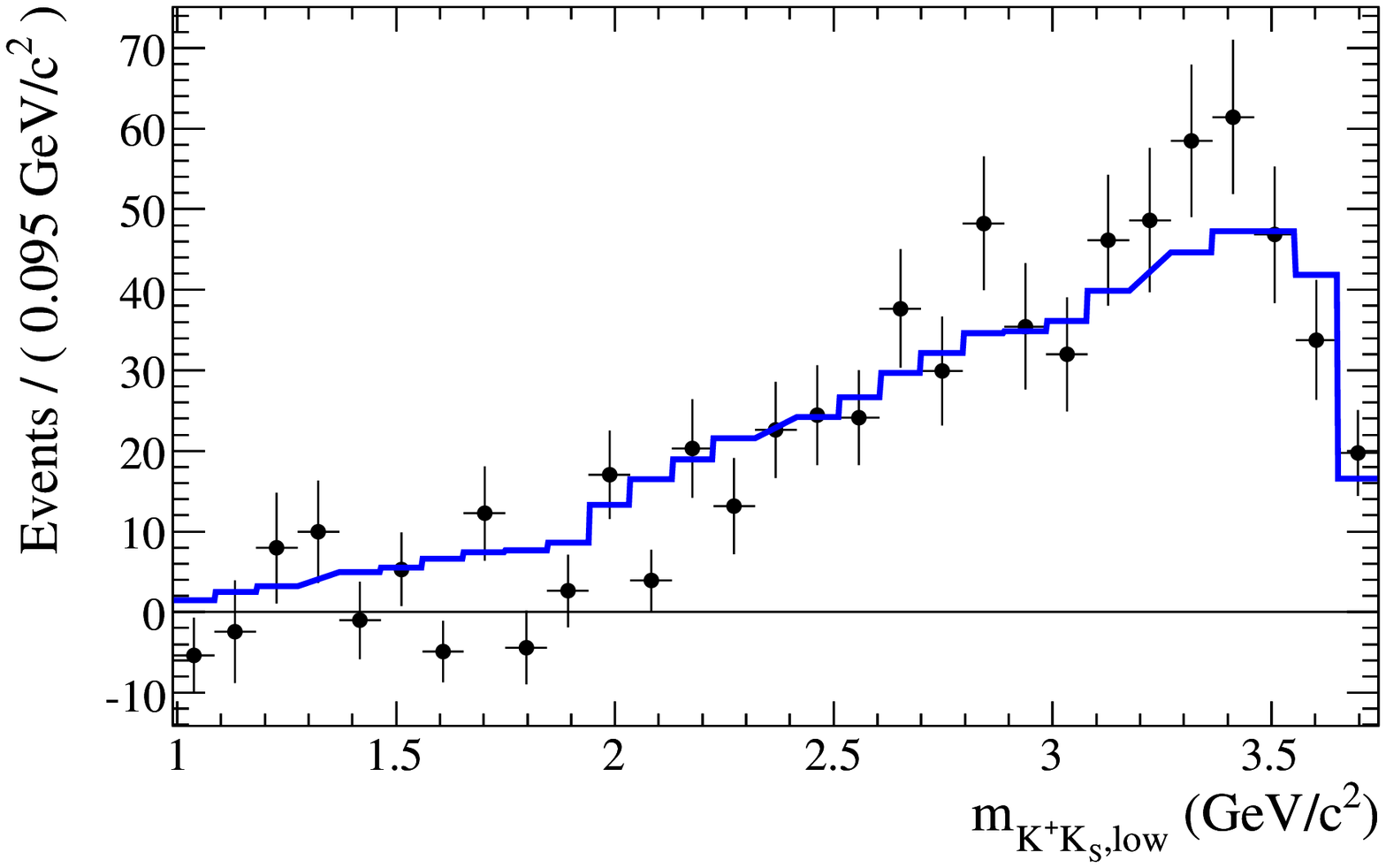}
\includegraphics[width=7.9cm,keepaspectratio]{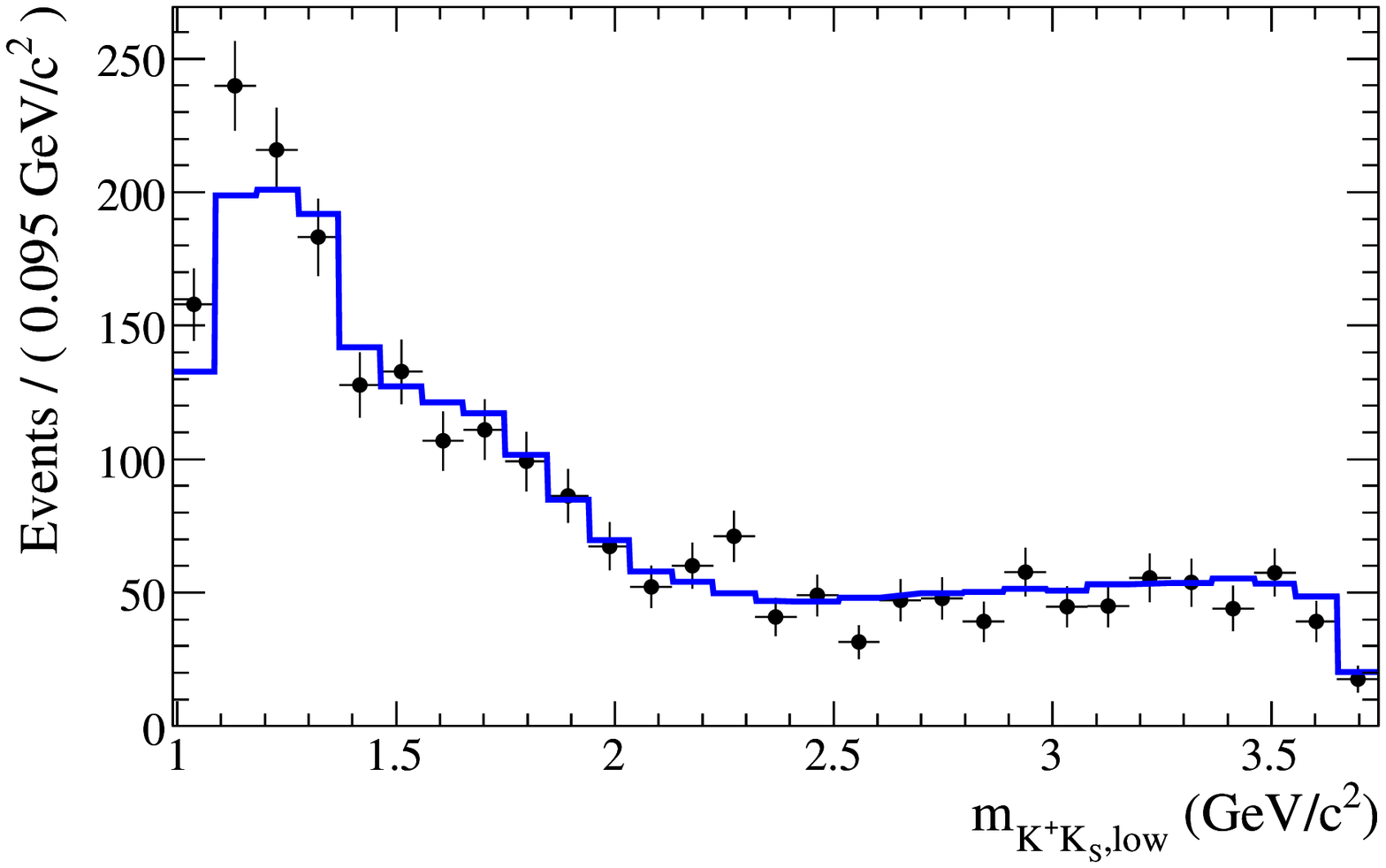}
\caption{\label{fig:kksks_DP}Distributions of $\mab=\mksksL$, $\mbc=\mkpkshiL$, and $\mac=\mkpksloL$,
 for signal-weighted (left) and background-weighted (right) \bkksks candidates in data. 
The event weighting is performed using the \splot method.  
 The fit model (histograms)
is shown superimposed over data (points). 
The two main peaks in the upper signal plot are the \fII/\ftwop  and \chiczero.
}
\end{figure*}

\begin{figure}[htbp]
\includegraphics[width=8.9cm,keepaspectratio]{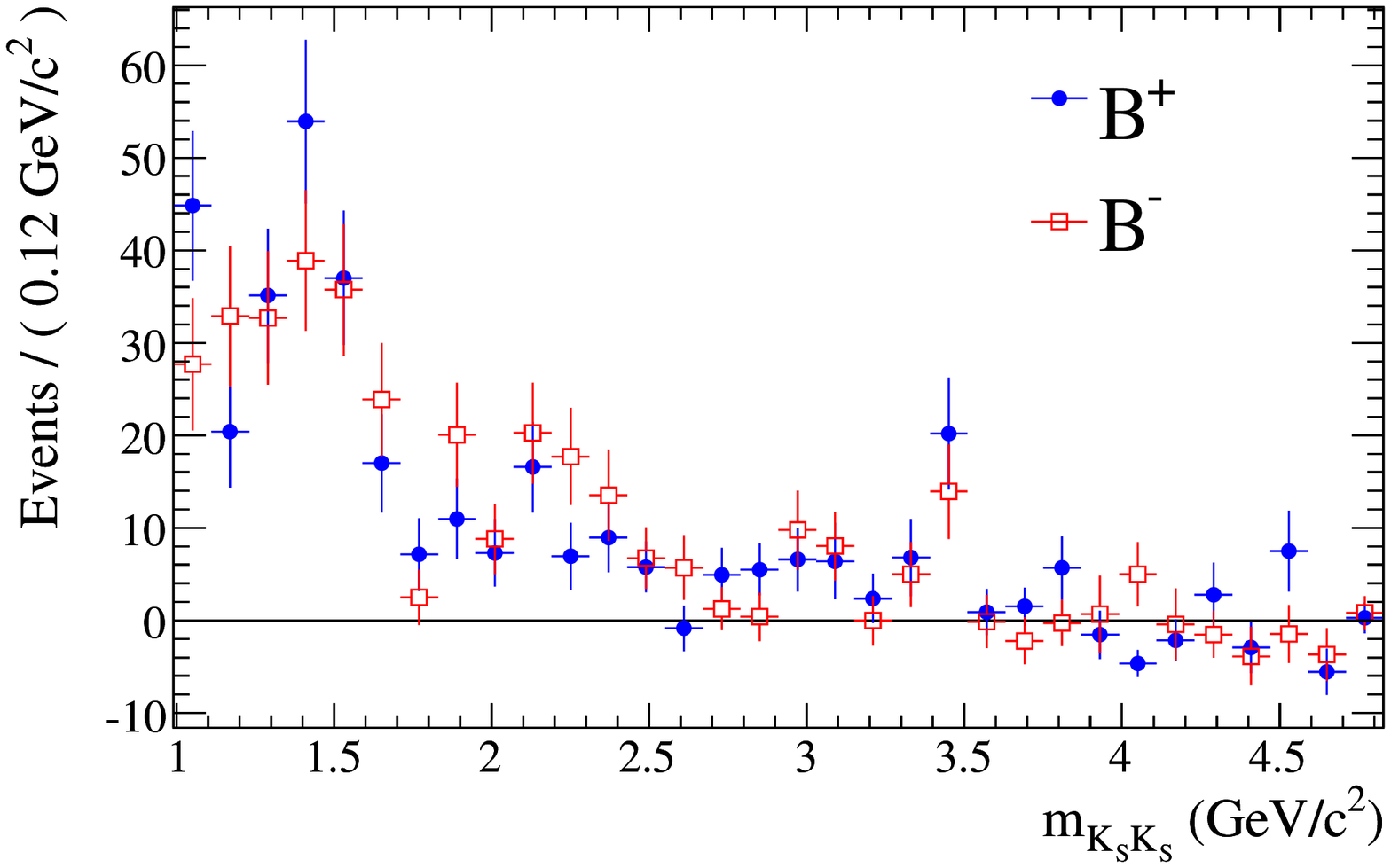}
\caption{\label{fig:kksks_mvar_BpBm}   
Signal-weighted \mab distribution for \bkksks candidates in data, plotted separately for \Bp and \Bm 
events. 
The event weighting is performed using the \splot method.  
Signal includes irreducible \BB backgrounds (class 4 in Table~\ref{tab:BBbkg_KKsKs}).
}
\end{figure}

The fit result for the global minimum solution is summarized in 
Tables~\ref{tab:isobarSummary_KKsKs_SolnI} and \ref{tab:fitresult_KKsKs_SolnI}.
The fit fraction matrix for the global 
mininum is given in the Appendix, and the correlation matrix of the 
isobar parameters is given in Ref.~\cite{epaps}.
The other minima all have
consistent values for the $\ftwop$ and $\chiczero$ fit fractions, but wide variations
in the fit fractions for the other states are seen.  In particular,
the fit fraction of the $\fI$ varies between $69\%$ and $152\%$ and the 
fit fraction of the $\fII$ varies between $3\%$ and $73\%$. 
This means the branching fractions of these states are very poorly constrained with
the current data.  However, 
the signal yields for the different solutions only vary between 636 and 
640 events.
We find a total inclusive branching fraction
of ${\cal B}(\bkksks)= ( 10.6 \pm 0.5 \pm 0.3 ) \times 10^{-6}$, or
${\cal B}(\bkksks)= ( 10.1 \pm 0.5 \pm 0.3 ) \times 10^{-6}$ if the $\chi_{c0}$ is excluded.

\begin{table}[htbp]
\center
\caption{
Isobar parameters for \bkksks, for the global minimum.
The NR coefficients are defined in Eq.~\eqref{eq:NRpoly}.  Phases are given in degrees. 
Only statistical uncertainties are given.
}
\begin{tabular}{ll|c}
\hline \hline
\multicolumn{2}{l|}{Parameter }   &  Value \\
\hline
\noalign{\vskip1pt} 
$\fI$$\Kpm$        &    $c$   &   $3.35\pm 0.22 $  \\
     &    $\phi$   &   $31\pm 9 $  \\
\hline
\noalign{\vskip1pt} 
$\fII$$\Kpm$        &    $c$   &   $0.20\pm 0.05 $  \\
     &    $\phi$   &   $-83\pm 18 $  \\
\hline
\noalign{\vskip1pt} 
$\ftwop$$\Kpm$        &    $c$   &   $0.00179\pm 0.00032 $  \\
     &    $\phi$   &   $-58\pm 12 $  \\
\hline
\noalign{\vskip1pt} 
$\fIII$$\Kpm$        &    $c$   &   $0.24\pm 0.07 $  \\
     &    $\phi$   &   $-22\pm 11 $  \\
\hline
\noalign{\vskip1pt} 
$\chiczero$$\Kpm$        &    $c$   &   $0.113\pm 0.017 $  \\
     &    $\phi$   &   $45\pm 60 $  \\
     &    $\delta$   &   $-12\pm 32 $  \\
\hline
NR   &    &   \\
          &   $b$   &    $-0.018\pm 0.023 $  \\
$a_{S0}$  &   $c$   & $  1.0 $ (fixed) \\
    &     $\phi$   &  $  0$ (fixed)    \\
$a_{S1}$        &    $c$   &    $1.00\pm 0.08 $  \\
     &    $\phi$   &    $129\pm 6 $  \\
$a_{S2}$        &    $c$   &    $0.51\pm 0.08 $  \\
     &    $\phi$   &    $-85\pm 8 $  \\
\hline  \hline
\end{tabular}
\label{tab:isobarSummary_KKsKs_SolnI}. 
\end{table}

\begin{table*}[htbp]
\center
\caption{Branching fractions (neglecting interference) for \bkksks.  
The ${\cal B}(\Bp\to R\Kp)$ column gives the branching fractions to intermediate
resonant states, corrected for secondary branching fractions obtained from 
Ref.~\cite{Nakamura:2010zzi}.
Central
values and uncertainties are for the global minimum only.  See the text
for discussion of the variations between the local minima.
}
\begin{tabular}{l|cc}
\hline \hline
\noalign{\vskip1pt} 
Decay mode &  ${\cal B}(\bkksks)\times \fitfrac_j~(10^{-6})$  &    ${\cal B}(\Bp\to R\Kp)~(10^{-6})$       \\
\hline
\noalign{\vskip1pt} 
$\fI$$\Kp$   &   $14.7\pm 2.8 \pm 1.8 $              &     \\
$\fII$$\Kp$   &   $0.42\pm 0.22 \pm 0.58 $           &  $20 \pm 10 \pm 27$  \\
$\ftwop$$\Kp$   &   $0.61\pm 0.21 ^{+0.12}_{-0.09} $ &  $2.8 \pm 0.9 ^{+0.5}_{-0.4}$  \\
$\fIII$$\Kp$   &   $0.48^{+0.40}_{-0.24} \pm 0.11 $  &   \\
$\chiczero$$\Kp$   &   $0.53\pm 0.10 \pm 0.04 $  &   $168 \pm 32 \pm 16$  \\
NR (S-wave)   &   $19.8\pm 3.7 \pm 2.5 $   &   \\
\hline  \hline
\end{tabular}
\label{tab:fitresult_KKsKs_SolnI}
\end{table*}

The global minimum has values of $\Acp = ( 4 \pm 5 \pm 2)\%$ and
$\Delta\phi = (-25 \pm 65  \pm 11)^{\circ}$.  The $\Acp$ for 
the other minima are between $2\%$ and $4\%$. 
A likelihood scan 
of $\Acp$ is shown in Fig.~\ref{fig:kksks_scan_Acp}. 
From the likelihood scan, we determine $\Acp = (4 ^{+4}_{-5} \pm 2)\%$.

\begin{figure}[htbp]
\includegraphics[width=8.9cm,keepaspectratio]{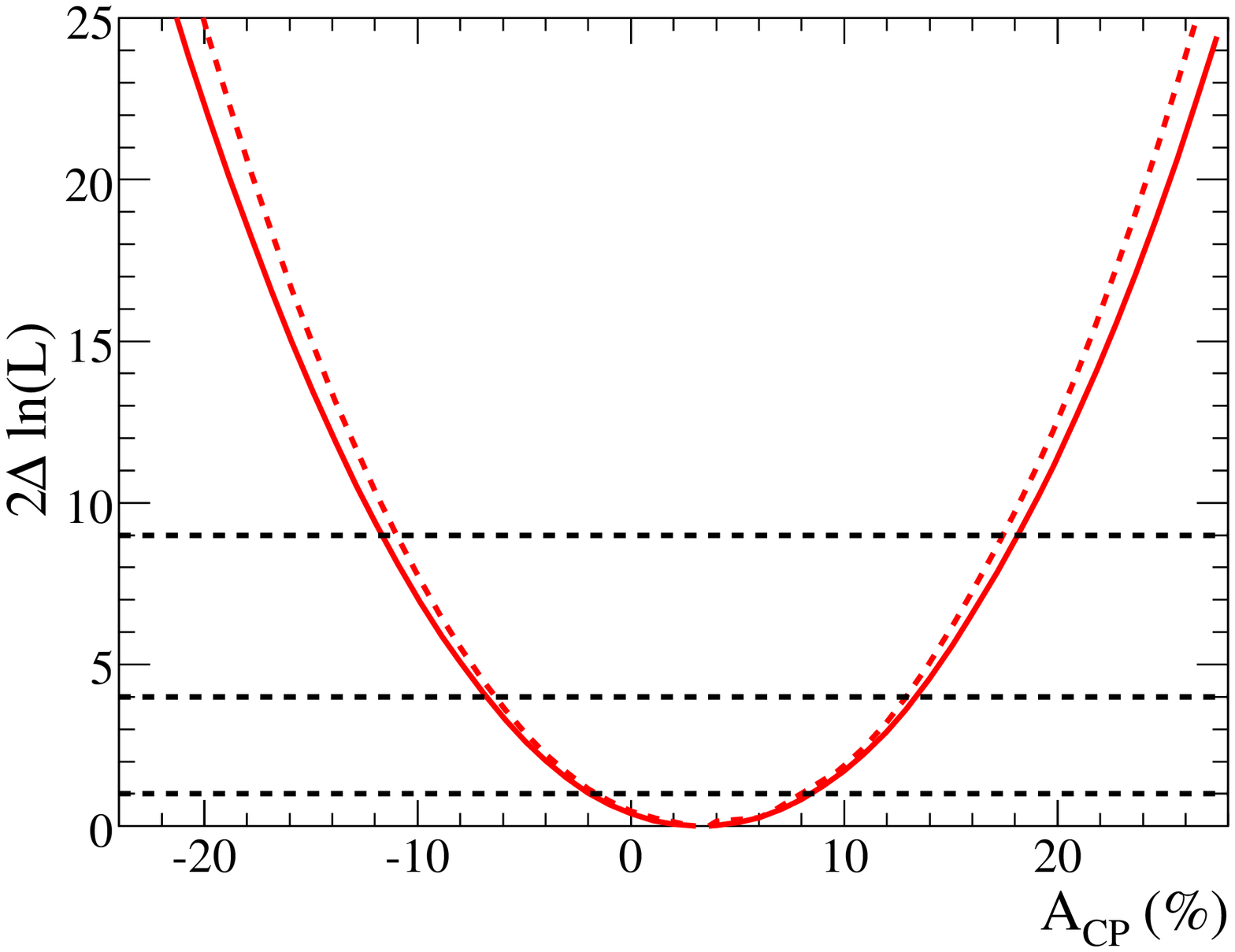}
\caption{\label{fig:kksks_scan_Acp}
Scan of $2\Delta \logL$, with (solid line) and without (dashed line)
systematic uncertainties,
as a function of $\Acp$ in \bkksks.}
\end{figure}

\subsection{\bkkks}
\label{sec:fitResultsKKKs}

The maximum-likelihood fit is performed simultaneously to 
$5627$ candidates in the \Kspp channel and 
$2910$ candidates in the \Kszz channel.  In the \Kspp
channel, 
we find $ 1419 \pm 43$ signal events (including $68 \pm 9$ signal-like \BB
background events, corresponding to categories 1-4 in Table~\ref{tab:BBbkg_KKKs}).
We also find $4178 \pm 71$ continuum events and  $29 \pm 28$ 
remaining $\BB$ events.

In the \Kszz channel, we find yields of $160 \pm 17$ signal events
(including $7 \pm 1$ signal-like \BB background events), $2703 \pm 55$ 
continuum events, and  $48 \pm 18$ remaining $\BB$ events.  All uncertainties 
are statistical only.

We vary three sets of $\betaeff$ and $\Acp$ values in the fit: one 
for the $\phiI$, another for the $\fI$, and a third that 
is shared by all the other charmless isobars in order to reduce the
number of fit parameters.  Note that this last set of isobars contains
both even-spin and odd-spin (P-wave NR) terms. Because of the sign flip in
Eq.~\eqref{eq:signflip},  the $\sin\deltamd\dt$-dependent \CP asymmetry 
(see Eq.~\eqref{eq:dalitz_plot_rate}) 
has opposite sign for the even-spin and odd-spin components.
We fix the $\betaeff$ of the $\chiczero$ to the SM value, and we fix its $\Acp(=-C)$ 
to 0.  

We perform hundreds of 
fits, each one with randomly chosen starting values for
the isobar parameters.  In addition to the global minimum, four
other local minima are found with values of $-2\logL$ within
9 units of the global minimum.  These different 
solutions all have consistent signal yields, but vary greatly 
for some isobar parameters.

Figure~\ref{fig:bkkkspp_projections} shows distributions of
\mes, \de,  and the \nn output for the \Kspp mode, and 
Fig.~\ref{fig:bkkkszz_projections} shows the same distributions
for the \Kszz mode.  Figure~\ref{fig:kkks_DP} shows the \mab, \mbc, and \mac
distributions for signal- and background-weighted events, 
for the \Kspp channel only.
Figure~\ref{fig:bkkks_deltat} shows the $\deltat$ distribution
and the time-dependent asymmetry for signal-weighted events, both for
the $\phiI$ region $(1.01 <\mab<1.03 \gevcc)$ and 
 the $\phiI$-excluded region.

\begin{figure*}[htbp]
\includegraphics[width=5.9cm,keepaspectratio]{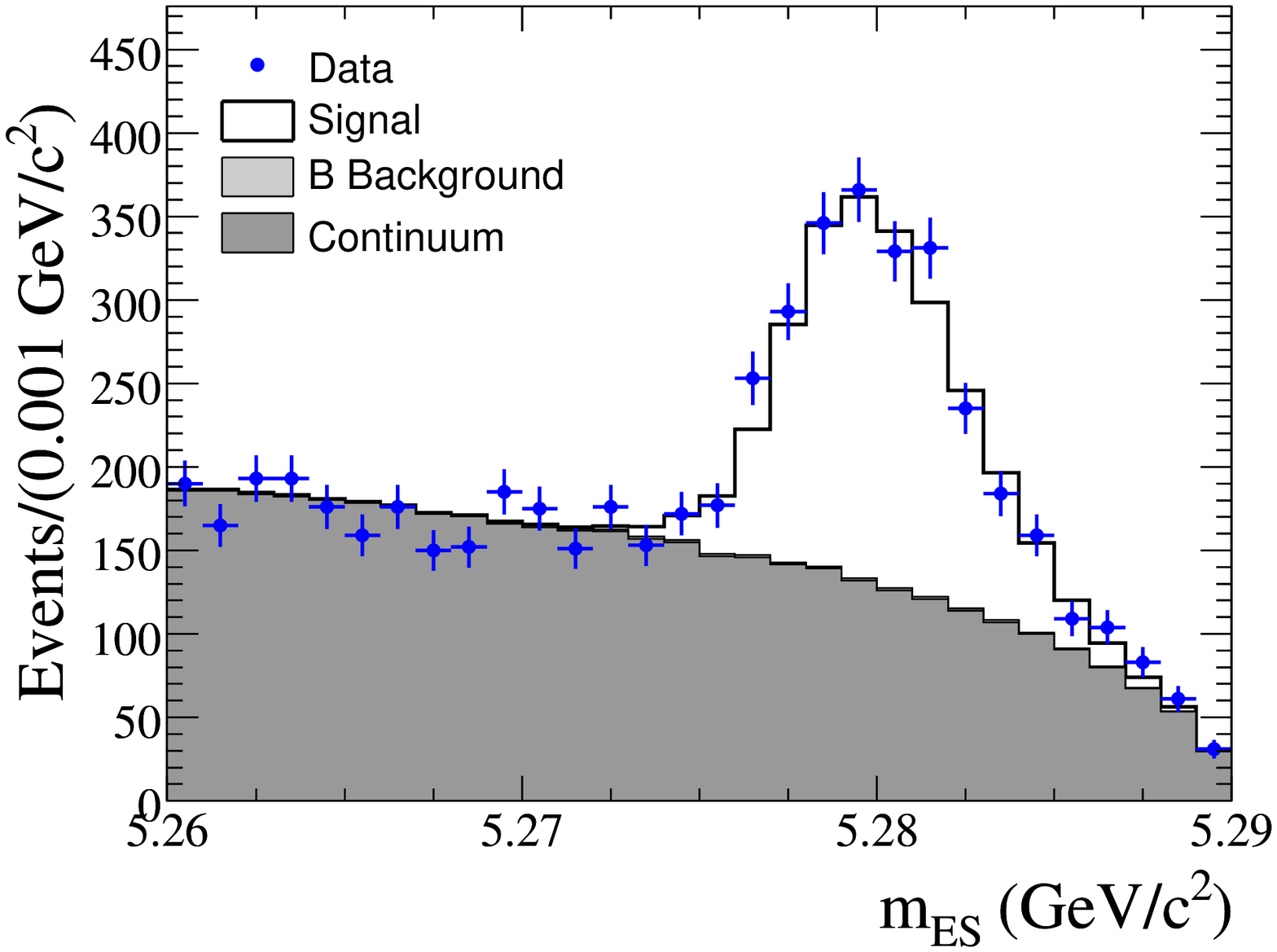}
\includegraphics[width=5.9cm,keepaspectratio]{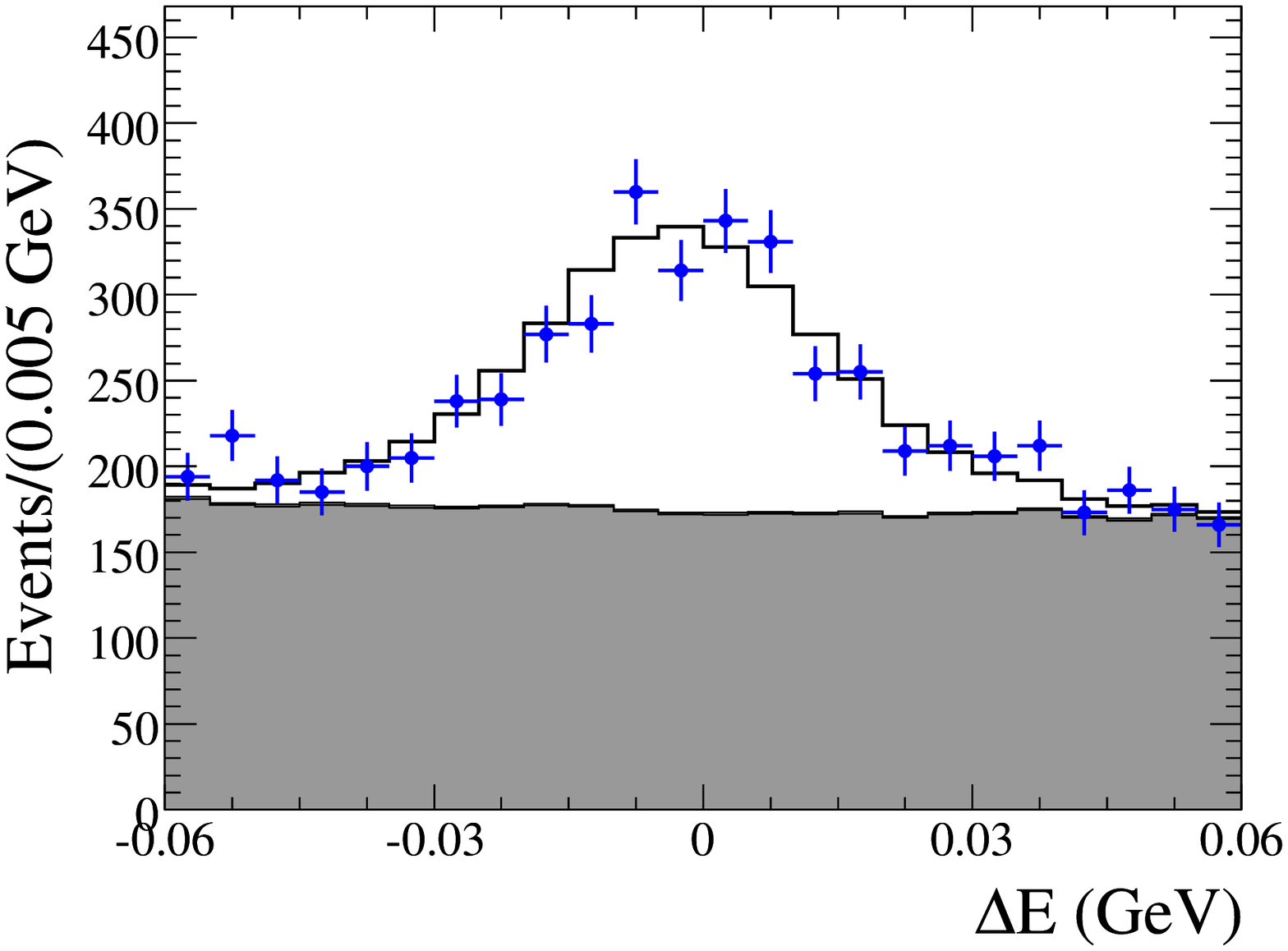}
\includegraphics[width=5.9cm,keepaspectratio]{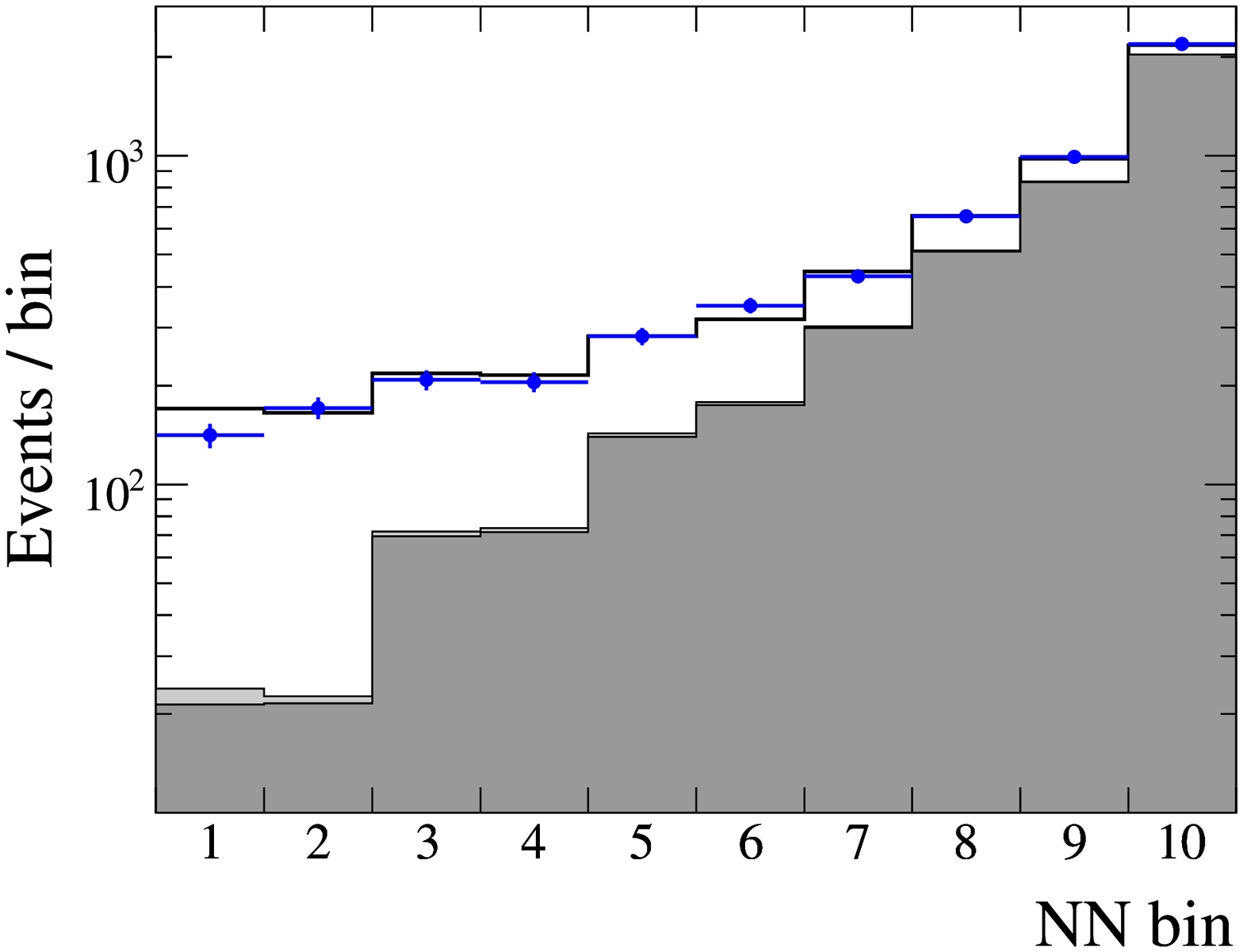}
  \caption{\label{fig:bkkkspp_projections}   Distributions of $\mes$ (left), 
	$\de$ (center), and $\nn$ output (right) for \bkkks, \Kspp.
    The NN output is shown in vertical log scale.}
\end{figure*}

\begin{figure*}[htbp]
\includegraphics[width=5.9cm,keepaspectratio]{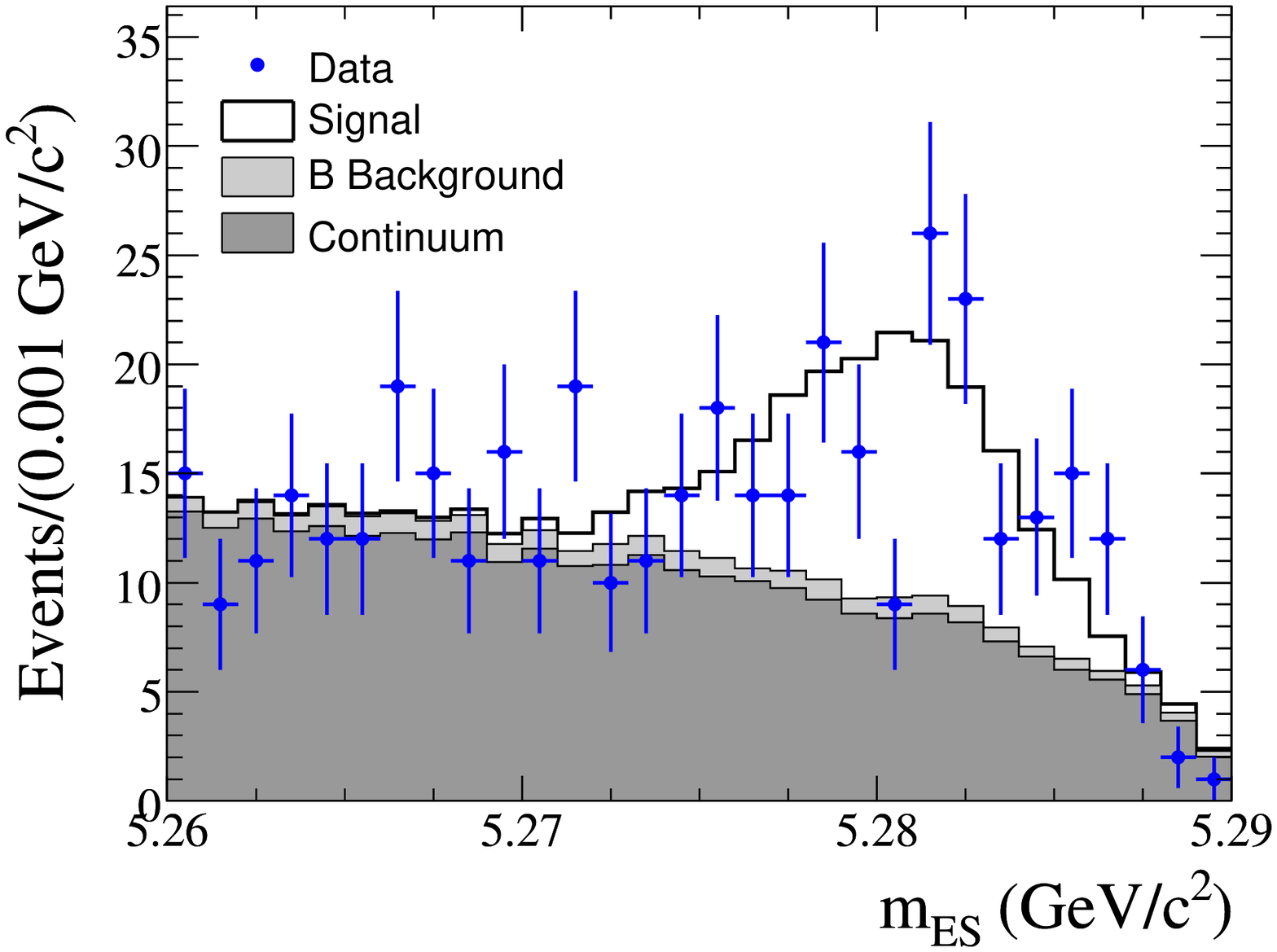}
\includegraphics[width=5.9cm,keepaspectratio]{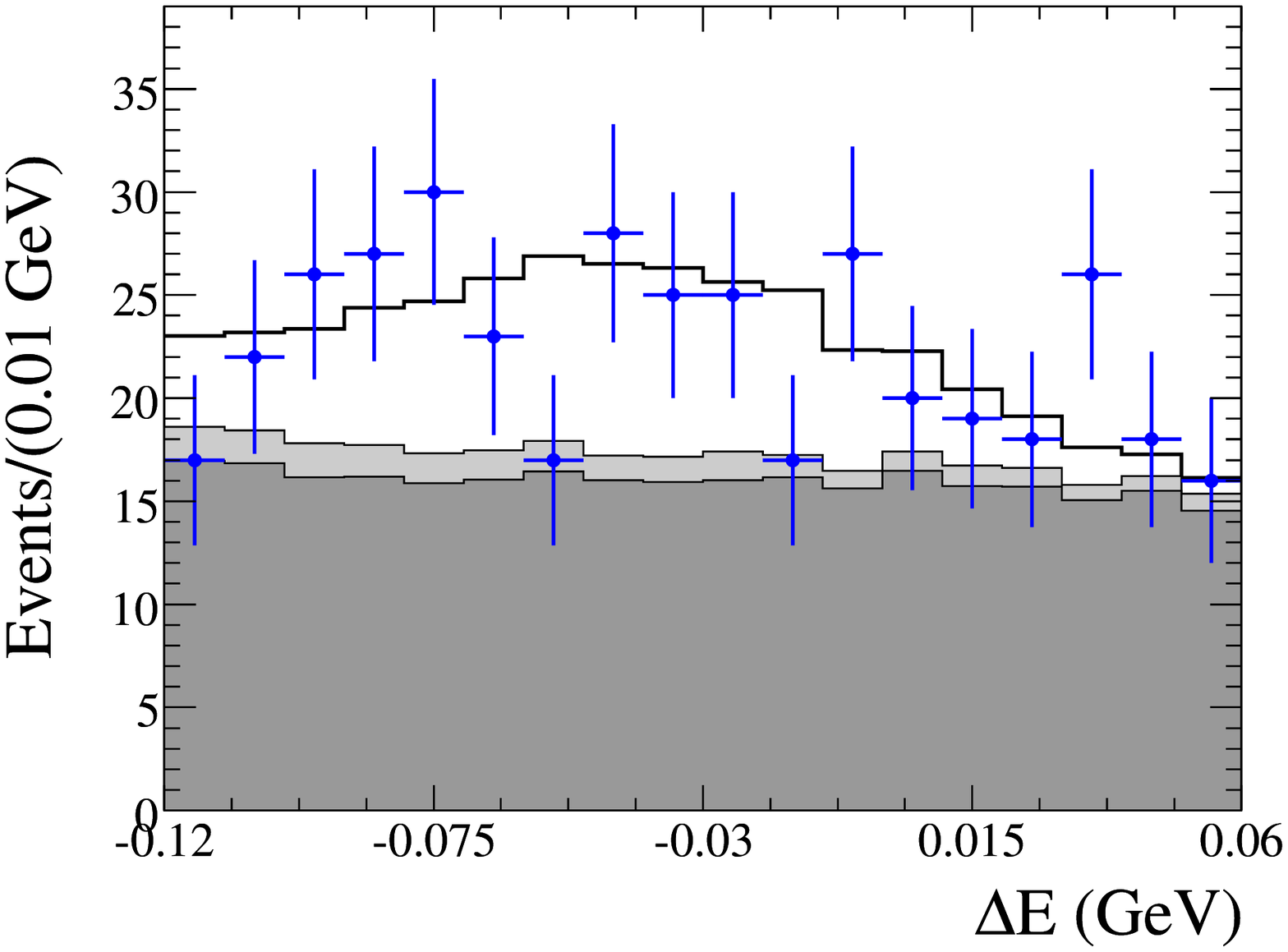}
\includegraphics[width=5.9cm,keepaspectratio]{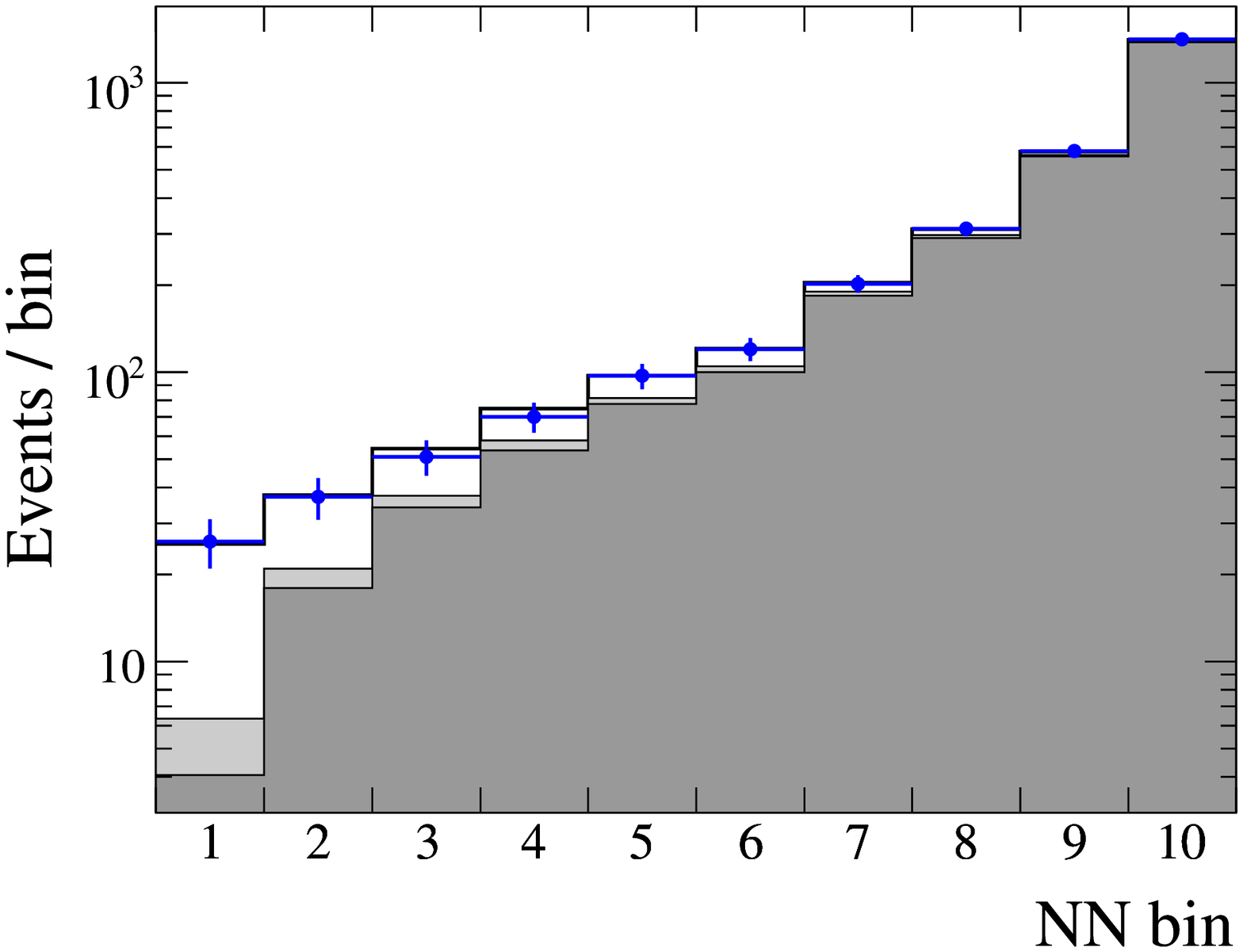}
  \caption{\label{fig:bkkkszz_projections}   Distributions of $\mes$ (left), 
	$\de$ (center), and $\nn$ output (right) for \bkkks, \Kszz.
  The signal in the \mes and \de plots has been enhanced by requiring the
\nn output be 6 or less. The NN output is shown in vertical log scale.  }
\end{figure*}

\begin{figure*}[htbp]
\includegraphics[width=7.9cm,keepaspectratio]{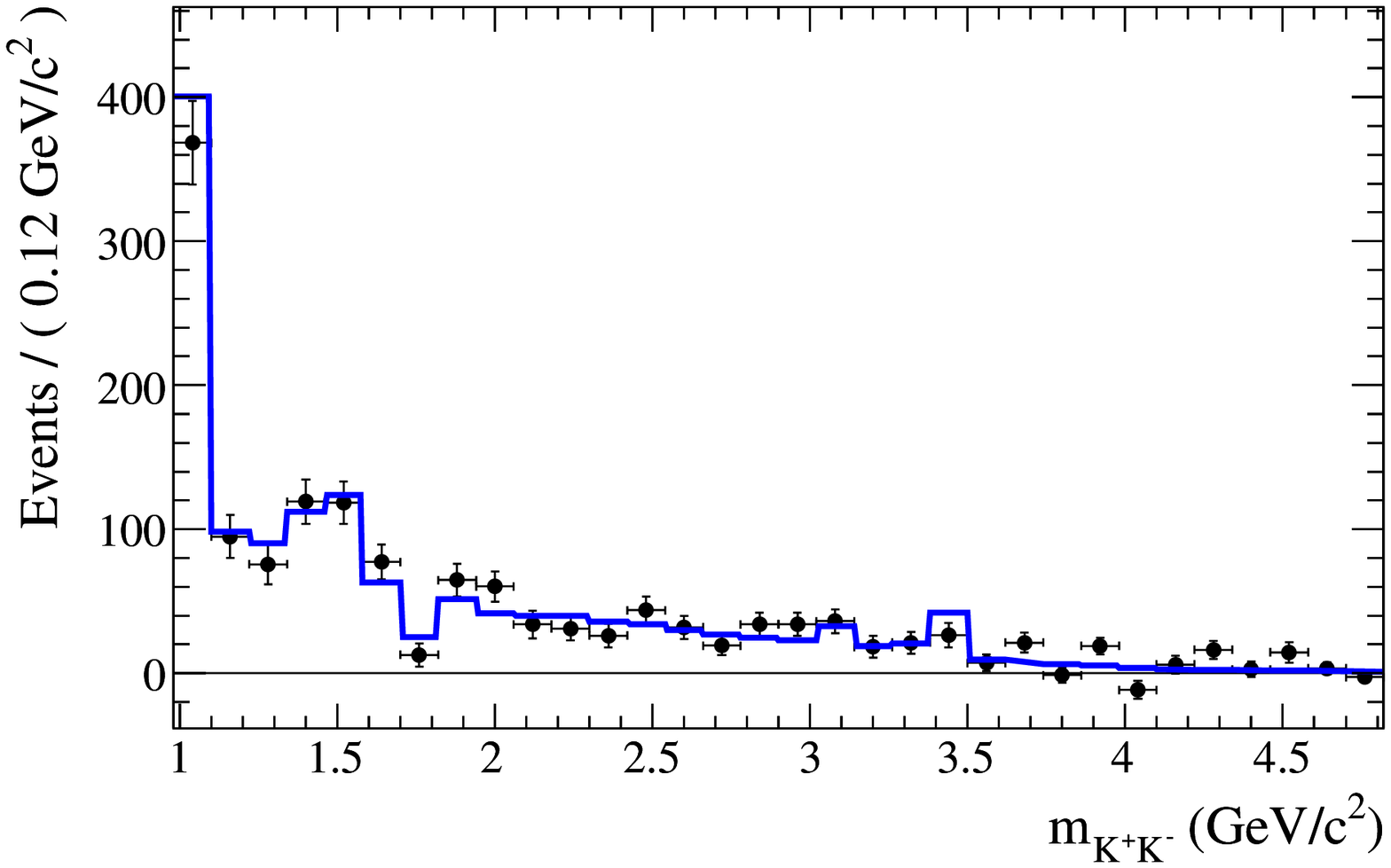}
\includegraphics[width=7.9cm,keepaspectratio]{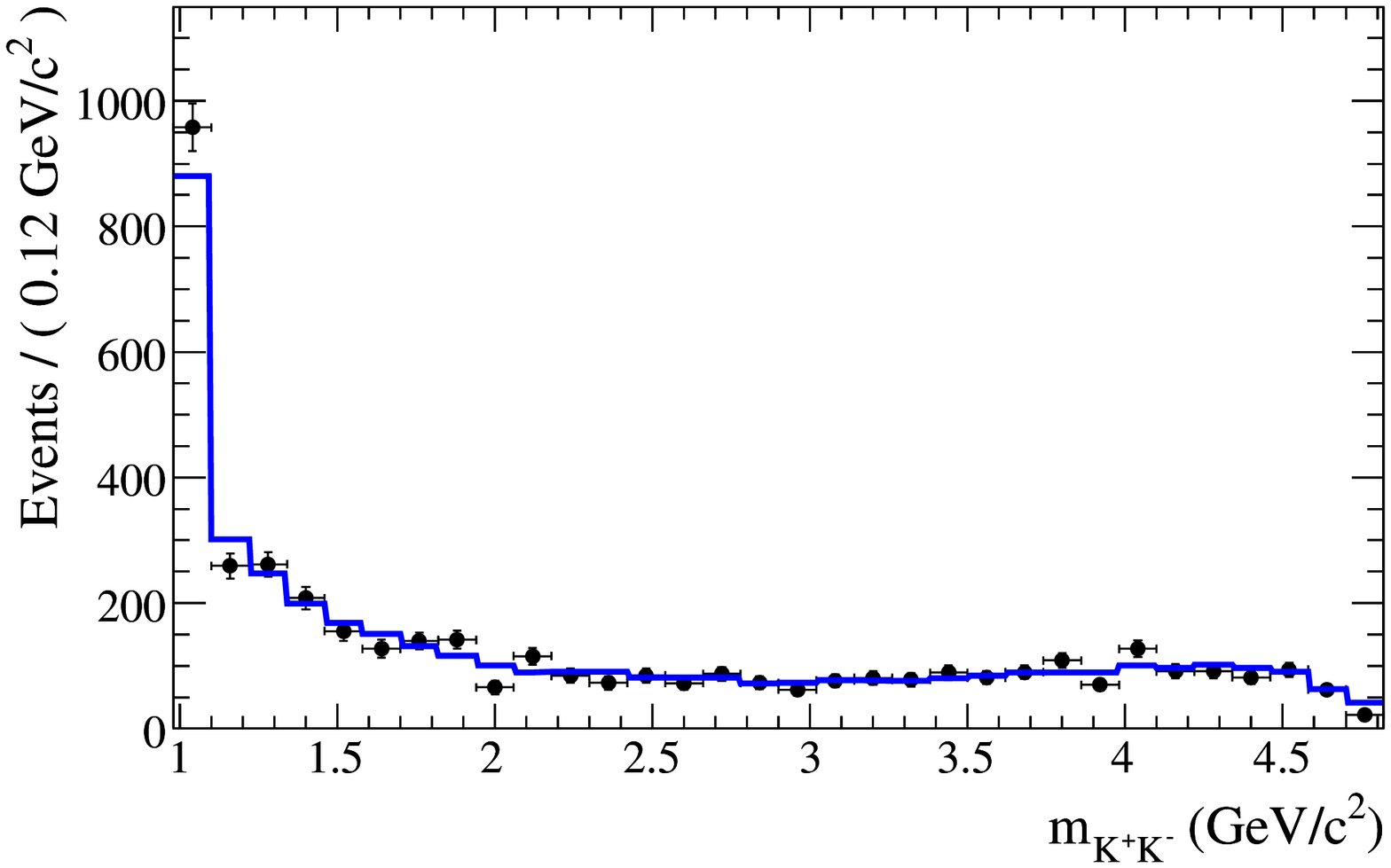}
\includegraphics[width=7.9cm,keepaspectratio]{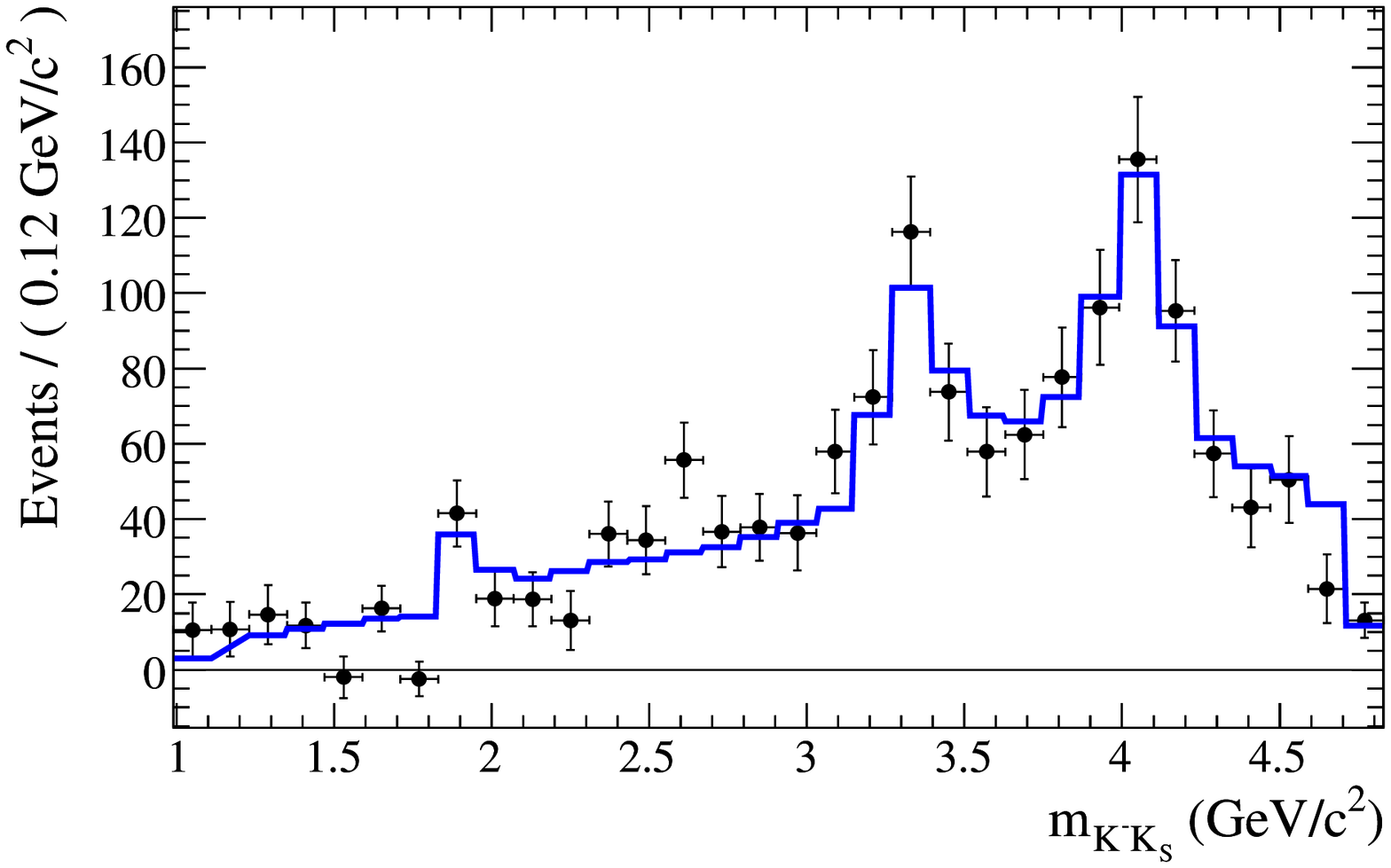}
\includegraphics[width=7.9cm,keepaspectratio]{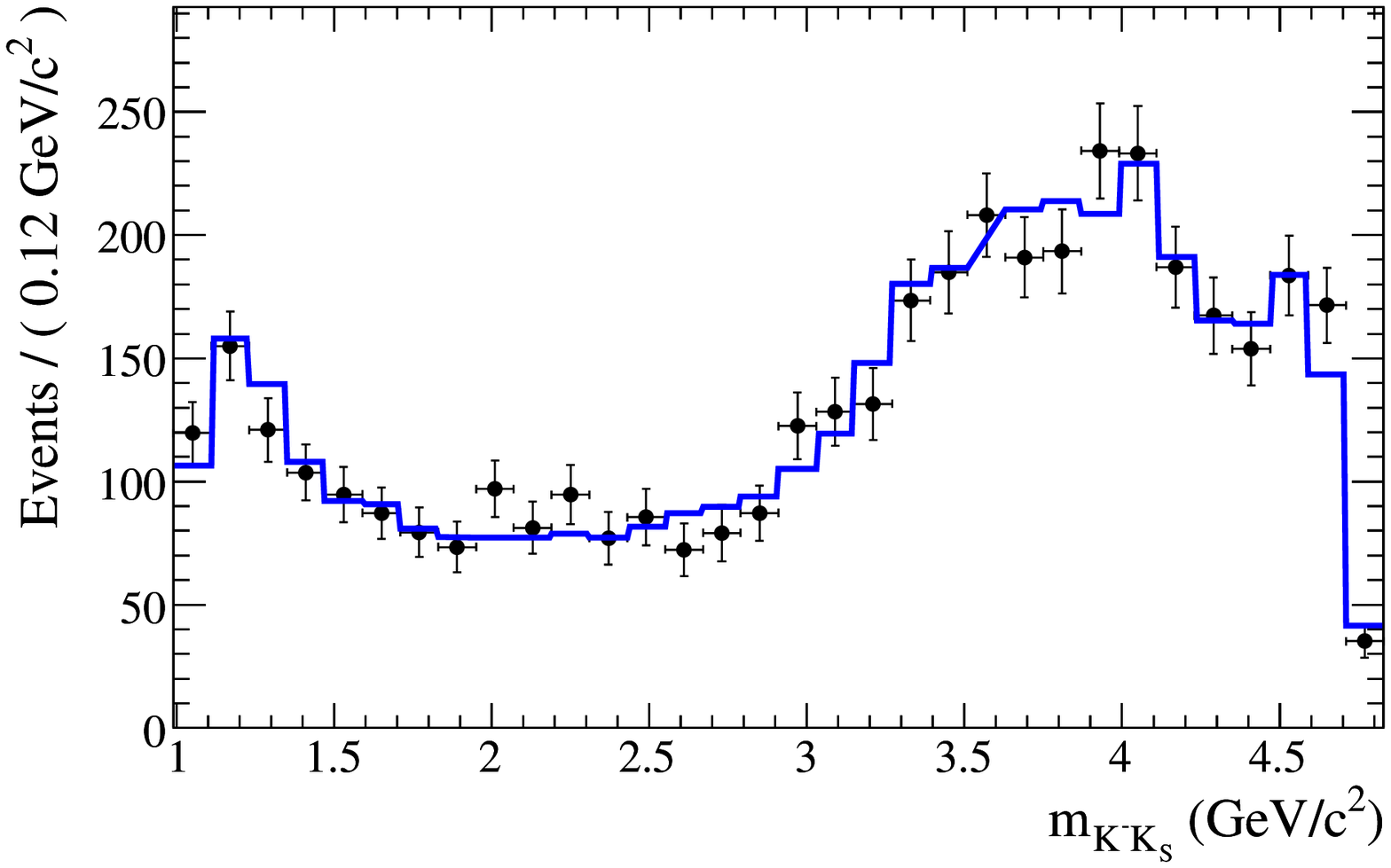}
\includegraphics[width=7.9cm,keepaspectratio]{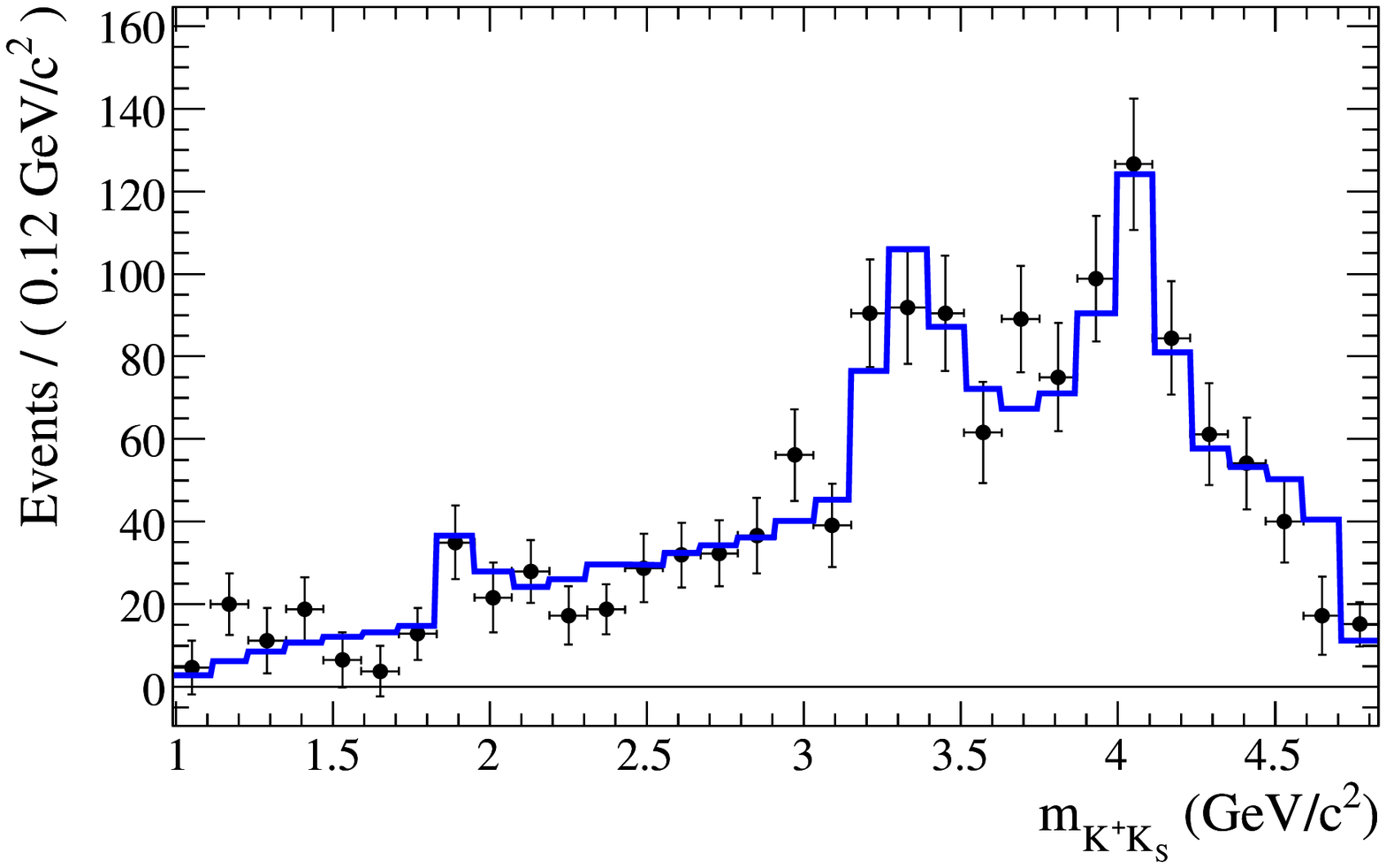}
\includegraphics[width=7.9cm,keepaspectratio]{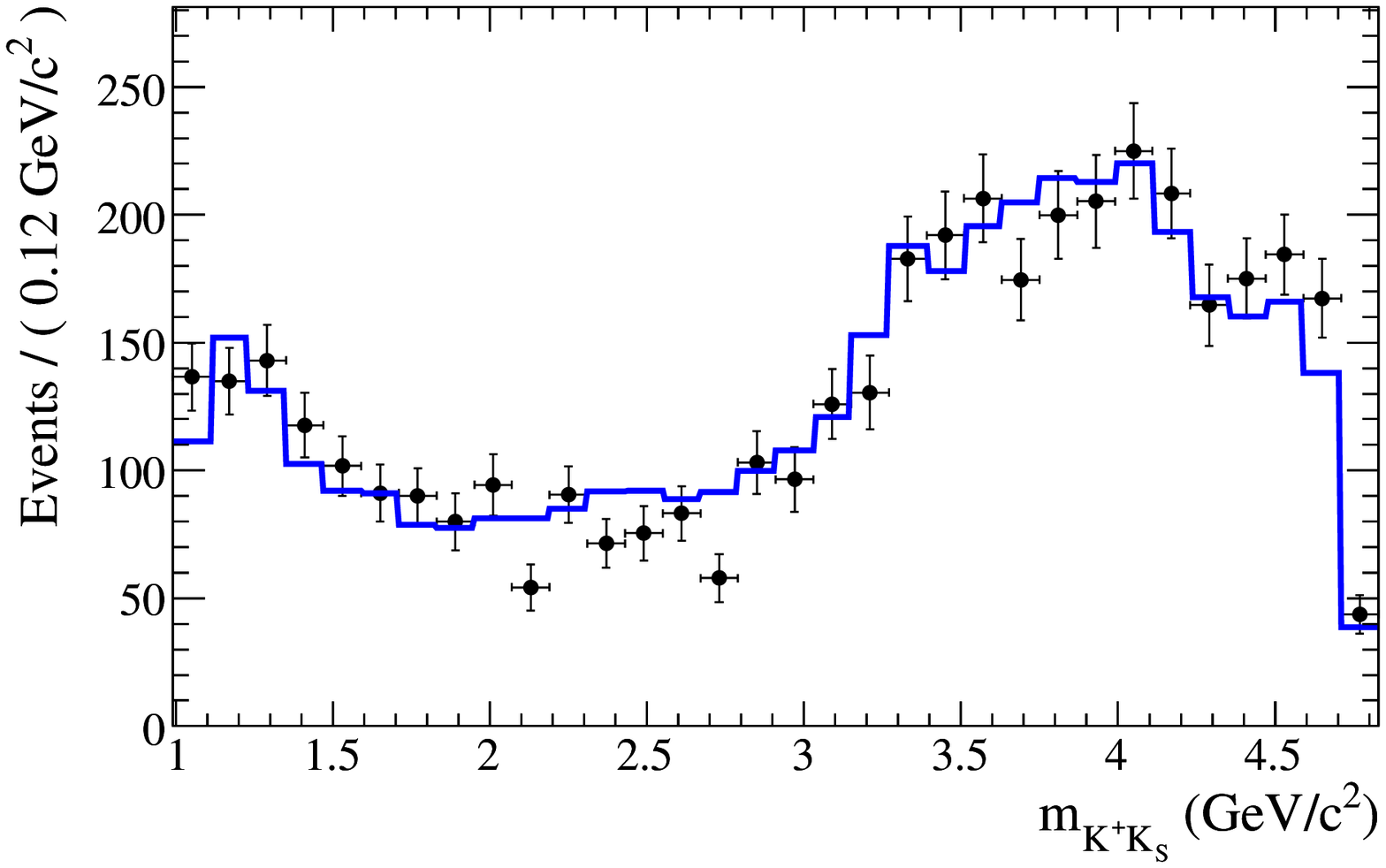}
\caption{\label{fig:kkks_DP}Distributions of $\mab=\mkkL$, $\mbc=\mkmksL$, and $\mac=\mkpksL$,
 for signal-weighted (left) and background-weighted (right) \bkkks candidates in data, \Kspp only.  
The event weighting is performed using the \splot method.  
The fit model (histograms)
is shown superimposed over data (points).
The two main peaks visible in the upper signal plot are due to the \phiI and \fII/\ftwop.
The left-most peak in the middle and lower signal plots is due to $\Dm/D_s^-$ (background).
The other horn-like peaks in those same plots are reflections from the \phiI. 
The upper background plot has a \phiI peak (mainly due to continuum).
}
\end{figure*}

\begin{figure*}[htbp]
\includegraphics[width=8.9cm,keepaspectratio]{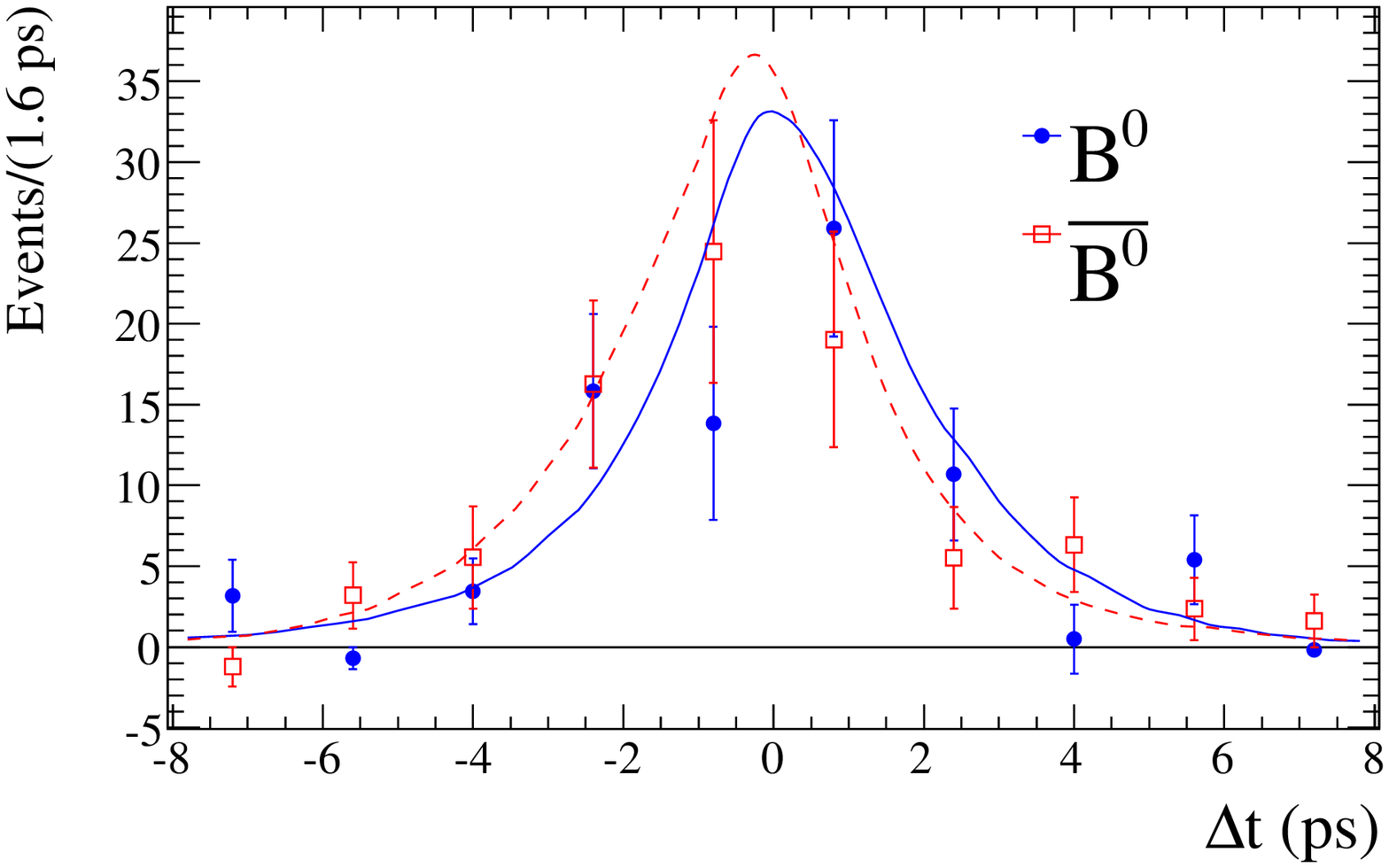}
\includegraphics[width=8.9cm,keepaspectratio]{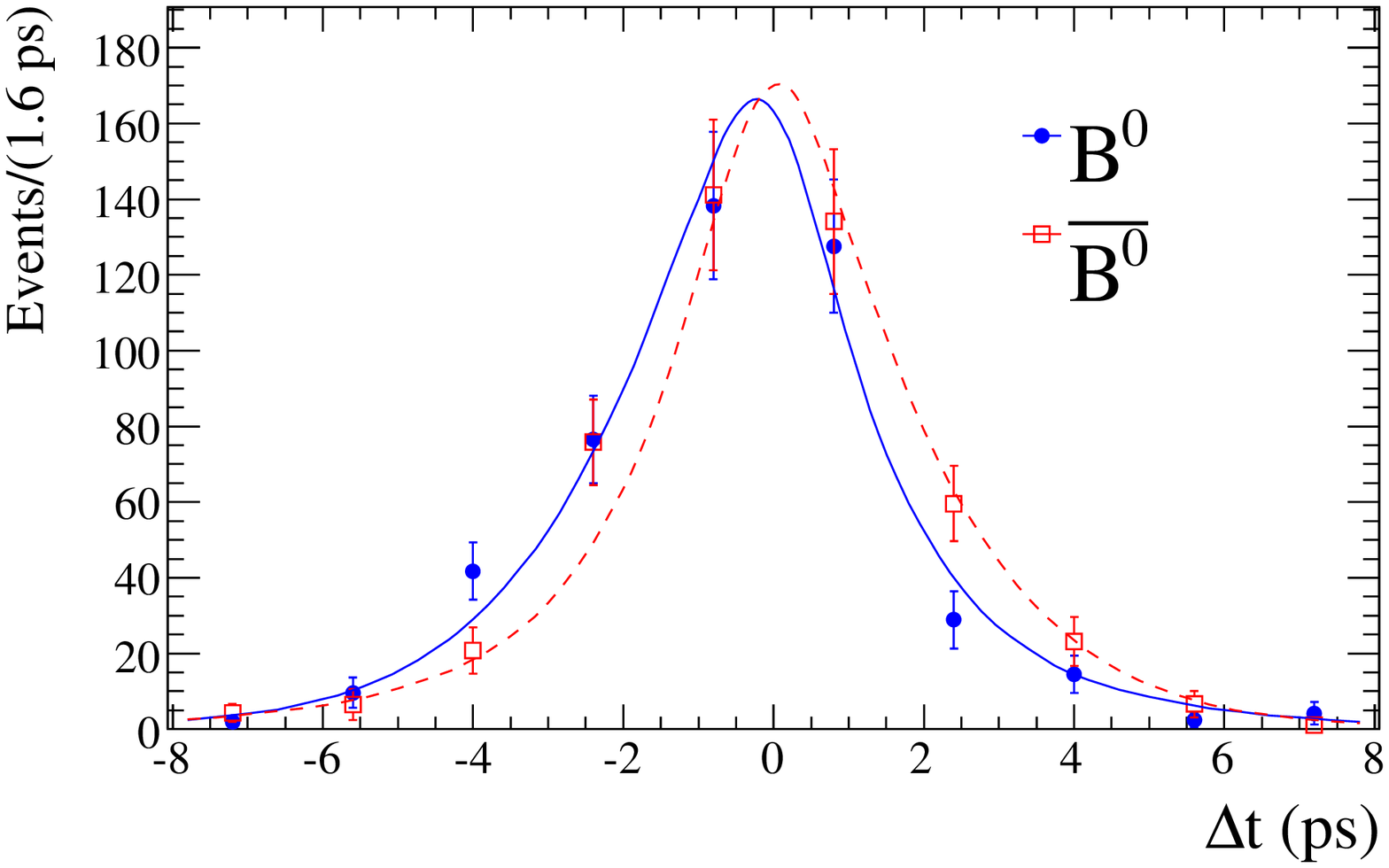}
\includegraphics[width=8.9cm,keepaspectratio]{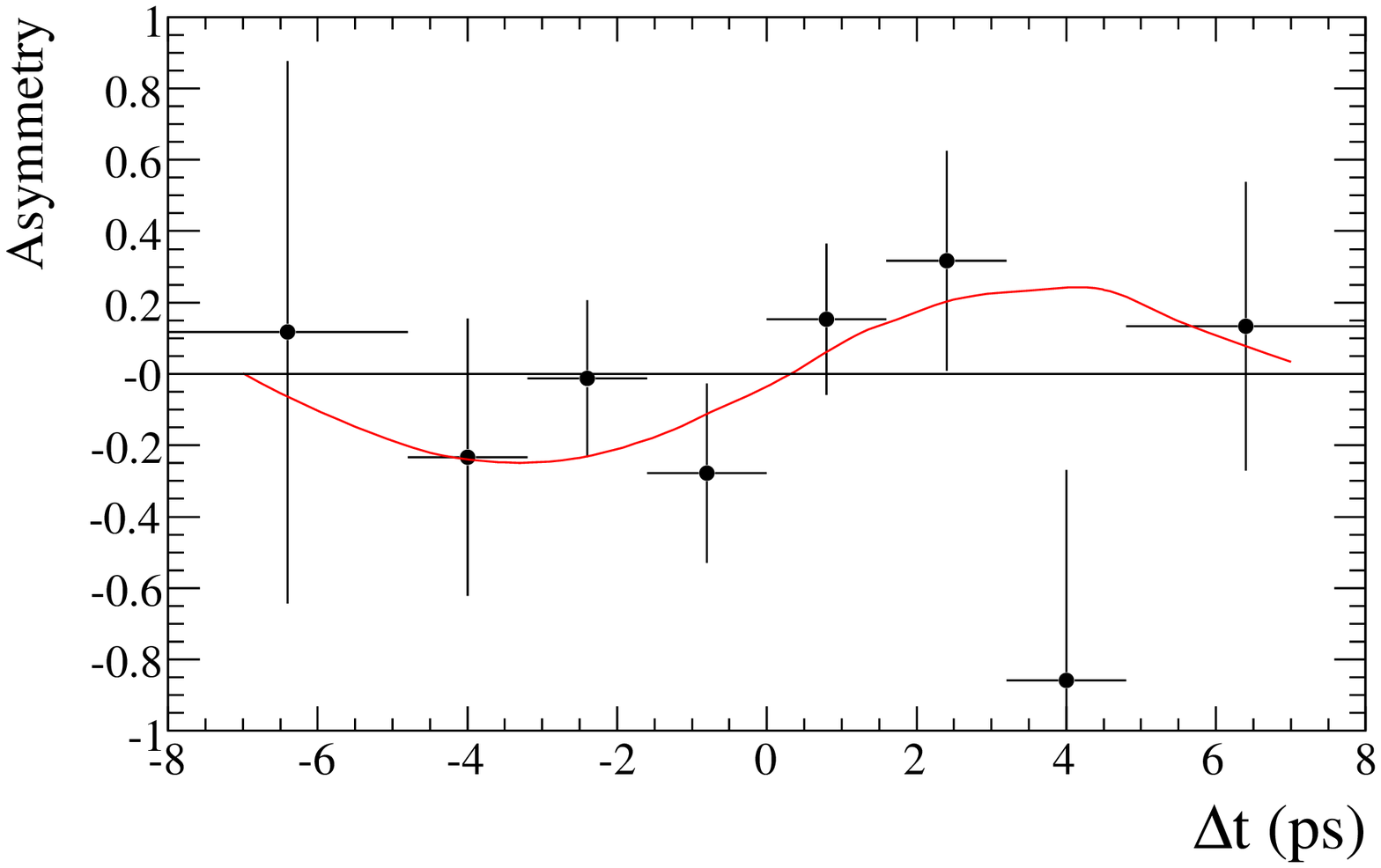}
\includegraphics[width=8.9cm,keepaspectratio]{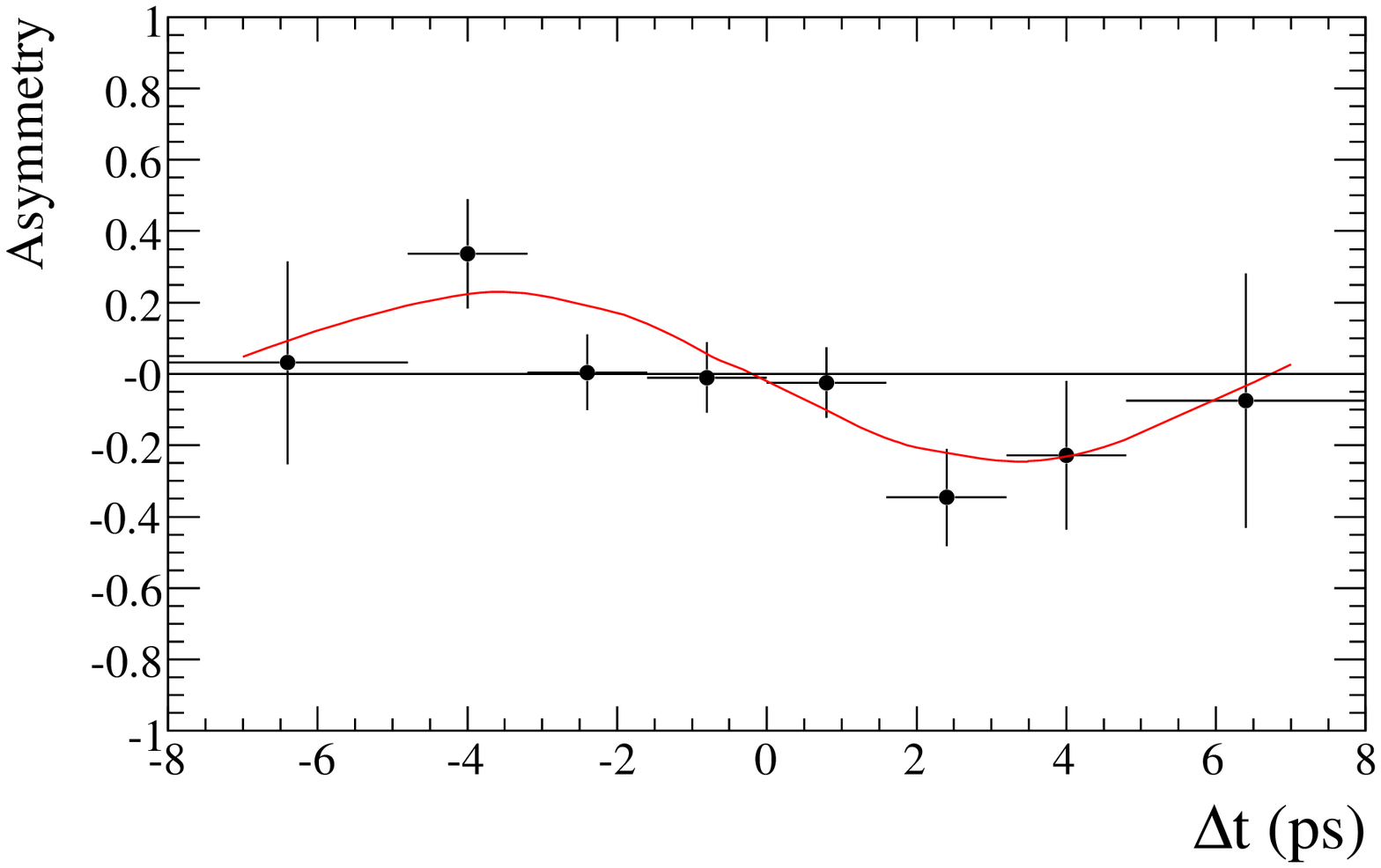}
\caption{\label{fig:bkkks_deltat}  
Top:  The \deltat distributions for \bkkks (\Kspp) signal events, in the $\phiI$
region $(1.01 <\mab<1.03 \gevcc)$ (left) and $\phiI$-excluded region (right).  
\Bz (\Bzb) tagged events are
shown as closed circles (open squares).  The fit model for \Bz (\Bzb) 
tagged events is shown by a solid (dashed) line.  The data points are signal-weighted using the
\splot method.
Bottom: The asymmetry $(N_{\Bz} - N_{\Bzb})/(N_{\Bz} + N_{\Bzb})$ as a function of \deltat, 
in the $\phiI$ region (left) and $\phiI$-excluded region (right).  The points represent
signal-weighted data, and the line is the fit model.
}
\end{figure*}

The \CP-conserving isobar parameters for the global minimum solution are summarized
in Table~\ref{tab:isobarSummary_KKKs_SolnI}, and the branching fractions are
given in Table~\ref{tab:fitresult_KKKs_SolnI}.
Table~\ref{tab:CPV_KKKs_SolnI} shows the values of the \CP-violating observables,
with the central values taken from the global minimum, and the errors taken from
likelihood scans.  (Note that the second minimum is separated from the global 
minimum by $-2\Delta\logL = 3.9$, so the likelihood scan is not impacted by the local
minima at the one standard deviation level.)  
In addition to $\betaeff$ and $\Acp$, we compute the quasi-two-body \CP-violating
parameter $S$, defined as
\beq
S_j  \equiv  - \frac{ 2~\I (e^{-2 i \beta} \overline{a}_j a_j^{*} ) }{ |a_j|^2 + |\overline{a}_j|^2 }
  =    \frac{1-b_j^2}{1+b_j^2} \sin ( 2 \betaeffj ) \,.
\eeq

The fit fraction matrix for the best solution is given 
in the Appendix, and the correlation matrix of the 
isobar parameters is given in Ref.~\cite{epaps}.  The correlation matrix for the
\CP-violating observables is given in Table~\ref{tab:corrMatrix_KKKs_SolnI}.

\begin{table}[htbp]
\center
\caption{\CP-conserving isobar parameters (defined in Eq.~\eqref{eq:isobarPars}) for \bkkks, 
for the global minimum. The NR coefficients are defined in Eq.~\eqref{eq:NRpoly}.  
Phases are given in degrees.  Only statistical uncertainties are given.
}
\begin{tabular}{ll|c}
\hline \hline
\multicolumn{2}{l|}{Parameter }   &  Value \\
\hline
\noalign{\vskip1pt} 
$\phiI$$\KS$        &    $c$   &    $0.039\pm 0.005 $  \\
     &    $\phi$   &    $20\pm 19 $  \\
\hline
\noalign{\vskip1pt} 
$\fI$$\KS$        &    $c$   &    $2.2\pm 0.5 $  \\
     &    $\phi$   &    $40\pm 16 $  \\
\hline
\noalign{\vskip1pt} 
$\fII$$\KS$        &    $c$   &    $0.22\pm 0.05 $  \\
     &    $\phi$   &    $17\pm 16 $  \\
\hline
\noalign{\vskip1pt} 
$\ftwop$$\KS$        &    $c$   &    $0.00080\pm 0.00028 $  \\
     &    $\phi$   &    $53\pm 23 $  \\
\hline
\noalign{\vskip1pt} 
$\fIII$$\KS$        &    $c$   &    $0.72\pm 0.11 $  \\
     &    $\phi$   &    $110\pm 11 $  \\
\hline
\noalign{\vskip1pt} 
$\chiczero$$\KS$        &    $c$   &    $0.144\pm 0.023 $  \\
     &    $\phi$   &    $-17\pm 29 $  \\
\hline
NR   & &  \\
$a_{S0}$  &   $c$   & $  1.0 $ (fixed) \\
    &     $\phi$   &  $  0$ (fixed)    \\
$a_{S1}$        &    $c$   &    $1.25\pm 0.25 $  \\
     &    $\phi$   &    $-149\pm 9 $  \\
$a_{S2}$        &    $c$   &    $0.58\pm 0.22 $  \\
     &    $\phi$   &    $56\pm 15 $  \\
$a_{P0}$        &    $c$   &    $1.22\pm 0.22 $  \\
     &    $\phi$   &    $65\pm 13 $  \\
$a_{P1}$        &    $c$   &    $0.28\pm 0.18 $  \\
     &    $\phi$   &    $-68\pm 28 $  \\
$a_{P2}$        &    $c$   &    $0.42\pm 0.16 $  \\
     &    $\phi$   &    $-131\pm 25 $  \\
\hline  \hline
\end{tabular}
\label{tab:isobarSummary_KKKs_SolnI}. 
\end{table}

\begin{table*}[htbp]
\center
\caption{Branching fractions (neglecting interference) for \bkkks. 
The ${\cal B}(\Bz\to R\Kz)$ column gives the branching fractions to intermediate
resonant states, corrected for secondary branching fractions obtained from
Ref.~\cite{Nakamura:2010zzi}.
In addition to quoting the overall NR branching fraction, we quote the 
S-wave and P-wave NR branching  fractions separately.  
Central values and uncertainties are for the global minimum only.  See the text
for discussion of the variations between the local minima.  
}
\begin{tabular}{l|cc}
\hline \hline
\noalign{\vskip1pt} 
Decay mode &  ${\cal B}(\bkkkz)\times \fitfrac_j~(10^{-6})$ &  ${\cal B}(\Bz\to R\Kz)~(10^{-6})$   \\
\hline
\noalign{\vskip1pt} 
$\phiI$$\Kz$   &   $3.48\pm 0.28 ^{+0.21}_{-0.14} $  &  $7.1 \pm 0.6 ^{+0.4}_{-0.3}$ \\
$\fI$$\Kz$   &   $7.0^{+2.6}_{-1.8} \pm 2.4 $         &   \\
$\fII$$\Kz$   &   $0.57^{+0.25}_{-0.19} \pm 0.12 $   &   $13.3 ^{+5.8}_{-4.4} \pm 3.2$ \\
$\ftwop$$\Kz$   &   $0.13^{+0.12}_{-0.08} \pm 0.16 $  &  $0.29 ^{+0.27}_{-0.18} \pm 0.36$   \\
$\fIII$$\Kz$   &   $4.4\pm 0.7 \pm 0.5 $               &   \\
$\chiczero$$\Kz$   &   $0.90\pm 0.18 \pm 0.06 $        &  $148 \pm 30 \pm 13$  \\
NR            &    $33\pm 5 \pm 9 $                    &   \\
NR (S-wave)   &   $30\pm 5 \pm 8 $                    & \\
NR (P-wave)   &   $3.1\pm 0.7 \pm 0.4 $               &   \\
\hline  \hline
\end{tabular}
\label{tab:fitresult_KKKs_SolnI}
\end{table*}

\begin{table*}[h]
\center
\caption{
\CP-violating parameters $\betaeff$, $\Acp$, and $S$ for \bkkks.  Central values correspond to the
global minimum.  Statistical uncertainties for $\betaeff$ and $\Acp$ are determined from likelihood scans.
}
\begin{tabular}{l|ccc}
\hline \hline
\noalign{\vskip1pt} 
Component  &  $\betaeff$ (deg)  &       $\Acp (=-C) (\%)$    & $S$   \\
\hline
\noalign{\vskip1pt} 
$\phiI$$\KS$   &    $21\pm 6 \pm 2 $ &    $-5\pm 18 \pm 5 $ &  $0.66 \pm 0.17 \pm 0.07$  \\
$\fI$$\KS$   &    $18\pm 6 \pm 4 $ &    $-28\pm 24 \pm 9 $  &  $0.55 \pm 0.18 \pm 0.12$ \\
Other   &    $20.3\pm 4.3 \pm 1.2 $ &    $-2\pm 9 \pm 3 $  &  $0.65 \pm 0.12 \pm 0.03$ \\
\hline  \hline
\end{tabular}
\label{tab:CPV_KKKs_SolnI}
\end{table*}

\begin{table*}[htbp]
\center
\caption{Statistical correlation matrix for the \CP-violating parameters $\betaeff$ and $\Acp$ for \bkkks. 
The matrix corresponds to the global minimum solution.}
\begin{tabular}{l|cccccc}
\hline \hline
  & $\betaeff(\phiI)$  & $\betaeff(\fI)$  & $\betaeff($Other$)$ &  $\Acp(\phiI)$  & $\Acp(\fI)$  & $\Acp($Other$)$ \\ 
 \hline
   $\betaeff(\phiI)$ &  $ 1.00 $   & $ 0.38 $   & $ 0.15 $   & $ 0.21 $   & $ -0.44 $   & $ -0.32 $    \\
   $\betaeff(\fI)$    &  &  $ 1.00 $   & $ 0.63 $   & $ -0.10 $   & $ 0.05 $   & $ -0.33 $    \\
   $\betaeff($Other$)$ &     &  &  $ 1.00 $   & $ -0.13 $   & $ 0.47 $   & $ 0.14 $   \\
   $\Acp(\phiI)$       &      &    &     &    $ 1.00 $   & $ -0.25 $   & $ -0.14 $    \\
   $\Acp(\fI)$    &  &     &       &                     &  $ 1.00 $   & $ 0.60 $    \\
   $\Acp($Other$)$ &     &  &      &     &                              & $ 1.00 $    \\
\hline  \hline
\end{tabular}
\label{tab:corrMatrix_KKKs_SolnI}
\end{table*}

The other minima all have consistent values for the $\phiI$, $\ftwop$, P-wave NR, 
and $\chiczero$ fit fractions, but there are large variations in the fit fractions 
for the other states.  Specifically, the fit fraction of the $\fI$ varies between $19\%$ and $41\%$, the 
fit fraction of the $\fII$ varies between $2\%$ and $51\%$, the 
fit fraction of the $\fIII$ varies between $2\%$ and $27\%$, and the S-wave NR 
fit fraction varies between $34\%$ and $120\%$.
The signal yields for the different solutions, however, 
exhibit negligible variation.
We calculate the inclusive branching fraction using only the yield in the \Kspp channel.
We find ${\cal B}(\bkkkz)= (26.5 \pm 0.9 \pm 0.8)  \times 10^{-6}$, or 
${\cal B}(\bkkkz)= (25.4 \pm 0.9 \pm 0.8) \times 10^{-6}$ if the $\chi_{c0}$ is excluded.

Likelihood scans for each of the \betaeff and \Acp are shown in 
Figs.~\ref{fig:bkkks_scan_phi}-\ref{fig:bkkks_scan_other}.
  $\betaeff({\rm other})$ is different from zero
with $4.3~\sigma$ significance.  We can also distinguish between 
$\betaeff$ and the trigonometric reflection $90^\circ - \betaeff$, due to the
sensitivity of the DP analysis to interference between S-wave and P-wave 
amplitudes.  We find that $\betaeff({\rm other})$ is favored over 
$90^\circ - \betaeff({\rm other})$ with  $4.8~\sigma$ significance.

\begin{figure*}[htbp]
\includegraphics[width=8.9cm,keepaspectratio]{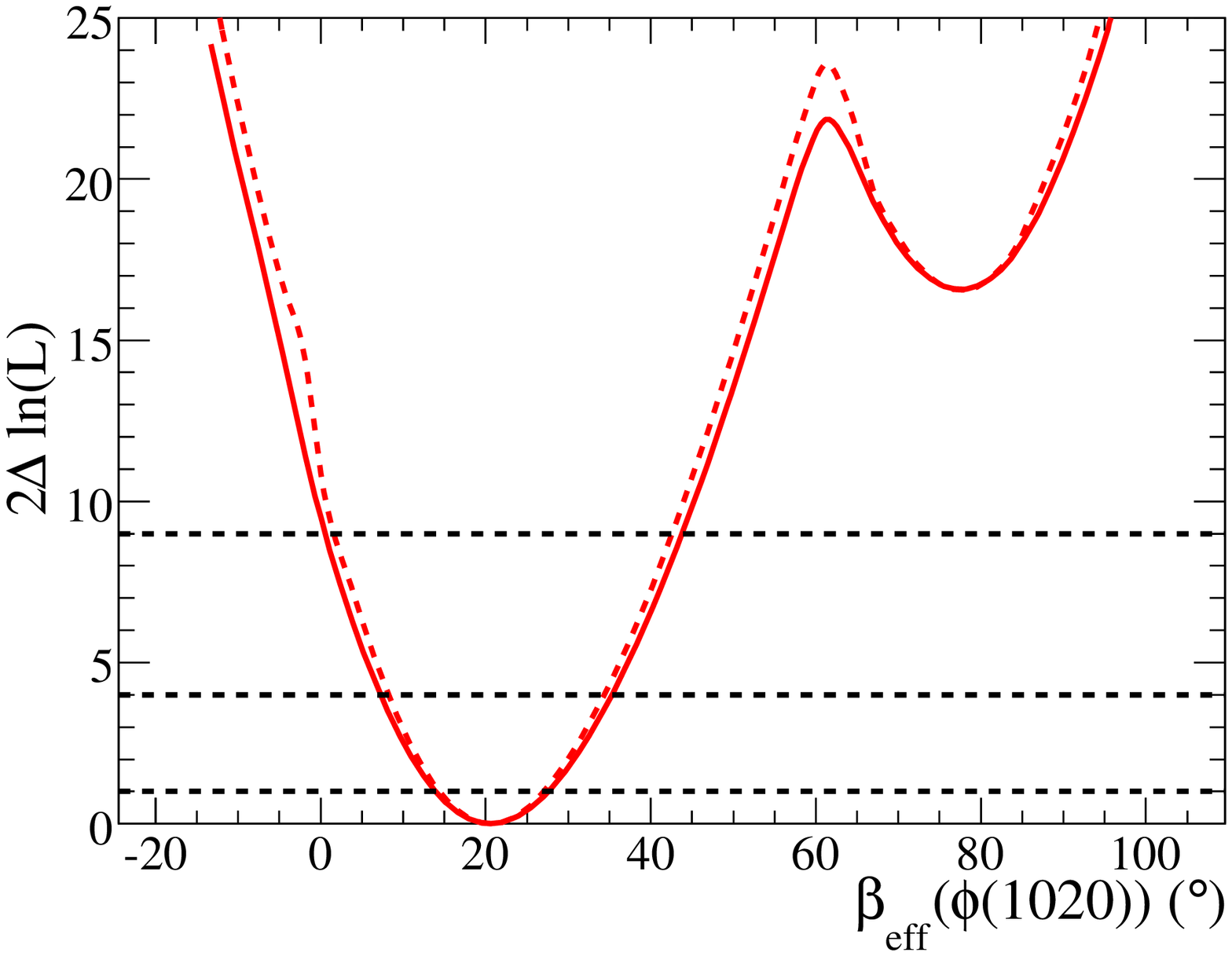}
\includegraphics[width=8.9cm,keepaspectratio]{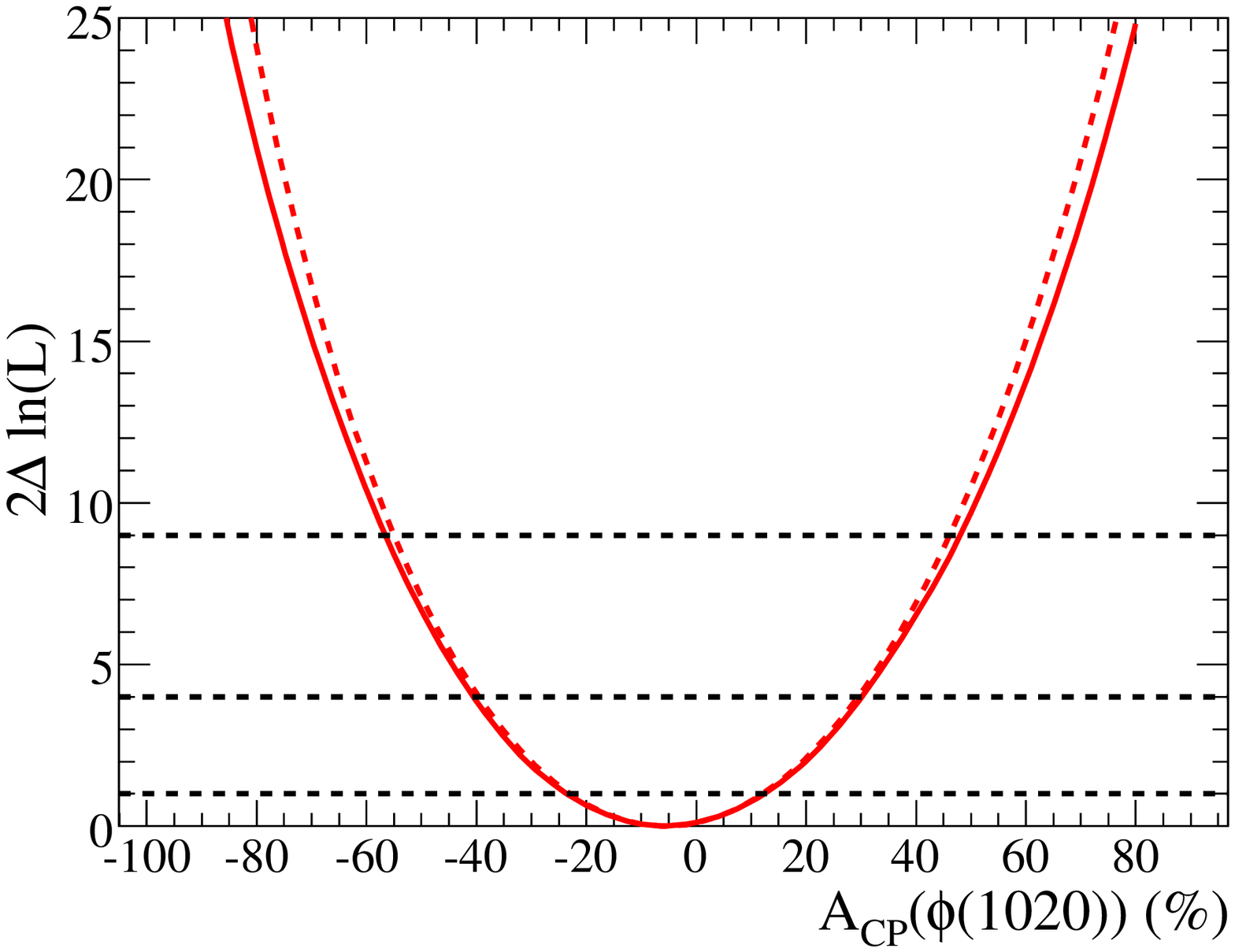}
\caption{\label{fig:bkkks_scan_phi}
Scan of $2\Delta \logL$, with (solid line) and without (dashed line)
systematic uncertainties,
as a function of $\betaeff$ (left) and $\Acp$ (right) for $\Bz\to\phiI\KS$.}
\end{figure*}

\begin{figure*}[htbp]
\includegraphics[width=8.9cm,keepaspectratio]{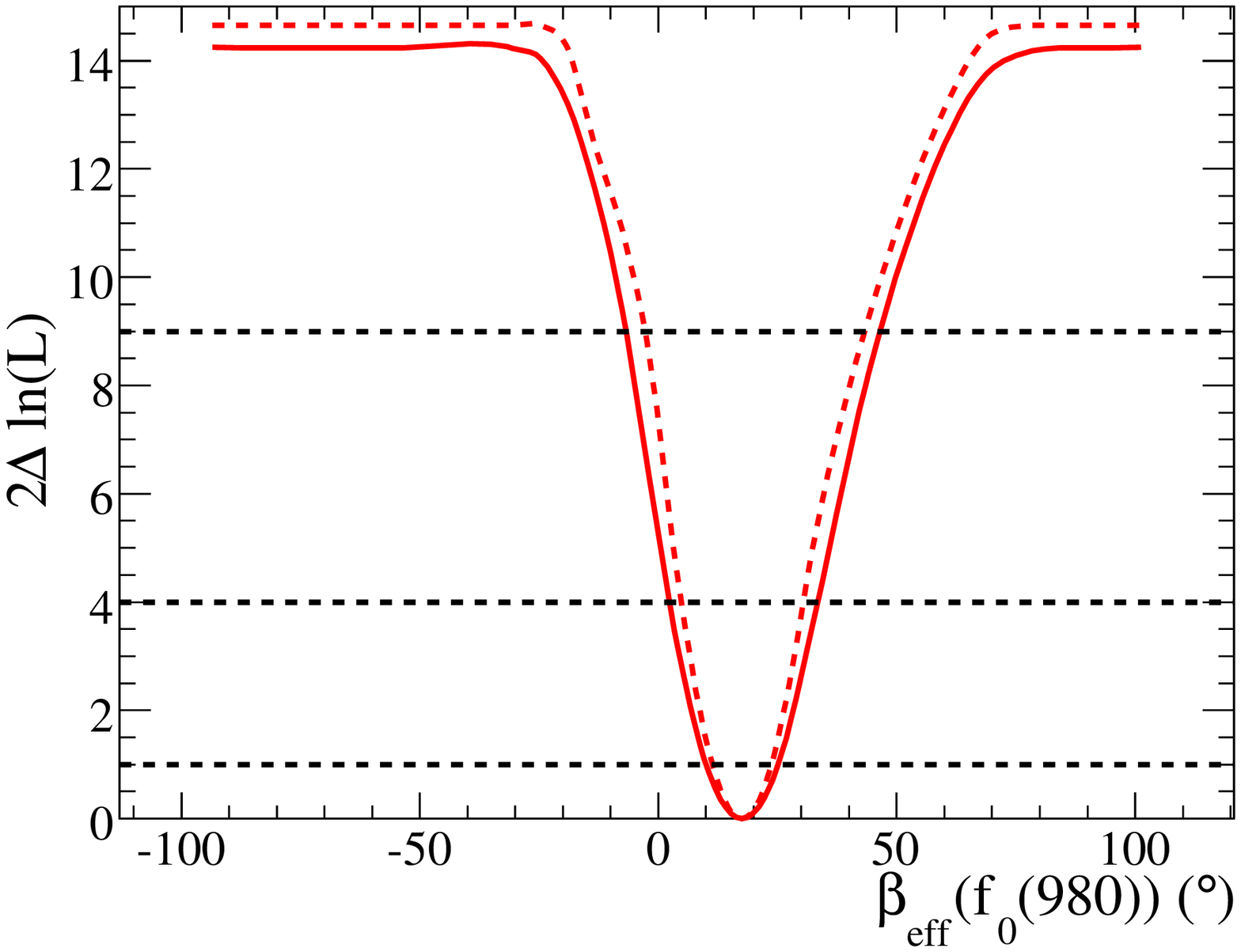}
\includegraphics[width=8.9cm,keepaspectratio]{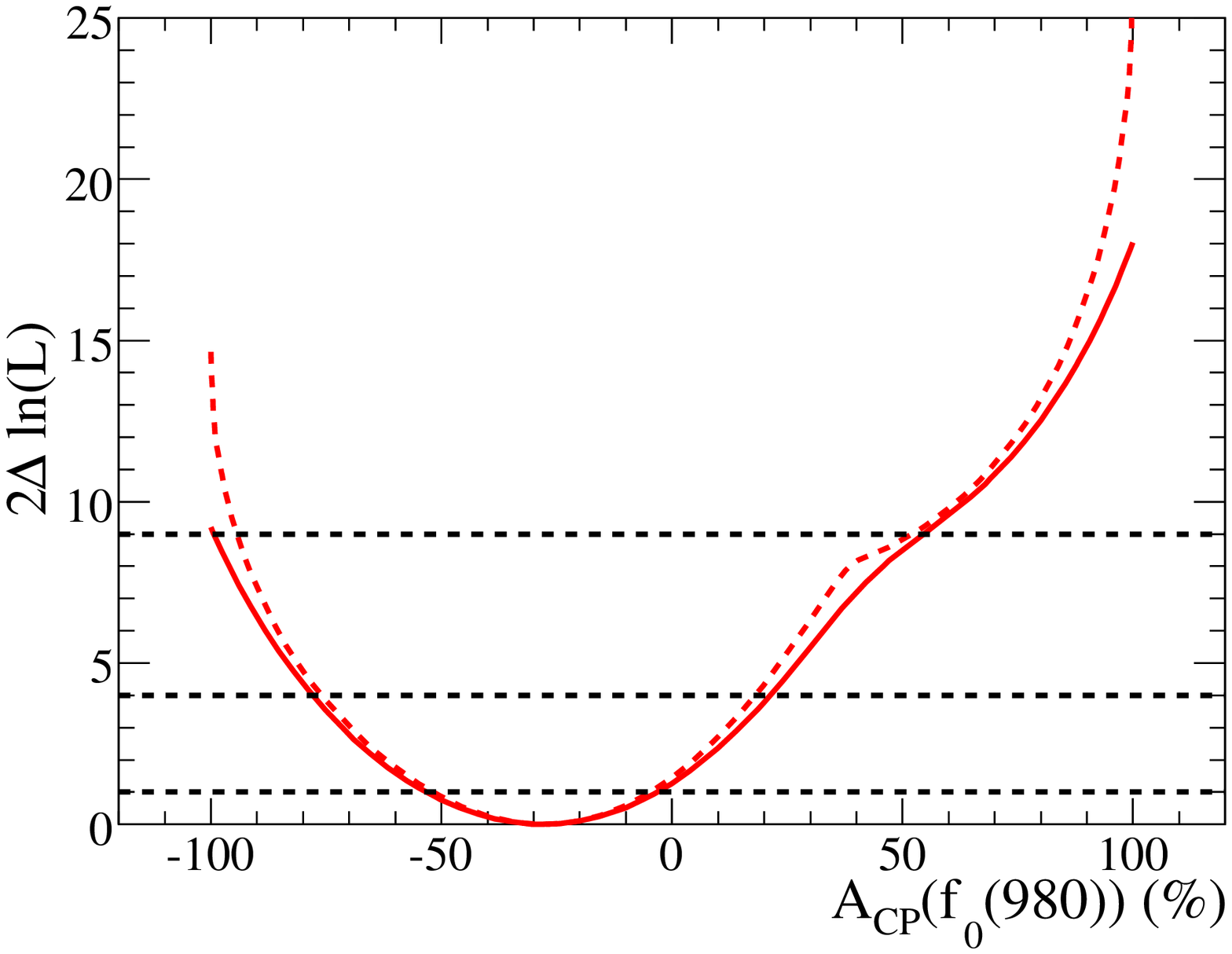}
\caption{\label{fig:bkkks_scan_f0}  
Scan of $2\Delta \logL$, with (solid line) and without (dashed line)
systematic uncertainties,
as a function of $\betaeff$ (left) and $\Acp$ (right) for $\Bz\to\fI\KS$.
The flat region of the $\betaeff$ scan is caused by the $\Acp(\fI)$ going to
$-100\%$ in this region, in which case $\betaeff$ becomes an irrelevant 
parameter.}
\end{figure*}

\begin{figure*}[htbp]
\includegraphics[width=8.9cm,keepaspectratio]{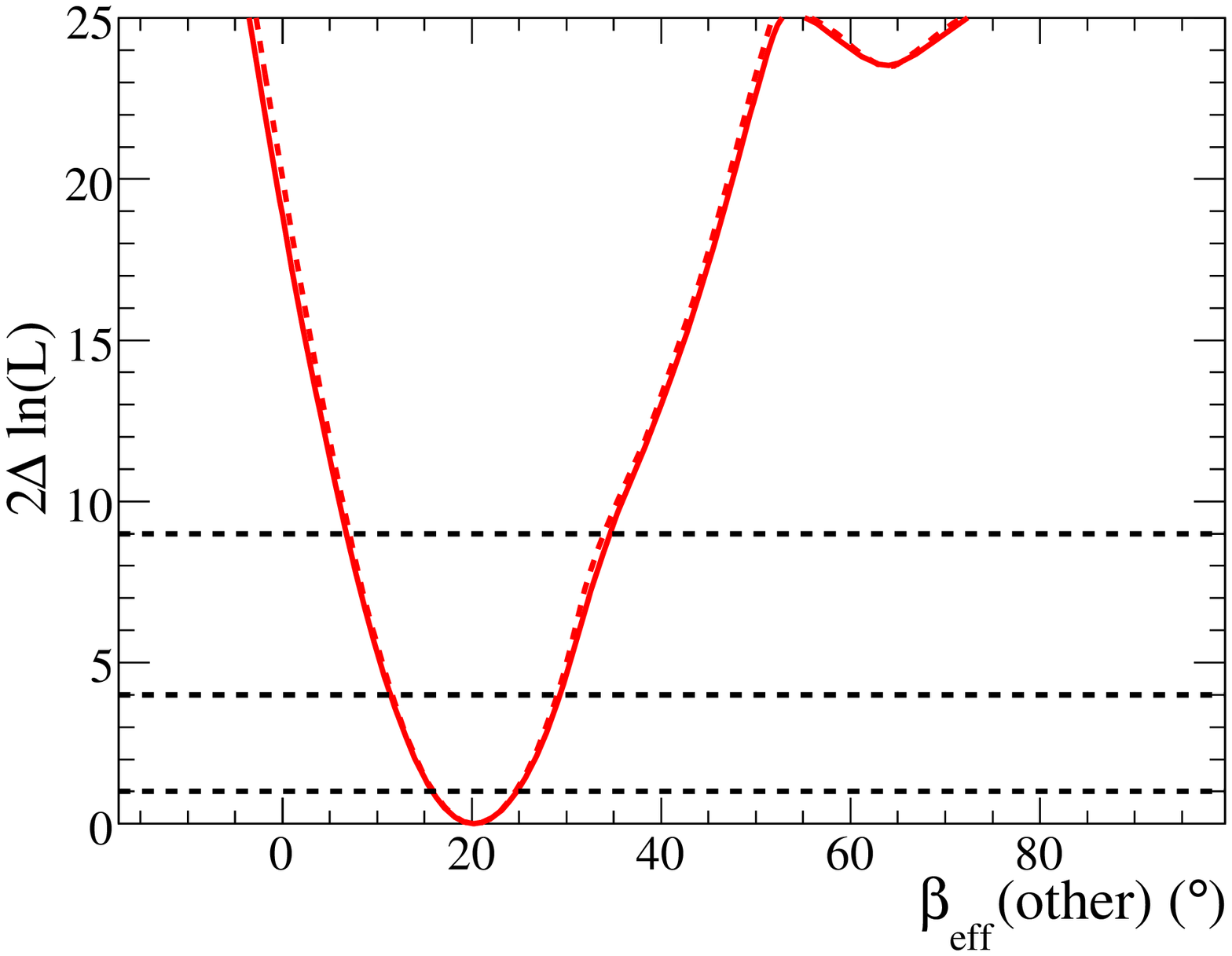}
\includegraphics[width=8.9cm,keepaspectratio]{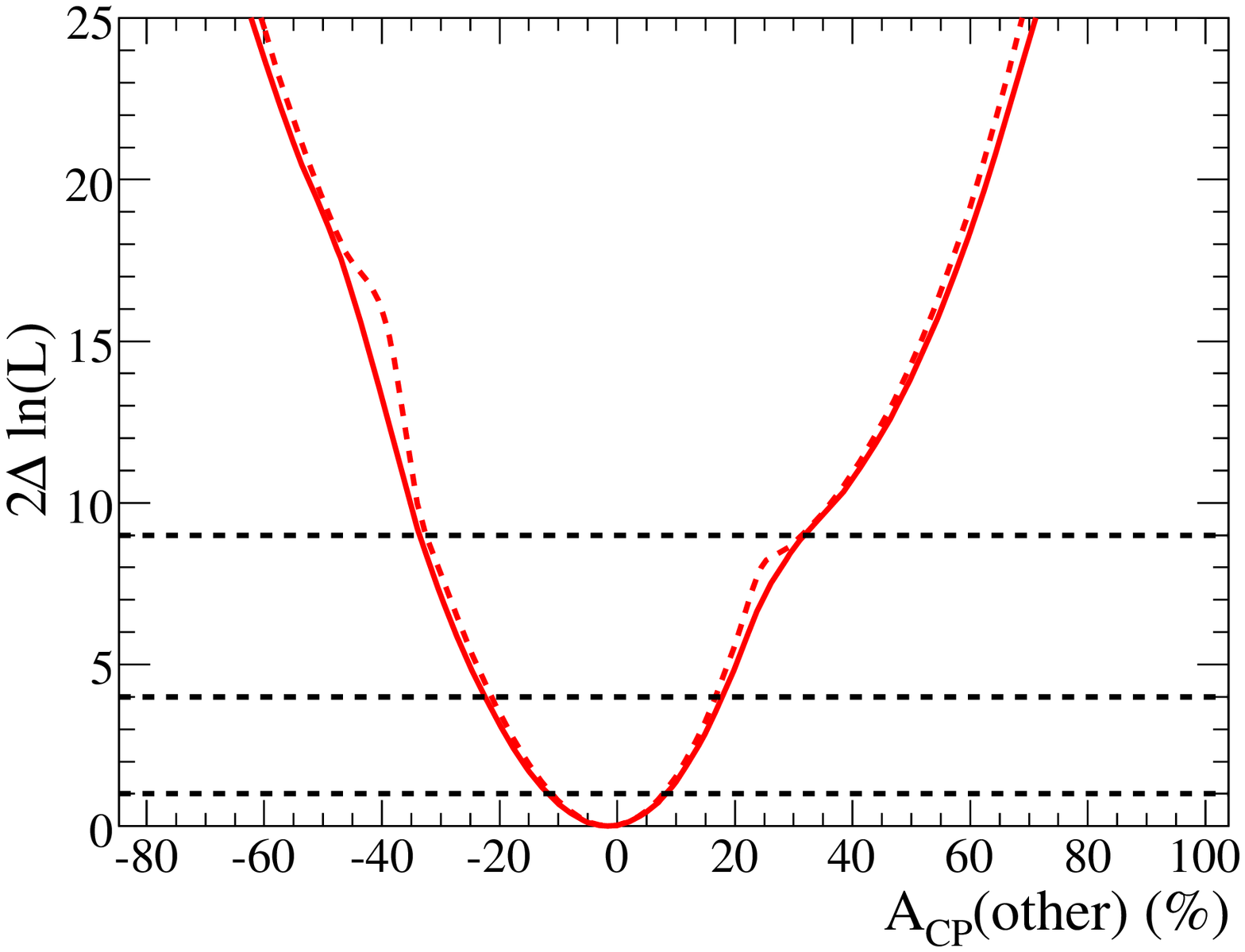}
\caption{\label{fig:bkkks_scan_other}  
Scan of $2\Delta \logL$, with (solid line) and without (dashed line)
systematic uncertainties,
as a function of $\betaeff$ (left) and $\Acp$ (right) for $\bkkks$, excluding
the $\phiI$, $\fI$, and $\chiczero$.}
\end{figure*}

\subsection{Interpretation}
\label{sec:interpretation}
The value we measure for $\Acp(\phi\Kp)$ is larger than the SM
prediction, while $\betaeff(\phi\KS)$ is in excellent agreement with
the SM.
We can use the measured $\Acp(\phi\Kp)$ and $\betaeff(\phi\KS)$ to put 
constraints on the amplitudes contributing to these decays.  We 
assume isospin symmetry, so that the amplitudes for $\Bp\to\phi\Kp$ and
$\Bz\to\phi\KS$ are the same.  We also assume that this amplitude, $\Amp$, can
be written as the sum of two amplitudes, $\Amp_1$ and $\Amp_2$, where $\Amp_1$
is the dominant penguin amplitude.  $\Amp_2$ is an arbitrary additional
amplitude with a different weak phase, which could be a tree, $u$-penguin,
or new physics amplitude.

Then 
\beqn
\Amp & = & \Amp_1 (1 + r e^{i (\strong + \weak)} ) \,,   \nonumber   \\
\Ampbar & = & \Amp_1 (1 + r e^{i (\strong - \weak)} ) \,,
\eeqn
where $r$ is the ratio $|\Amp_2 / \Amp_1|$, and $\strong$ and $\weak$ are the
relative strong and weak phases, respectively, between $\Amp_2$ and $\Amp_1$.
The \CP asymmetries in this case are
\beq
\Acp(\phi\Kp) = \frac{2 r \sin{\weak} \sin{\strong}}{1 + 2 r \cos{\weak}\cos{\strong} + r^2}
\eeq
and
\beq
\betaeff(\phi\KS) =  \beta + \frac{1}{2}\arctan\Big(\frac{2 r \sin\weak\cos\strong + r^2 \sin (2\weak)} {1 + 2 r \cos \weak \cos\strong  + r^2 \cos(2\weak)} \Big)\,.
\eeq
Note that $\Acp(\phi\KS) = \Acp(\phi\Kp)$ under our assumptions.  However, since
the experimental precision on $\Acp(\phi\KS)$ is very poor compared to
$\Acp(\phi\Kp)$,  we only include the more precise $\Acp(\phi\Kp)$ measurement
in our analysis.
By combining the likelihood scans of $\Acp(\phi\Kp)$ and $\betaeff(\phi\KS)$, we can 
put constraints on $r$, $\strong$, and $\weak$.  
Figure~\ref{fig:pheno} shows
the resulting constraints in the $r$-$\weak$, $r$-$\strong$, 
and $\strong$-$\weak$ planes.

\begin{figure*}[htbp]
\includegraphics[width=8.9cm,keepaspectratio]{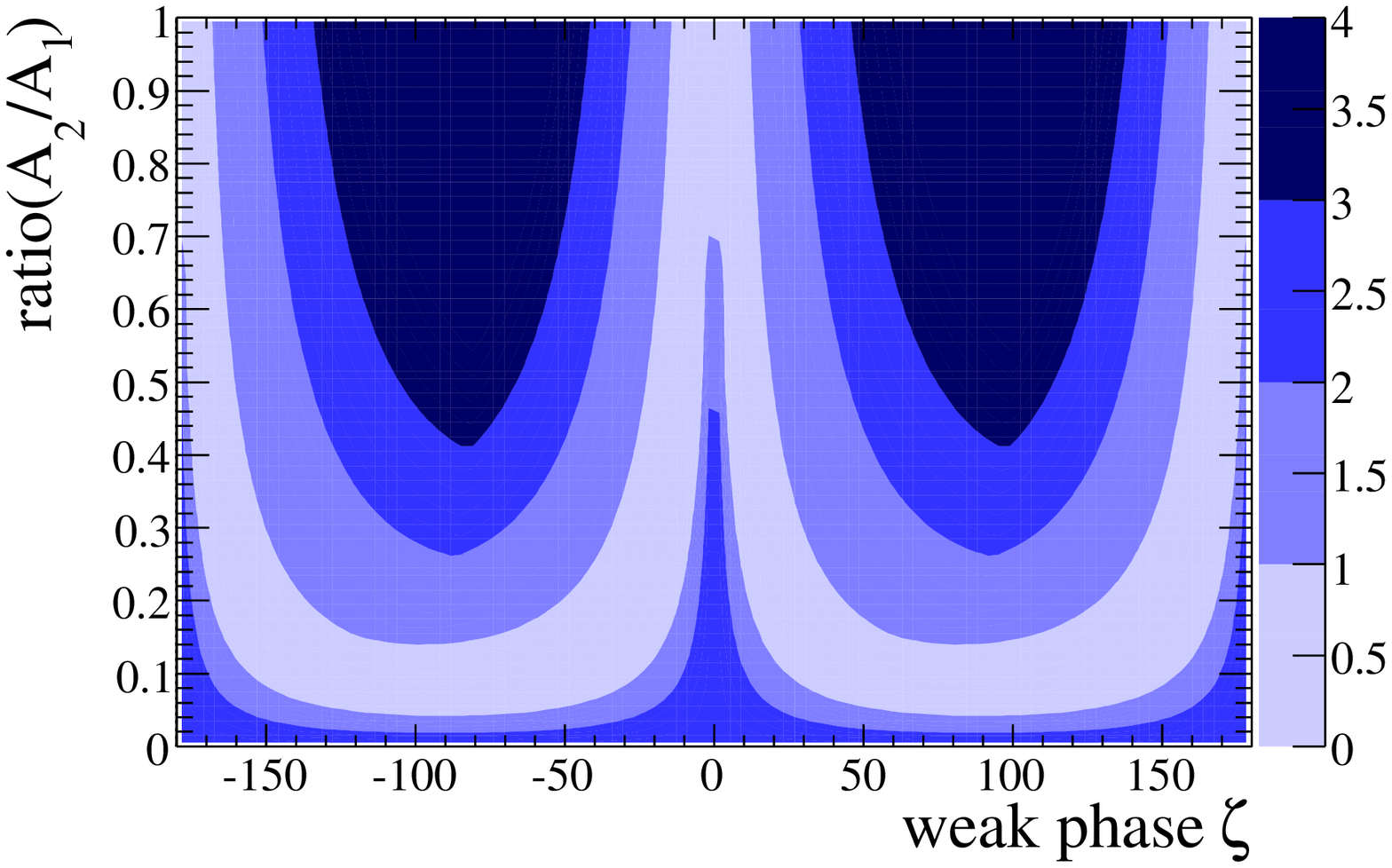}
\includegraphics[width=8.9cm,keepaspectratio]{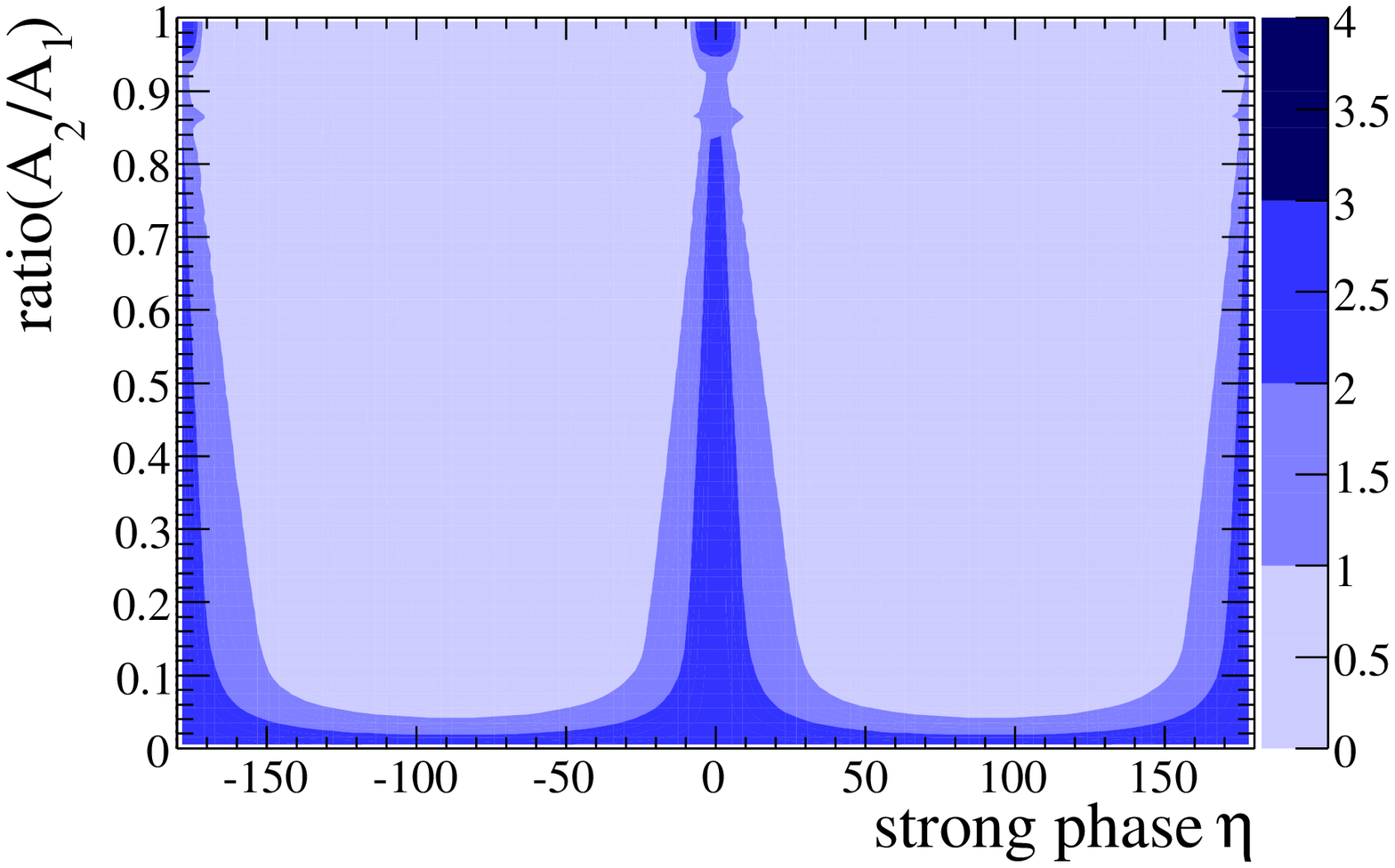}
\includegraphics[width=8.9cm,keepaspectratio]{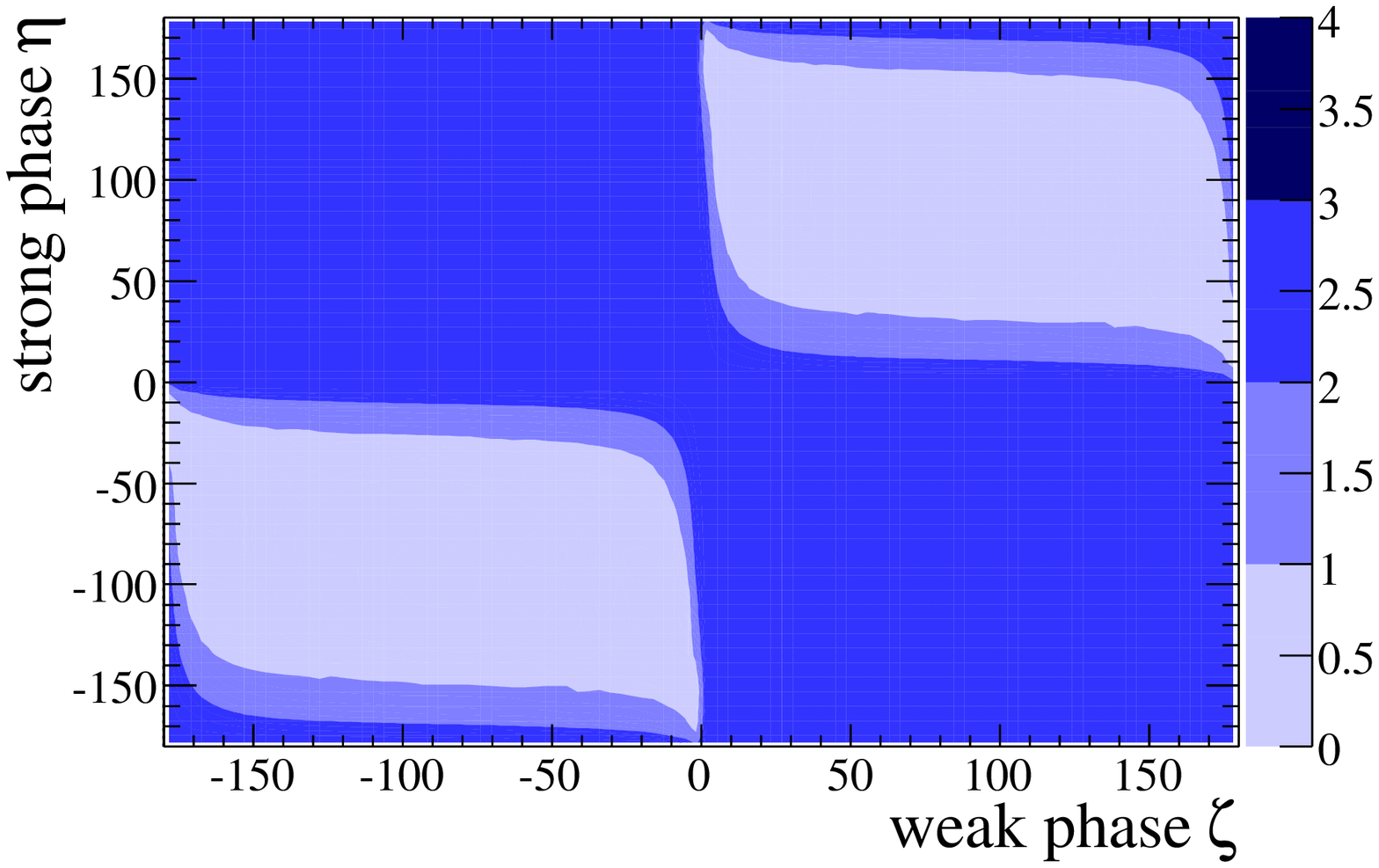}
\caption{\label{fig:pheno}  
Constraints on the amplitude ratio $|\Amp_2 / \Amp_1|$  and the relative strong and weak phases, 
$\strong$ and $\weak$, between $\Amp_1$ and $\Amp_2$, for $\Bp\to\phi\Kp$ and $\Bz\to\phi\KS$ decays.
The shaded areas, from light to dark, show the 
1$\sigma$, 2$\sigma$, $3\sigma$, and $4\sigma$ allowed regions.
}
\end{figure*}

The non-zero value of $\Acp(\phi\Kp)$ leads to $r=0$ being disfavored
by 2.8$\sigma$, with
a value of approximately $0.1$ favored for most values of $\weak$.  
There is little constraint on $\weak$ and $\strong$, except 
that values of $0$ or $\pm 180^\circ$ are disfavored (because 
 $\Acp(\phi\Kp)$ is non-zero), and the first and third quadrants
of the $\strong$-$\weak$ plane are favored (because  $\Acp(\phi\Kp)$
is positive).

%% file: Systematics.tex
\section{SYSTEMATIC UNCERTAINTIES}
\label{sec:Systematics}

The systematic uncertainties for \bkkkboth, \bkksks, and \bkkks parameters are summarized
in Tables~\ref{tab:KKKch_systematics}, \ref{tab:KKsKs_systematics}, 
and \ref{tab:KKKs_systematics}, respectively.  For each
decay mode, the systematic uncertainties are assessed only 
for the best solution.

We vary the masses and widths of the resonances in the signal
model by their errors as given in
Table~\ref{tab:model}. In addition, we vary the Blatt-Weisskopf radii
of any non-scalar resonances, and change the Blatt-Weisskopf radius
of the \B meson from 0 to $1.5~(\gevc)^{-1}$.  We take the observed 
differences
in any fit parameters as systematic uncertainties (listed
in the ``Lineshape'' column in 
Tables~\ref{tab:KKKch_systematics}-\ref{tab:KKKs_systematics}).

We vary any \BB background yields that are fixed in
the nominal fit.  If the \BB class contains only a single decay
mode, the yield is varied according to the uncertainty on the
world average of its branching fraction.  If the \BB class contains
multiple decay modes, then we vary its yield by $50\%$.  The \CP-asymmetries
of the \BB background classes are also varied, either by the
uncertainty on the world average or by a conservative estimate.  
Systematic uncertainties are also assigned due to the limited
sizes of the \BB MC samples, which affects the \BB PDF shapes.  We also
vary signal and continuum background PDF parameters that are fixed
in the nominal fits.  This includes the parameters of the 
$\deltat$ resolution function and the mistag rate.
An additional
systematic uncertainty is contributed by the limited size of
the data sideband sample used to create the continuum DP PDFs.  These 
systematic uncertainties are listed under ``Fixed PDF Params'' in 
Tables~\ref{tab:KKKch_systematics}-\ref{tab:KKKs_systematics}.

Biases in the fit procedure are studied by performing hundreds
of pseudo-experiments using MC events passed through a
GEANT4 detector simulation.  We do not correct for any observed 
biases, but instead assign  systematic uncertainties, listed under
``Fit Bias'' in 
Tables~\ref{tab:KKKch_systematics}-\ref{tab:KKKs_systematics}.

We also study the effect of additional resonances that are
not included in our nominal isobar models (see 
Sec.~\ref{sec:modelDetermination}).  We test for the
$f_0(1370)$, $a_0^0(1450)$, $f_2(1270)$, $f_2(2010)$, and
$f_2(2300)$ in each mode.  We also test for the
$\phi(1680)$ in \bkkkboth and \bkkks, and the $a_0^{\pm}(980)$ 
and $a_0^{\pm}(1450)$ in \bkksks and \bkkks.  
These resonances are modeled by RBW lineshapes, 
except for the $a_0^{\pm}(980)$, which is modeled by a  Flatt\'e
lineshape.  We first fit to data including these additional resonances
in the model.  Then, using this fit result, we generate a large
number of data-sized simulated datasets.  We then fit to these simulated datasets
with and without the additional resonances in the signal model, and
take the observed differences as a systematic uncertainty.  This
is listed as ``Add Resonances'' in 
Tables~\ref{tab:KKKch_systematics}-\ref{tab:KKKs_systematics}.  
In \bkkkboth, the addition of the 
$f_0(1370)$ causes Solution II to be the global mininum rather than 
Solution I, so we do not assign a systematic uncertainty for it.

Additional systematic uncertainties are listed as ``Other'' in
Tables~\ref{tab:KKKch_systematics}-\ref{tab:KKKs_systematics}.
Systematic uncertainties are assessed for tracking efficiency, \KS
reconstruction, and \Kpm PID.  We also compute a systematic uncertainty
due to the limited sizes of the MC samples used to calculate the 
signal efficiency as a function of DP position.  We 
assign a $1\%$ systematic uncertainty due to possible detector charge 
asymmetries
not properly modeled in the detector simulation.
For the \CP-violating parameters in \bkkks, we assign a systematic uncertainty
due to the interference between CKM-favored and CKM-suppressed 
tag-side \B decays~\cite{Long:2003wq}.

\begin{table*}[htbp]
\begin{center}
\caption{ \label{tab:KKKch_systematics}
        Summary of systematic uncertainties for \bkkkboth parameters. Errors on phases, \Acp's, and branching fractions
are given in degrees, percent, and units of $10^{-6}$, respectively.}
\begin{tabular}{l|cccccc}
\hline \hline
Parameter      &    Lineshape  &    Fixed PDF Params  &    Other  &    Add Resonances  &    Fit Bias  &    Total  \\
\hline
$\Delta\phi$($\phiI$)  & $3$   & $1$   & $0$   & $2$   & $2$   & $4$  \\
$\Delta\phi$($\fI$)  & $2$   & $1$   & $0$   & $6$   & $1$   & $6$  \\
$\Delta\phi$($\ftwop$)  & $1$   & $0$   & $0$   & $3$   & $1$   & $3$  \\
$\Delta\phi$($\chiczero$)  & $1$   & $1$   & $0$   & $1$   & $1$   & $2$  \\
\Acp($\phiI$)  & $0.2$   & $0.2$   & $1.0$   & $0.3$   & $0.7$   & $1.3$  \\
\Acp($\fI$)  & $3$   & $1$   & $1$   & $2$   & $1$   & $4$  \\
\Acp($\ftwop$)  & $1$   & $1$   & $1$   & $3$   & $1$   & $4$  \\
\Acp(NR)  & $1.1$   & $0.4$   & $1.0$   & $0.8$   & $0.7$   & $1.9$  \\
${\cal B}$($\phiI$)  & $0.20$   & $0.04$   & $0.11$   & $0.14$   & $0.08$   & $0.29$  \\
${\cal B}$($\fI$)  & $1.2$   & $0.1$   & $0.3$   & $2.5$   & $0.4$   & $2.8$  \\
${\cal B}$($\fII$)  & $0.06$   & $0.02$   & $0.02$   & $0.52$   & $0.02$   & $0.52$  \\
${\cal B}$($\ftwop$)  & $0.05$   & $0.01$   & $0.02$   & $0.07$   & $0.10$   & $0.13$  \\
${\cal B}$($\fIII$)  & $0.08$   & $0.04$   & $0.03$   & $0.49$   & $0.05$   & $0.50$  \\
${\cal B}$($\chiczero$)  & $0.01$   & $0.01$   & $0.03$   & $0.02$   & $0.04$   & $0.06$  \\
${\cal B}$(NR)  & $1.0$   & $0.2$   & $0.5$   & $7.4$   & $0.3$   & $7.6$  \\
${\cal B}$(NR (S-wave))  & $13$   & $2$   & $1$   & $23$   & $2$   & $27$  \\
${\cal B}$(NR (P-wave))  & $10$   & $2$   & $1$   & $25$   & $3$   & $27$  \\
${\cal B}$(Total)  & $0.0$   & $0.2$   & $0.8$   & $0.1$   & $0.4$   & $0.9$  \\
${\cal B}$(Charmless)  & $0.0$   & $0.2$   & $0.8$   & $0.1$   & $0.3$   & $0.9$  \\
\hline \hline
\end{tabular}
\end{center}
\end{table*}

\begin{table*}[htbp]
\begin{center}
\caption{ \label{tab:KKsKs_systematics}
        Summary of systematic uncertainties for \bkksks parameters. Errors on \Acp and branching fractions
are given in percent and units of $10^{-6}$, respectively.}
\begin{tabular}{l|cccccc}
\hline \hline
Parameter      &    Lineshape  &    Fixed PDF Params  &    Other  &    Add Resonances  &    Fit Bias  &    Total  \\
\hline
\Acp  & $0$   & $0$   & $1$   & $0$   & $1$   & $2$  \\
${\cal B}$($\fI$)  & $1.4$   & $0.3$   & $0.3$   & $1.0$   & $0.4$   & $1.8$  \\
${\cal B}$($\fII$)  & $0.05$   & $0.03$   & $0.01$   & $0.57$   & $0.04$   & $0.58$  \\
${\cal B}$($\ftwop$)  & $0.06$   & $0.02$   & $0.02$   & $0.07$   & $0.03$   & $0.10$  \\
${\cal B}$($\fIII$)  & $0.06$   & $0.04$   & $0.01$   & $0.02$   & $0.08$   & $0.11$  \\
${\cal B}$($\chiczero$)  & $0.01$   & $0.01$   & $0.01$   & $0.00$   & $0.03$   & $0.04$  \\
${\cal B}$(NR (S-wave))  & $1.3$   & $0.6$   & $0.4$   & $2.0$   & $0.2$   & $2.5$  \\
${\cal B}$(Total)  & $0.0$   & $0.2$   & $0.2$   & $0.0$   & $0.0$   & $0.3$  \\
${\cal B}$(Charmless)  & $0.0$   & $0.2$   & $0.2$   & $0.0$   & $0.0$   & $0.3$  \\
\hline \hline
\end{tabular}
\end{center}
\end{table*}

\begin{table*}[htbp]
\begin{center}
\caption{ \label{tab:KKKs_systematics}
        Summary of systematic uncertainties for \bkkks parameters. Errors on angles, \Acp's, and branching fractions
are given in degrees, percent, and units of $10^{-6}$, respectively.}
\begin{tabular}{l|cccccc}
\hline \hline
Parameter      &    Lineshape  &    Fixed PDF Params  &    Other  &    Add Resonances  &    Fit Bias  &    Total  \\
\hline
$\betaeff$($\phiI$)  & $2$   & $1$   & $0$   & $2$   & $0$   & $2$  \\ 
$\betaeff$($\fI$)  & $1$   & $1$   & $0$   & $4$   & $0$   & $4$  \\ 
$\betaeff$({\rm other})  & $0.7$   & $0.4$   & $0.2$   & $0.8$   & $0.4$   & $1.2$  \\ 
\Acp($\phiI$)  & $2$   & $2$   & $2$   & $2$   & $3$   & $5$  \\ 
\Acp($\fI$)  & $6$   & $3$   & $2$   & $5$   & $2$   & $9$  \\ 
\Acp({\rm other})  & $1$   & $1$   & $1$   & $2$   & $1$   & $3$  \\ 
${\cal B}$($\phiI$)  & $0.13$   & $0.05$   & $0.08$   & $0.05$   & $0.03$   & $0.18$  \\ 
${\cal B}$($\fI$)  & $1.3$   & $0.3$   & $0.1$   & $2.0$   & $0.1$   & $2.4$  \\ 
${\cal B}$($\fII$)  & $0.04$   & $0.02$   & $0.02$   & $0.10$   & $0.03$   & $0.12$  \\ 
${\cal B}$($\ftwop$)  & $0.02$   & $0.01$   & $0.00$   & $0.15$   & $0.02$   & $0.16$  \\ 
${\cal B}$($\fIII$)  & $0.3$   & $0.1$   & $0.1$   & $0.4$   & $0.1$   & $0.5$  \\ 
${\cal B}$($\chiczero$)  & $0.02$   & $0.02$   & $0.02$   & $0.01$   & $0.04$   & $0.06$  \\ 
${\cal B}$(NR(Total))  & $2$   & $1$   & $1$   & $8$   & $1$   & $9$  \\ 
${\cal B}$(NR (S-wave))  & $2$   & $1$   & $1$   & $8$   & $1$   & $8$  \\ 
${\cal B}$(NR (P-wave))  & $0.1$   & $0.2$   & $0.1$   & $0.3$   & $0.1$   & $0.4$  \\ 
${\cal B}$(Total)  & $0.0$   & $0.4$   & $0.7$   & $0.0$   & $0.1$   & $0.8$  \\ 
${\cal B}$(Charmless)  & $0.1$   & $0.4$   & $0.6$   & $0.0$   & $0.2$   & $0.8$  \\ 
\hline \hline
\end{tabular}
\end{center}
\end{table*}

%% file: summary.tex
\section{SUMMARY}
\label{sec:summary}
We have performed amplitude analyses of the decays \bkkkboth and \bkksks,
and a time-dependent amplitude analysis of \bkkks, using a data
sample of  approximately $470\times 10^6$ \BB decays.  

For \bkkkboth, we find two solutions separated by $5.6$ units of $-2\logL$.
The favored solution has a direct \CP asymmetry in $\Bp\to\phiI\Kp$ of 
$\Acp= (12.8\pm 4.4 \pm 1.3)\%$. A likelihood scan shows that
$\Acp$ differs from 0 by $2.8\sigma$, including systematic 
uncertainties.  This can be compared with
the SM expectation of  $\Acp= (0.0 $-$ 4.7) \%$.  For \bkkks, we find five solutions,
and determine $\betaeff(\phi\KS) = (21\pm 6 \pm 2)^\circ$ from a likelihood scan. 
Excluding the $\phiI\KS$ and $\fI\KS$ contributions, we  measure 
$\betaeff = (20.3\pm 4.3 \pm 1.2)^\circ$
for  the remaining $\bkkks$ decays, and 
exclude the trigonometric reflection $90^\circ - \betaeff$ at $4.8\sigma$,
including systematic uncertainties.
For \bkksks, there is insufficient data to fully constrain the many complex amplitudes in
the DP model.  However, from a likelihood scan we measure an overall direct 
\CP asymmetry of $\Acp = (4 ^{+4}_{-5} \pm 2)\%$.
By combining the $\Acp(\phi\Kpm)$ and $\betaeff(\phi\KS)$ results and assuming isospin symmetry,
we place constraints on the possible SM and NP amplitudes contributing to these decays.

We also study the DP structure of the three \bkkkgeneric modes, by means of an 
angular-moment analysis. This includes the first ever DP analysis of \bkksks. 
To describe the large  nonresonant contributions seen in the three \bkkk modes, we introduce a polynomial
model that includes explicit S-wave and P-wave terms and allows 
for phase motion.
We conclude that the hypothetical particle dubbed the $f_X(1500)$
is not a single scalar resonance, but instead can be described by the sum of the
well-established resonances $\fII$, $\ftwop$, and $\fIII$.

%% file: pubboard/acknowledgements.tex
We are grateful for the 
extraordinary contributions of our \pep2\ colleagues in
achieving the excellent luminosity and machine conditions
that have made this work possible.
The success of this project also relies critically on the 
expertise and dedication of the computing organizations that 
support \babar.
The collaborating institutions wish to thank 
SLAC for its support and the kind hospitality extended to them. 
This work is supported by the
US Department of Energy
and National Science Foundation, the
Natural Sciences and Engineering Research Council (Canada),
the Commissariat \`a l'Energie Atomique and
Institut National de Physique Nucl\'eaire et de Physique des Particules
(France), the
Bundesministerium f\"ur Bildung und Forschung and
Deutsche Forschungsgemeinschaft
(Germany), the
Istituto Nazionale di Fisica Nucleare (Italy),
the Foundation for Fundamental Research on Matter (The Netherlands),
the Research Council of Norway, the
Ministry of Education and Science of the Russian Federation, 
Ministerio de Ciencia e Innovaci\'on (Spain), and the
Science and Technology Facilities Council (United Kingdom).
Individuals have received support from 
the Marie-Curie IEF program (European Union) and the A. P. Sloan Foundation (USA).

%% file: Appendix.tex
\section*{APPENDIX}
\label{sec:appendix}

We give tables of the interference fit fractions $\fitfrac_{jk}$, defined in Eq.~\eqref{eq:interfitfrac}.

\begin{table*}[htbp]
\center
\caption{Values of the interference fit fractions $\fitfrac_{jk}$ for \bkkkboth, Solution I.  The diagonal terms
$\fitfrac_{jj}$ are the ordinary fit fractions $\fitfrac_{j}$, which sum to $272\%$. The NR component is split
into S-wave and P-wave parts for these calculations. Values are given in percent.}
\begin{tabular}{l|cccccccc}
\hline \hline
  & $\phiI$  & $\fI$  & $\fII$  & $\ftwop$  & $\fIII$  & $\chiczero$  & NR (S-wave)  & NR (P-wave) \\
 \hline
   $\phiI$ &  $ 12.9 $   & $ -0.1 $   & $ 0.0 $   & $ 0.0 $   & $ 0.1 $   & $ -0.0 $   & $ -7.4 $   & $ 8.2 $   \\
   $\fI$    &  &  $ 27.2 $   & $ -4.7 $   & $ -0.0 $   & $ -5.4 $   & $ -1.0 $   & $ -0.8 $   & $ -3.7 $   \\
   $\fII$    &     &  &  $ 2.1 $   & $ 0.0 $   & $ 2.3 $   & $ 0.1 $   & $ 3.1 $   & $ -0.8 $   \\
   $\ftwop$    &     &     &  &  $ 2.0 $   & $ 0.1 $   & $ -0.0 $   & $ -0.0 $   & $ 0.7 $   \\
   $\fIII$    &     &     &     &  &  $ 3.2 $   & $ -0.1 $   & $ -13.5 $   & $ 4.9 $   \\
   $\chiczero$    &     &     &     &     &  &  $ 3.2 $   & $ 3.3 $   & $ -1.8 $   \\
   NR (S-wave)    &     &     &     &     &     &  &  $ 151.4 $   & $ -155.0 $   \\
   NR (P-wave)    &     &     &     &     &     &     &  &  $ 69.4 $   \\
\hline  \hline
\end{tabular}
\label{tab:interfracs_KKKch_SolnI}
\end{table*}

\begin{table*}[htbp]
\center
\caption{Values of the interference fit fractions $\fitfrac_{jk}$ for \bkkkboth, Solution II.  The diagonal terms
$\fitfrac_{jj}$ are the ordinary fit fractions $\fitfrac_{j}$, which sum to $174\%$.  The NR component is split
into S-wave and P-wave parts for these calculations. Values are given in percent.}
\begin{tabular}{l|cccccccc}
\hline \hline
  & $\phiI$  & $\fI$  & $\fII$  & $\ftwop$  & $\fIII$  & $\chiczero$  & NR (S-wave)  & NR (P-wave) \\
 \hline
   $\phiI$ &  $ 12.3 $   & $ -0.3 $   & $ -0.1 $   & $ -0.0 $   & $ -0.1 $   & $ -0.1 $   & $ -1.5 $   & $ 5.1 $   \\
   $\fI$    &  &  $ 12.5 $   & $ 1.5 $   & $ 0.1 $   & $ 3.9 $   & $ 0.6 $   & $ -40.6 $   & $ -10.2 $   \\
   $\fII$    &     &  &  $ 2.6 $   & $ -0.0 $   & $ 2.3 $   & $ 0.1 $   & $ -3.5 $   & $ -0.0 $   \\
   $\ftwop$    &     &     &  &  $ 1.5 $   & $ 0.0 $   & $ -0.0 $   & $ -0.3 $   & $ 0.7 $   \\
   $\fIII$    &     &     &     &  &  $ 2.5 $   & $ -0.0 $   & $ -11.6 $   & $ -2.4 $   \\
   $\chiczero$    &     &     &     &     &  &  $ 3.6 $   & $ -1.5 $   & $ 0.5 $   \\
   NR (S-wave)    &     &     &     &     &     &  &  $ 91.1 $   & $ -17.2 $   \\
   NR (P-wave)    &     &     &     &     &     &     &  &  $ 48.2 $   \\
\hline  \hline
\end{tabular}
\label{tab:interfracs_KKKch_SolnII}
\end{table*}

\begin{table*}[htbp]
\center
\caption{Values of the interference fit fractions $\fitfrac_{jk}$ for \bkksks, for the global minimum.  The diagonal terms
$\fitfrac_{jj}$ are the ordinary fit fractions $\fitfrac_{j}$, which sum to $345\%$.  Values are given in percent.}
\begin{tabular}{l|cccccc}
\hline \hline
  & $\fI$  & $\fII$  & $\ftwop$  & $\fIII$  & $\chiczero$  & NR (S-wave) \\
 \hline
   $\fI$ &  $ 139.0 $   & $ -19.2 $   & $ 0.0 $   & $ -12.4 $   & $ -1.0 $   & $ -217.0 $   \\
   $\fII$    &  &  $ 4.0 $   & $ -0.0 $   & $ 4.1 $   & $ 0.2 $   & $ 9.5 $   \\
   $\ftwop$    &     &  &  $ 5.7 $   & $ -0.0 $   & $ -0.0 $   & $ -0.0 $   \\
   $\fIII$    &     &     &  &  $ 4.5 $   & $ 0.1 $   & $ -9.2 $   \\
   $\chiczero$    &     &     &     &  &  $ 5.0 $   & $ -0.0 $   \\
   NR (S-wave)    &     &     &     &     &  &  $ 186.5 $   \\
\hline  \hline
\end{tabular}
\label{tab:interfracs_KKsKs}
\end{table*}

\begin{table*}[htbp]
\center
\caption{Values of the interference fit fractions $\fitfrac_{jk}$ for \bkkks, for the global minimum.  The diagonal terms
$\fitfrac_{jj}$ are the ordinary fit fractions $\fitfrac_{j}$, which sum to $188\%$.   The NR component is split
into S-wave and P-wave parts for these calculations. 
Values are given in percent.}
\begin{tabular}{l|cccccccc}
\hline \hline
  & $\phiI$  & $\fI$  & $\fII$  & $\ftwop$  & $\fIII$  & $\chiczero$  & NR (S-wave)  & NR (P-wave) \\
 \hline
   $\phiI$ &  $ 13.1 $   & $ 0.0 $   & $ 0.0 $   & $ 0.0 $   & $ 0.0 $   & $ 0.0 $   & $ 0.0 $   & $ 0.2 $   \\
   $\fI$    &  &  $ 26.3 $   & $ 0.1 $   & $ -0.0 $   & $ 14.4 $   & $ -0.7 $   & $ -81.2 $   & $ 0.0 $   \\
   $\fII$    &     &  &  $ 2.1 $   & $ -0.0 $   & $ 5.3 $   & $ -0.1 $   & $ -0.7 $   & $ 0.0 $   \\
   $\ftwop$    &     &     &  &  $ 0.5 $   & $ -0.0 $   & $ 0.0 $   & $ 0.0 $   & $ 0.0 $   \\
   $\fIII$    &     &     &     &  &  $ 16.7 $   & $ -0.2 $   & $ -27.0 $   & $ 0.0 $   \\
   $\chiczero$    &     &     &     &     &  &  $ 3.4 $   & $ 1.6 $   & $ 0.0 $   \\
   NR (S-wave)    &     &     &     &     &     &  &  $ 114.5 $   & $ 0.0 $   \\
   NR (P-wave)    &     &     &     &     &     &     &  &  $ 11.7 $   \\
\hline  \hline
\end{tabular}
\label{tab:interfracs_KKKs}
\end{table*}

%% file: prd.bbl
\begin{thebibliography}{28}
\bibitem[{\citenamefont{{Charge-conjugate decays are implied throughout, unless
  otherwise indicated.}}()}]{ChargeConj}
\bibinfo{author}{\bibnamefont{{Charge-conjugate decays are implied throughout,
  unless otherwise indicated.}}}
\end{thebibliography}
